\documentclass[11pt]{article}
\usepackage{amssymb}
\usepackage{graphicx}
\usepackage{amsmath}
\usepackage{a4}
\usepackage{amsfonts}

\newtheorem{theorem}{Theorem}[section]

\newtheorem{definition}[theorem]{Definition}
\newtheorem{example}[theorem]{Example}

\newtheorem{lemma}[theorem]{Lemma}

\newtheorem{proposition}[theorem]{Proposition}
\newtheorem{remark}[theorem]{Remark}

\newenvironment{proof}[1][Proof]{\textbf{#1.} }{\ \rule{0.5em}{0.5em}}
\input{tcilatex}

\begin{document}

\title{A geometric study of many body systems }
\author{Eldar Straume \\
Department of Mathematical Sciences,\\
Norwegian University of Science and Technology (NTNU)\\
N-7491 Trondheim, Norway\\
e-mail: eldars@math.ntnu.no}
\maketitle

\begin{abstract}
An n-body system is a labelled collection of n point masses in a Euclidean
space, and their congruence and internal symmetry properties involve a rich
mathematical structure which is investigated in the framework of equivariant
Riemannian geometry. Some basic concepts are n-configuration, configuration
space, internal space, shape space, Jacobi transformation and weighted root
system. The latter is a generalization of the root system of SU(n), which
provides a bookkeeping for expressing the mutual distances of the point
masses in terms of the Jacobi vectors. Moreover, its application to the
study of collinear central n-configurations yields a simple proof of
Moulton's enumeration formula.  A major topic is the study of matrix spaces
representing the shape space of n-body configurations in Euclidean k-space,
the structure of the m-universal shape space and its O(m)-equivariant linear
model. This also leads to those \textquotedblleft orbital
fibrations\textquotedblright\ where SO(m) or O(m) act on a sphere with a
sphere as orbit space. A few of these examples are encountered in the
literature, e.g. the special case $S^{5}/O(2)\approx S^{4}$ was analyzed
independently by Arnold, Kuiper and Massey in the 1970's.

\ 
\end{abstract}

\tableofcontents

\section{Introduction}

An \emph{n-body system} is defined to be a labelled collection of $n$ point
masses (or particles) $P_{i}$ of mass $m_{i}$ in Euclidean 3-space $\mathbb{R%
}^{3}$, and it is of general importance to find appropriate mathematical
models to describe and analyze such a system. We have in mind few-body
systems as well as many-body systems, ranging from different areas such as
celestial and quantum mechanics or quantum chemistry (n-body problem,
n-atomic molecules etc.). Despite the diversity of the applications they all
share a fundamental underlying mathematical structure, in terms of kinematic
concepts and internal space geometry, and the role of the mass distribution.
Here we shall focus attention on these basic structures, in a modern
geometric and topological setting with orthogonal transformation groups in
the forefront. \ This approach also establishes similar results for point
masses in any Euclidean spaces $\mathbb{R}^{d},d\geq 2$.

In this introductory section we first give an overview of the paper, which
take up several different topics. Then, in the following subsections we
introduce some basic concepts and constructions which we shall return to
later. Here the presentation is rather informal or expository, with some
comments on the history.

\subsection{A brief overview}

Our approach is to combine symmetry and kinematic geometric principles in
the framework of equivariant geometry, exhibiting the importance of the
classical orthogonal transformation groups and the associated orbit space
constructions. This enables us to investigate in a unifying way the notions
of congruence, internal configuration space and internal symmetry group.
Internal symmetries of n-body systems are investigated in Section 2 from a
categorical viewpoint based on pure geometrical principles. As we shall
explain, for various reasons they should be referred to as Jacobi
transformations, but in the physics literature they are also encountered as 
\emph{kinematic rotations} or \emph{democracy transformations}, cf. \cite%
{Littlejohn1}.

Section 3 is largely devoted to the study of the topology of shape spaces in
general, and here it is natural to consider n-body systems in higher
dimensional Euclidean spaces $\mathbb{R}^{d}$ as well. We shall exploit the
fact that congruence and internal symmetry for point masses in $\mathbb{R}%
^{d}$ combine together to a tensor product representation of some $%
O(d)\times O(m)$, acting on the matrix space $M(d,m)\simeq \mathbb{R}^{dm}$
by matrix multiplication. In short, a typical shape space is the orbit space 
$M^{\ast }=S^{p}/O(d)$, where $S^{p}$ is the unit sphere of $M(d,m)$. Now $%
M^{\ast }$ inherits the action of $O(m)$ as its symmetry group, but on the
other hand, $M^{\ast }$ also naturally embeds as an $O(m)$-invariant subset
of the linear space $Sym^{0}(m)$ of symmetric matrices with zero trace. Of
particular interest are the cases where $S^{p}/H\approx S^{q}$ is also a
sphere, $H=SO(m)$ or $O(m)$. They yield an infinite family of "orbital
fibrations" $S^{p}\rightarrow S^{q}$, which for $q=1,2,4$ are the Hopf
fibrations, cf. Section 3.4.2.

In the study of many body systems, we apply the orbit space reduction to the
configuration manifold rather than to its cotangent bundle, as in the
reduction method of Marsden-Weinstein which, for example, Iwai \cite{Iwai1}
applies to the Hamiltonian system describing classical molecular dynamics.
On the other hand, R. Littlejohn and his collaborators (cf. e.g.\cite%
{Littlejohn2}, \cite{Littlejohn1}, \cite{Littlejohn3}) have investigated the
gauge fields that arise on the reduced (i.e. internal) configuration space,
and our work is directly related to the geometric framework of their
investigations.

In Section 4 we determine the geometric invariants of n-body systems, namely
polynomial functions on the configuration space which are both congruence
invariants and internal symmetry (or democracy) invariants. The topic is
certainly well understood in classical invariant theory, but we are also
seeking symmetrized expressions for centered n-body configurations, that is,
their center of mass is fixed at the origin.

In Section 5 we introduce the notion of the \emph{weighted root system} of
an n-body system with a given mass distribution. This is a geometric
invariant which generalizes the notion of a root system of Cartan type $%
A_{n-1}$ in classical Lie theory, and its underlying structure is actually
inherent in various contexts. For example, it encodes the data of the
relative positions of the binary collision varieties in the configuration
space or shape space. We shall introduce it as a bookkeeping device for
expressing the \textit{mutual distances} between the n bodies in terms of
the Jacobi vectors.

In Section 6 we present, as a nice application of the weighted root system,
a simple proof of Moulton's classification of collinear \emph{central
n-configurations} (cf. \cite{Moulton}) which is also well adapted for
numerical computations. Recall that these are the configurations
characterizing n collinear masses capable of a rigid uniform rotation under
the mutual gravitational forces.

The present paper is essentially a preprint (with the same title) of the
author from April 2002, prompted by the paper Hsiang \cite{Hsiang1} and the
succeeding joint work \cite{Hsiang3}. Our Sections 5-6 recall the contents
of \cite{Hsiang3}, but in the present paper the topology of general shape
spaces, the universal shape space and its linear model, is a major topic. It
turns out that some few-body shape spaces have an interesting history in the
literature. For example, it was surprising to find that the quotient space
of complex projective plane $\mathbb{C}P^{2}$ modulo complex conjugation is
topologically a 4-sphere, namely $S^{5}/O(2)\approx S^{4}.$ Several
independent and different proofs of this fact had already been published
(cf. Arnold \cite{Arnold1}, \cite{Arnold2}, Kuiper\cite{Kuiper}, Massey\cite%
{Massey}). \ Now we also find that our Section 3 has some overlapping with
the more recent paper Atiyah-Berndt \cite{Atiyah}.

\subsection{n-configuration space, internal space and shape space}

The location of an n-body system is conveniently represented by its \emph{%
n-configuration }$\mathbf{X\ =(a}_{1},\mathbf{a}_{2},..,\mathbf{a}_{n})$
where $\mathbf{a}_{i}$ is the position vector of $P_{i}$. The n-tuple $%
\mathbf{X}$ is regarded as a vector in the \emph{free} \emph{n-configuration}
\emph{space} 
\begin{equation}
\text{ }\ \hat{M}_{n}=\mathbb{R}^{3}\times \mathbb{R}^{3}\times ....\times 
\mathbb{R}^{3}\ \simeq \mathbb{R}^{3n}\text{,}  \label{totalconf}
\end{equation}%
namely a Euclidean 3n-space with an orthogonal splitting reflecting the
individual positions of the n point masses, which may possibly coincide. The
mass distribution $(m_{1},m_{2},..,m_{n})$ is tacitly assumed to be fixed
unless otherwise stated, and in general we assume $m_{i}>0$. The \emph{%
centered n-configuration} \emph{space} is the subspace of dimension $3n-3$ 
\begin{equation}
M_{n}\subset \hat{M}_{n}:\sum m_{i}\mathbf{a}_{i}=0,  \label{centered}
\end{equation}%
consisting of those n-configurations with its center of mass at the origin. $%
\hat{M}_{n}$ has the following mass dependent (Jacobi) \emph{kinematic metric%
}, namely the inner product of $\mathbf{X}$ and $\mathbf{Y=(b}_{1},\mathbf{b}%
_{2},..,\mathbf{b}_{n})$ is 
\begin{equation}
\left\langle \mathbf{X,Y}\right\rangle =\sum m_{i}\mathbf{a}_{i}\cdot 
\mathbf{b}_{i}\text{ }  \label{innerprod1}
\end{equation}%
The isometry group of $\hat{M}_{n}$ is the associated Euclidean group 
\begin{equation}
E(\hat{M}_{n})=\hat{M}_{n}\hat{\times}O(\hat{M}_{n})\simeq \mathbb{R}^{3n}%
\hat{\times}O(3n)=E(3n)  \label{Euclidean}
\end{equation}%
which we have expressed in the usual way as a semidirect product of the
subgroups of translations and orthogonal transformations respectively.

Two n-configurations $\mathbf{X}$ and $\mathbf{Y}$ are regarded as \emph{%
congruent} if their i-th components $\mathbf{a}_{i}$ and $\mathbf{b}_{i}$
differ by the same Euclidean motion in $\mathbb{R}^{3}$ for each i. Thus the
congruence relation is defined by the natural (diagonal) $E(3)$-action on $%
\hat{M}_{n}$, by which $E(3)$ (resp. $O(3))$ is embedded as a subgroup of $%
E(3n)$ (resp. $O(3n)).$ The space $\bar{M}_{n}$ of congruence classes is the
(\emph{congruence) moduli space}, also referred to as the \emph{internal
space. }It may be viewed as the result of the two-step orbit space
constructionb 
\begin{equation}
\hat{M}_{n}\rightarrow \frac{\hat{M}_{n}}{\mathbb{R}^{3}}\simeq
M_{n}\rightarrow \frac{M_{n}}{O(3)}=\frac{\hat{M}_{n}}{E(3)}=\bar{M}_{n}
\label{2-step}
\end{equation}%
where in the first step the translation-reduced space is identified with $%
M_{n}$. The second step is the orbit space construction of the
transformation group ($O(3),M_{n})$, from which it follows that $\bar{M}_{n}$
is a stratified manifold of dimension $3n-6$ (cf. Section 3).

Next, let us divide \textquotedblright congruence\textquotedblright\ into
\textquotedblright shape\textquotedblright\ and \textquotedblright
size\textquotedblright\ and use the squared norm function $I$ (cf. (\ref%
{kinemat1})), namely the polar moment of inertia, as a measure of the size
of an n-configuration (or of its congruence class). Then $\bar{M}_{n}$ has
the structure of a cone, where each ray emanates from the cone vertex (or
base point) $O=(I=0)$ and the ray represents a fixed shape. The \emph{shape
space }is the space of rays, which we may regard as the orbit space 
\begin{equation}
\frac{\bar{M}_{n}-\left\{ 0\right\} }{\mathbb{R}^{+}}=\frac{M_{n}-\left\{
0\right\} }{O(3)\times\mathbb{R}^{+}}\simeq M_{n}^{\ast}  \label{shape0}
\end{equation}
where $O(3)\times\mathbb{R}^{+}$ is the group of similarity transformations
of Euclidean 3-space with the induced (diagonal) action on $M_{n}$. However,
each ray has a unique point where $I=1$, and therefore it is more convenient
to identify $M_{n}^{\ast}$ with the subset of $\bar{M}_{n}$ consisting of
classes of unit size $I=1$. Thus, the internal space $\bar{M}_{n}$ is
naturally a cone over the shape space $M_{n}^{\ast}$.

\subsection{Geometrization, symmetry and reduction}

\bigskip The viewpoint that kinematics is a geometric discipline has a long
history. For an n-body motion $t\rightarrow\mathbf{X(}t\mathbf{)}$, the
fundamental kinematic quantities are 
\begin{align}
I & =\left\vert \mathbf{X}\right\vert ^{2}=\sum m_{i}\left\vert \mathbf{a}%
_{i}\right\vert ^{2}  \notag \\
T & =\frac{1}{2}\left\vert \mathbf{\dot{X}}\right\vert ^{2}=\frac{1}{2}\sum
m_{i}\left\vert \mathbf{\dot{a}}_{i}\right\vert ^{2}  \label{kinemat1} \\
\mathbf{\Omega} & =\mathbf{X\times\dot{X}=}\sum m_{i}\mathbf{a}_{i}\times%
\mathbf{\dot{a}}_{i}  \notag \\
\mathbf{p} & \text{ }\mathbf{=}\sum m_{i}\mathbf{\dot{a}}_{i}  \notag
\end{align}
These are the moment of inertia, kinetic energy, angular and linear
momentum, respectively. In particular, the \emph{hyperradius }$\rho=\sqrt{I}$%
\ is the norm of $\mathbf{X}$ in the n-configuration space (\ref{totalconf})
with respect to the inner product (\ref{innerprod1}), and using $T$ the same
metric may be presented as the \emph{kinematic Riemannian metric}\textit{\ } 
\begin{equation}
ds^{2}=2Tdt^{2}=\sum m_{i}(dx_{i}^{2}+dy_{i}^{2}+dz_{i}^{2})  \label{metric1}
\end{equation}

Furthermore, dynamics was incorporated in this geometric setting by the
classical geometrization procedure which dates at least back to the early
19th century. Let us briefly recall the basic idea behind this, namely
Jacobi's reformulation of Lagrange's least action principle, which goes as
follows. When dynamics is taken into account and the above n-body motion is
due to a force field $\nabla U$ derived from the potential energy $-U$, the
total energy $h=T-U$ is conserved and the metric (\ref{metric1}) should be
conformally modified to the following\ \emph{dynamical metric} \emph{\ } 
\begin{equation}
ds_{h}^{2}=(U+h)ds^{2}  \label{dynmetric}
\end{equation}
depending on a fixed energy level $h.$ Then the trajectories of total energy 
$h$ can be recovered as the geodesics of this metric. This applies, for
example, to physically important n-body systems with potential functions
(such as the Newtonian or Coulomb potential) depending on the pairwise
distances $r_{ij}$ $=\left\vert \mathbf{a}_{i}-\mathbf{a}_{j}\right\vert $.
Therefore, they are invariant under the Euclidean group $E(3)$, and
consequently both linear and angular momentum of the motion are conserved.

Since the early days it has been an important issue how to fully utilize
conservation laws. Sometimes these are first integrals associated with
symmetry groups, and the reduction of integration problems using continuous
(or infinitesimal) symmetries dates back to Sophus Lie's work in the 1870's.

Clearly, the most effective usage of the invariance of linear momentum $%
\mathbf{p}$ is to choose an inertial frame at the center of mass, thereby
reducing the configuration space $\hat{M}_{n}$ to its subspace $M_{n}$ (\ref%
{centered}) and hence the associated linear momentum vanishes. A much harder
problem is to further use the invariance of $\mathbf{\Omega}$, which is
related to the congruence action of $O(3)$ (or $SO(3))$ on $M_{n}$ and its
cotangent bundle. Namely, if $T^{\ast}$ denotes the cotangent bundle
construction, consider the two ways of reducing the phase space 
\begin{equation*}
T^{\ast}(\frac{M_{n}}{O(3)})=T^{\ast}\bar{M}_{n}\text{ , \ \ \ \ }\frac{%
T^{\ast}(M_{n})}{O(3)}
\end{equation*}
to dimension $6n-12$ and $6n-9$, respectively. This suggests that the orbit
space reduction is most effective at the level of the configuration space
rather than the phase space.

In this paper we shall focus attention on two main issues. Firstly, it is
important to construct appropriate coodinates for $M_{n}$, which is actually
a problem with no canonical or generally \textquotedblright
best\textquotedblright\ solution. However, from the representation theory of 
$O(3)$ and natural guidelines suggested by the splitting or invariance of
the kinematic quantities, we are led to a natural approach whose origin may,
in fact, be ascribed to Jacobi. Secondly, we inquire about the structure of
the internal space $\bar{M}_{n}$, as an orbit space of $M_{n}$ modulo $O(3).$
It has the natural \emph{kinematic metric} 
\begin{equation}
d\bar{s}^{2}=2\bar{T}dt^{2}  \label{metric1b}
\end{equation}
where $\bar{T}=T-T^{\omega}$ and $T^{\omega}$ is the kinetic energy due to
purely rotational motion. Indeed, this metric is Riemannian and coincides
with the induced orbital distance metric (see e.g. \cite{Straume1}), and the
reduction map $M_{n}\rightarrow\bar{M}_{n}$ is a (stratified) Riemannian
submersion. As a consequence, the geodesics in $\bar{M}_{n}$ are the image
curves of those geodesics (i.e., linear motions of constant speed) in ($%
M_{n},ds^{2})$ with vanishing angular momentum.

On the other hand, the appropriate reduction of the dynamical equations in $%
M_{n}$ to the level of $\bar{M}_{n}$ is generally a hard problem even today.
In $M_{n}$ the geometrization procedure yields the conformal modification of
the kinematic metric (\ref{metric1}) leading to the dynamical metric (\ref%
{dynmetric}). Similarly, one may search for a similar geometrization
procedure at the internal space level, which should yield a dynamical
Riemannian metric on $\bar{M}_{n}$ of type 
\begin{equation}
d\bar{s}_{h,\Omega}^{2}=F(U,h,\mathbf{\Omega)}d\bar{s}^{2}
\label{dynmetric2}
\end{equation}
in analogy with (\ref{dynmetric}). However, this only works in special
cases, for example, for the (classical) planar n-body problem, but we leave
this topic here.

Actually, in his study of celestial mechanics Jacobi himself abandoned the
above Riemannian geometric approach in favor of the increasingly successful
Hamiltonian formalism where, at the phase space level (symplectic geometry),
ideas involving symmetry, conservation laws and integrability questions have
been continuously developed up to present time. On the other hand, with the
modern techniques of equivariant geometry, Lie transformation groups and
related reduction theory and quotient constructions, the framework of
Riemannian geometry and modern differential geometric techniques are now
more readily applicable for the study of n-body problems. We also refer to
Littlejohn-Reinsch\cite{Littlejohn1} for a general discussion with many
references to the gauge kinematic and dynamics of many particle systems. For
real historical background information on the geometrization of physics we
also propose the survey article L\"{u}tzen\cite{Lutzen} on the 19th century
interactions between mechanics and differential geometry.

\subsection{Congruence and internal symmetry in n-body spaces}

Let us explain the interaction of the notions of congruence\emph{\ }and
internal symmetry\emph{\ }for n-body systems, where by congruence we mean $%
O(3)$-congruence and assume the translational degrees of freedom have been
eliminated. For this purpose we introduce a notion somewhat more general
than a free m-configuration space (\ref{totalconf}), namely an \emph{m-body
space} 
\begin{equation*}
M\simeq\mathbb{R}^{3m}:(O(3),m\rho_{3})
\end{equation*}
is a Euclidean 3m-dimensional space with an orthogonal representation of
type $(O(3),m\rho_{3})$, that is, $m$ copies of the standard representation $%
\rho_{3}$ of $O(3)$. However, particles and mass distributions are not
mentioned, and there is no specific decomposition into 3-dimensional
invariant subspaces $\mathbb{R}^{3}$ as in (\ref{totalconf}). As an example,
we have in mind the centered n-configuration space $M_{n}$ (\ref{centered}),
where $m=n-1\ $and the original mass distribution is disguised in the
Euclidean metric (\ref{innerprod1}).

On the other hand, associated with $\mathbb{R}^{3m}$ is the totality $%
\mathcal{D(}\mathbb{R}^{3m})$ of all orthogonal, $O(3)$-invariant and
ordered decompositions 
\begin{equation}
\mathbb{R}^{3m}=V_{1}\oplus V_{2}\oplus \text{ . . .}\oplus V_{m}\text{ , \ }%
V_{i}\simeq \mathbb{R}^{3}.  \label{gauge}
\end{equation}%
Then, for a specific choice of decomposition we can identify $\mathbb{R}%
^{3m} $ with the free m-configuration space $\hat{M}_{m}$ as in (\ref%
{totalconf}) and associate a particle $P_{i}$ with position vector $\mathbf{a%
}_{i}$ to each summand $V_{i}$. Moreover, the metric on $\mathbb{R}^{3m}$
viewed as a metric on $\hat{M}_{m}\ $should have the form (\ref{innerprod1}%
), so we also need a mass distribution. However, by suitably scaling of the
vectors $\mathbf{a}_{i}$ we are actually free to choose any mass
distribution since a vector in $\mathbb{R}^{3m}$ written as an
m-configuration $(\mathbf{b}_{1}\mathbf{,b}_{2}\mathbf{,..,b}_{m})$
determines position vector $\mathbf{a}_{i}$ and mass $m_{i}$ modulo the
constraint $\mathbf{b}_{i}=\sqrt{m_{i}}\mathbf{a}_{i}$ for each $i$ (see
Section 2.1).

In any case, $\mathbb{R}^{3m}$ has the given \emph{conguence group} $%
O(3)\subset O(3m)$, and we define the \emph{internal }(or inner) symmetries
to be those transformations $\varphi\in O(3m)$ which commute with $O(3)$,
that is, $\varphi\psi=$ $\psi\varphi$ for each $\psi\in O(3).$ By standard
representation theory (cf. Schur's lemma) these\ $\varphi$ constitute a
subgroup isomorphic to $O(m),$ which we shall refer to as the (internal) 
\emph{symmetry group. }The two subgroups of $O(3m)$ intersect at $\mathbb{Z}%
_{2}=\left\{ \pm Id\right\} $ and hence combine to a subgroup 
\begin{equation*}
\frac{O(3)\times O(m)}{\mathbb{Z}_{2}}\rightarrow O(3m)
\end{equation*}
which is also described mathematically by the tensor product $%
\rho_{3}\otimes\rho_{m}$ of their standard representations.

We also point out that $O(m)$ acts naturally on the set of decompositions (%
\ref{gauge}). This action is, in fact, transitive and has isotropy group of
type $O(1)^{m}$, and hence establishes a 1-1-correspondence 
\begin{equation*}
\mathcal{D(}\mathbb{R}^{3m})\simeq\frac{O(m)}{O(1)^{m}}
\end{equation*}

To be more explicit, let us fix some decomposition $(\ref{gauge})$ and
choose orthonormal bases to identify each $V_{i}\ $with $\mathbb{R}^{3}$,
and hence identify $\mathbb{R}^{3m}$ in (\ref{gauge}) with the space $M(3,m)$
of real $3\times m$- matrices $\ $ 
\begin{equation}
X=(\mathbf{x}_{1}\mathbf{,x}_{2}\mathbf{,..,x}_{m})\   \label{X}
\end{equation}
with the standard Euclidean norm square 
\begin{equation}
\left\vert X\right\vert ^{2}=trace(X^{t}X)=\sum_{i=1}^{m}\left\vert \mathbf{x%
}_{i}\right\vert ^{2}\ ,  \label{norm1}
\end{equation}
where $\mathbf{x}_{i}\in$ $V_{i}$ is the i-th column vector of $X.$ Then the
action of $\psi\in O(3)$ and $\varphi\in O(m)$ is just matrix multiplication
on the left and right side respectively, inducing a joint left action on
matrices by 
\begin{equation}
(\psi,\varphi)X=\psi X\varphi^{-1}\text{ (matrix multiplication)}
\label{action}
\end{equation}
Thus congruence and symmetry combined together is the following tensor
product representation 
\begin{equation}
(O(3)\times O(m),\rho_{3}\otimes\rho_{m},M(3,m)),  \label{tensor}
\end{equation}
whose orbit structure will be analyzed by combining two consecutive orbit
space constructions 
\begin{equation}
M(3,m)\rightarrow\frac{M(3,m)}{O(3)}=\bar{M}(3,m)\rightarrow\frac{\bar {M}%
(3,m)}{O(m)}=\frac{M(3,m)}{O(3)\times O(m)}\simeq C(\Delta )\ \ 
\label{twostep}
\end{equation}

Of primary interest is the first orbit space, namely\ the \emph{internal} 
\emph{space} $\bar{M}=\bar{M}(3,m)\ $with its induced metric structure. The
symmetry group $O(m)$ descends faithfully to an induced transformation group
on $\bar{M}$ which is, in fact, its isometry group. Thus, the last step in (%
\ref{twostep}) yields the final orbit space which is geometrically the
Riemannian cone over a spherical triangle $\Delta $ (resp. a circular arc if 
$m=2):$ 
\begin{equation}
C(\Delta )=\left\{ 
\begin{array}{cc}
\mathbb{R}^{3}/B_{3} & \text{if }m\geq 3 \\ 
\mathbb{R}^{2}/B_{2} & \text{if }m=2%
\end{array}%
\right. \   \label{orbitorbitspace}
\end{equation}%
Of course, this is a Euclidean cone. However, as indicated it is also the
fundamental domain of the Weyl group $B_{3\text{ }}($resp. $B_{2}$ if $m=2)$
in classical Lie theory, and this tells us that $C(\Delta )$ embeds
isometrically into $M(3,m)$ as a cross section of the transformation group (%
\ref{tensor}). Namely, $C(\Delta )$ hits every orbit of that group at a
unique point and, moreover, it is perpendicular to every orbit.

We refer to Section 3 for further analysis of the above orbit spaces, in a
more general setting involving all matrix spaces $M(d,m)$ with the
transformation groups $O(d)\times O(m),$ for any $d\geq1$.

\subsection{Jacobi vectors and the centered configuration space}

In dynamics it is the conservation of linear momentum $\mathbf{p}$ that
enables one to reduce the n-body problem to an (n-1)-body problem plus a
trivial 1-body problem for the\ motion of the center of mass. To explain
this, consider the canonical orthogonal and $O(3$)-invariant decomposition
of the free n-configuration space 
\begin{equation}
\hat{M}_{n}=M_{n}\oplus \Delta \mathbb{R}^{3}\text{ , cf. }(\ref{totalconf})
\label{splitting}
\end{equation}%
where the subspace $M_{n}\simeq \mathbb{R}^{3n-3}$ is the \emph{centered}
n-configuration space\ (\ref{centered}) and the "diagonal" $\Delta \mathbb{R}%
^{3}=\left\{ \mathbf{(a},\mathbf{a},..,\mathbf{a})\right\} $ is its
orthogonal complement. For a motion $\mathbf{X(}t\mathbf{)}$ in $\hat{M}_{n}$
the vector $\mathbf{a=a(}t\mathbf{)}$ is, indeed, the center of mass of $%
\mathbf{X(}t\mathbf{)}$. Now, conservation of $\mathbf{p}$ means our
inertial frame of reference will remain inertial if we translate the frame
so that $\mathbf{a}$ becomes the new origin. Therefore, with respect to the
new frame, the motion will take place in the summand $M_{n}$ in (\ref%
{splitting}). This simple reduction is a key step in the integration of the
classical Kepler problem, where $n=2$ and $M_{2}\ \simeq \mathbb{R}%
^{3}\simeq \hat{M}_{1}$ is the configuration space of a ficticious\ 1-body
system.

However, for $n>2$ $M_{n}$ is just an (n-1)-body space and there is no
canonical way of further decomposing $M_{n}$ to become the configuration
space of $n-1$ ficticious particles with appropriate position vectors and
mass distribution, cf. (\ref{gauge}). This fact is clearly reflected by the
variety of types of coordinates for $M_{n}$ which can be found in the
literature. Recall, for example, the efforts of Lagrange, Jacobi and
Delaunay who constructed their own \textquotedblright
good\textquotedblright\ coordinates to study the classical $3$-body problem
(cf. e.g. Marchal $\cite{Marchal}).$

Of particular interest to us is Jacobi's approach, which by repeated
applications generalizes to $n$ bodies, but again there is no canonical way
of doing so. Anyhow, our interpretation of his basic idea is that a solution
of the above splitting problem amounts to the construction of a
transformation $\emph{\ }$ \emph{\ } 
\begin{align}
\Psi & :\ M_{n}\rightarrow\hat{M}_{n-1}\ \   \label{coordinate map} \\
\ & :\mathbf{(a}_{1},\mathbf{a}_{2},..,\mathbf{a}_{n})\text{ }\rightarrow(%
\mathbf{x}_{1}\mathbf{,x}_{2}\mathbf{,..,x}_{n-1})\ \text{ }  \notag
\end{align}
with the \textquotedblright appropriate\textquotedblright\ properties (see
below), connecting the (n-1)-body space $M_{n}$ (with the Jacobi metric (\ref%
{innerprod1})) to a\emph{\ standard model}, namely the free configuration
space $\hat{M}_{n-1}$ with all masses equal to 1.

By viewing the vectors $\mathbf{x}_{i}\in\mathbb{R}^{3}$ as the columns of a
matrix (\ref{X}) we may identify $\hat{M}_{n-1}$ with the matrix space 
\begin{equation}
M(3,n-1)=\mathbb{R}_{\ }^{3}\otimes\mathbb{R}^{n-1}=\ \sum_{i=1}^{n-1}\oplus%
\mathbb{R}_{i}^{3}=\sum_{j=1}^{3}\oplus\mathbb{R}_{j}^{n-1}  \label{decomp1}
\end{equation}
with the norm as in (\ref{norm1}), congruence group $O(3)$, symmetry group\ $%
O(n-1)$ and their joint tensor product representation of $O(3)\times O(n-1)$%
, see (\ref{action}), (\ref{tensor}). In (\ref{decomp1}) we have also
indicated that the column and row vectors of a matrix belong to two
different Euclidean spaces, having the standard action of $O(3)$ and $O(n-1)$
respectively.

Now, what are those \textquotedblright appropriate\textquotedblright\
properties Jacobi transformations such as $\Psi$ should satisfy ? In short,
the answer is that $\Psi$ is just an $O(3)$-equivariant isometry, that is, 
\begin{equation}
i)\text{ }\Psi(g\mathbf{X)}=g\Psi(\mathbf{X)}\text{, }g\in O(3),\text{ \ }ii)%
\text{\ }\left\vert \mathbf{X}\right\vert =\left\vert \Psi (\mathbf{X)}%
\right\vert \text{\ \ }  \label{isometry1}
\end{equation}
Let $\Psi$ in (\ref{coordinate map}) be a given transformation of this type.
It associates to a centered n-configuration\textbf{\ }$\mathbf{X}$ $=\mathbf{%
(a}_{1},\mathbf{a}_{2},..,\mathbf{a}_{n})$ its \emph{Jacobi vector matrix\ }$%
\mathbf{\ }$%
\begin{equation*}
X=\Psi(\mathbf{X)}=(\mathbf{x}_{1},\mathbf{x}_{2},..,\mathbf{x}_{n-1})\in
M(3,n-1)
\end{equation*}
whose column vectors $\mathbf{x}_{i}$ will be referred to as the
corresponding \emph{Jacobi vectors. }On the other hand, if $\Psi^{\prime}$
is another Jacobi transformation, then the composition $\Psi^{\prime}\circ%
\Psi^{-1}=\varphi$ is still an $O(3)$-equivariant isometry. Consequently, we
can write $\Psi ^{\prime}=\varphi\circ\Psi$ where 
\begin{equation}
\varphi:M(3,n-1)\rightarrow M(3,n-1)  \label{isometry2}
\end{equation}
is a Jacobi transformation of the standard model (\ref{decomp1}), namely an
orthogonal transformation commuting with $O(3).$ In other words, $\varphi$
belongs to the symmetry group $O(n-1)$ (acting on the matrices $X$ by right
multiplication). This explains the non-uniqueness of $\Psi$! Briefly, by
knowing only one of the $\Psi^{\prime}s$ we obtain all of them by composing $%
\Psi$ with any $\varphi\in O(n-1)$.

So far, however, we have not constructed a single $\Psi$ in (\ref{coordinate
map}), but we have just seen that this is all we need to do. We refer to
Section 2.2 for the explicit construction of our \emph{stan}$\emph{dard}$
choice $\Psi=\Psi_{0}$ of Jacobi transformation and hence also our preferred
choice of Jacobi vectors\emph{\ }$\mathbf{x}_{i}$ as distinguished linear
combinations of the vectors $\mathbf{a}_{j}$.

\begin{remark}
Quantities which are invariant under the transformation group $%
(O(n-1),M_{n}) $ are also insensitive to the different orderings of the $n$
bodies. Therefore, this group is sometimes referred to as the \emph{democracy%
} group. Elements $\varphi\in SO(n-1)$ are also named \emph{kinematic
rotations}. See e.g. $\cite{Littlejohn1}.$
\end{remark}

\section{Jacobi transformations from a categorical viewpoint}

Loosely speaking, there is the category of m-body spaces whose objects are
the Euclidean spaces $M\simeq \mathbb{R}^{3m}$ with an orthogonal
transformation group of type $(O(3),m\rho _{3})$, and the morphisms will be
called \emph{Jacobi transformations, }namely they are the $O(3)$-equivariant
isometries $M\rightarrow M^{\prime }$ between the spaces. \ Here the matrix
space $M(3,m)=\hat{M}_{m}$ is a distinguished object which also serves as
our \emph{standard model} (\ref{decomp1}). Therefore, for a given m-body
space $M$ a \textquotedblright good\textquotedblright\ coordinate system for 
$M$ amounts to the choice of an appropriate Jacobi transformation \ 
\begin{equation*}
\Psi :M\rightarrow M(3,m)
\end{equation*}%
and the columns of the matrix $X\in M(3,m)$ will be the Jacobi vectors with
respect to $\Psi $. We are primarily interested in the centered
n-configuration space $M=M_{n}$ (where $m=n-1)$.

In the following two subsections we shall characterize \emph{Jacobi
transformations} in several equivalent ways, and we start with the classical
Jacobi construction which also justifies our usage of terminology.

\subsection{Transformations of n-body systems and Jacobi's approach}

In the Introduction we actually started (less formally) with the category of
n-body systems, whose objects are n-tuples $(P_{1},P_{2},...,P_{n})$ of
point masses $P_{i}=(\mathbf{a}_{i},m_{i})$ with $\mathbf{a}_{i}$ and $m_{i}$
as position vector and mass respectively. In this setting, what should be
the appropriate morphisms 
\begin{equation}
\Phi :(P_{1},...,P_{n})\rightarrow (P_{1}^{\prime },...,P_{n}^{\prime })%
\text{ \ ? }  \label{p-morphism}
\end{equation}%
After all, the usefulness of such transformations with the desired
properties, to simplify the further analysis, is well documented in both
classical and quantum mechanics. First of all, we propose to consider
transformations 
\begin{equation}
\Phi :%
\begin{array}{c}
(\mathbf{a}_{1},....,\mathbf{a}_{n})\rightarrow (\mathbf{a}_{1}^{\prime
},....,\mathbf{a}_{n}^{\prime }) \\ 
(m_{1},...,m_{n})\rightarrow (m_{1}^{\prime },...,m_{n}^{\prime })%
\end{array}%
\   \label{morphism}
\end{equation}%
where the first map in (\ref{morphism}) is an invertible linear
transformation on\ $\hat{M}_{n}=\mathbb{R}^{3n}$ which may possibly depend
on the masses $m_{i}$, with the choice of "new" masses $m_{i}^{\prime }$
constrained in some way. But $\Phi $ also transforms the motions of n-body
system and hence also the first three basic kinematic quantities in (\ref%
{kinemat1}), so let us demand that $\Phi $ preserves them, namely 
\begin{align}
I& =\ \sum m_{i}\left\vert \mathbf{a}_{i}\right\vert ^{2}=\sum m_{j}^{\prime
}\left\vert \mathbf{a}_{j}^{\prime }\right\vert ^{2}  \notag \\
2T& =\sum m_{i}\left\vert \mathbf{\dot{a}}_{i}\right\vert ^{2}=\sum
m_{j}^{\prime }\left\vert \mathbf{\dot{a}}_{j}^{\prime }\right\vert ^{2}
\label{kinemat2} \\
\mathbf{\Omega }& =\sum m_{i}\mathbf{a}_{i}\times \mathbf{\dot{a}}_{i}=\sum
m_{j}^{\prime }\mathbf{a}_{j}^{\prime }\times \mathbf{\dot{a}}_{j}^{\prime }
\notag
\end{align}

Motivated by an idea of Jacobi,\textit{\ }we define the \emph{(i,j)-basic
Jacobi transformation} for $i\neq j$ \textit{\ } 
\begin{equation}
\rho_{ij}:(P_{1},...,P_{n})\rightarrow(P_{1}^{\prime},...,P_{n}^{\prime})
\label{Jacobi1}
\end{equation}
by demanding $P_{k}^{\prime}=P_{k},m_{k}^{\prime}=m_{k}$ for $k\neq i,j$,
and 
\begin{align}
P_{i}^{\prime} & =(\mathbf{a}_{i}^{\prime},m_{i}^{\prime})=(\mathbf{a}_{i}-%
\mathbf{a}_{j},\frac{m_{i}m_{j}}{m_{i}+m_{j}})  \label{Jacobi2} \\
P_{j}^{\prime} & =(\mathbf{a}_{j}^{\prime},m_{j}^{\prime})=(\frac {m_{i}%
\mathbf{a}_{i}+m_{j}\mathbf{a}_{j}}{m_{i}+m_{j}},m_{i}+m_{j})  \notag
\end{align}
In particular, $P_{j}^{\prime}$ is the center of mass of $P_{i}\ $and $P_{j}$%
, and $m_{i}^{\prime}$ is their reduced mass 
\begin{equation}
\mu_{ij}=\frac{m_{i}m_{j}}{m_{i}+m_{j}}  \label{redmass}
\end{equation}
Among the basic Jacobi transformations let us also include the following
mass normalizing transformation 
\begin{equation}
\rho_{0}:(\mathbf{a}_{i},m_{i})\rightarrow(\sqrt{m_{i}}\mathbf{a}%
_{i},1),\forall i  \label{Jacobi3}
\end{equation}
and the permutation transformations 
\begin{equation}
\bar{\sigma}:(P_{1},...,P_{n})\rightarrow(P_{\sigma(1)},...,P_{\sigma(n)})
\label{permutation}
\end{equation}
which permute the vectors $\mathbf{a}_{i}$ and masses $m_{i}$ covariantly,
namely there is one for each permutation $\sigma$ of $\{1,2,..,n\}$. It is
straightforward to check that the above transformations, and hence all their
compositions, have the invariance property (\ref{kinemat2}).

Next, we inquire whether the above special transformations already generate
all those $\Phi$ with the property (\ref{kinemat2}). This is affirmatively
settled in the following subsection, but first we shall make some clarifying
observations. Note that velocity vectors $\mathbf{\dot{a}}_{i}$ of a motion
are transformed in the same way as the position vectors $\mathbf{a}_{i}$ and
may well be regarded as vectors $\mathbf{b}_{i}$ independent of the $\mathbf{%
a}_{i}$'s. In particular, invariance of $I$ means the same as invariance of $%
T,$ namely that $\Phi$ is an isometry. Therefore, the remaining issue is the
nature of the invariance of $\mathbf{\Omega.}$

Consider the $3n\times 3n$-matrix of a linear transformation $\Phi ,$
regarded as an $n\times n$-matrix 
\begin{equation}
\left[ \Phi \right] =\left( 
\begin{array}{ccc}
B_{11} & B_{12} & ..... \\ 
B_{21} & B_{22} & ..... \\ 
.... & ... & ....%
\end{array}%
\right)  \label{blocks}
\end{equation}%
where each block $B_{ij}$ is a $3\times 3$-matrix. Such an $\Phi $ expresses
each $\mathbf{a}_{j}^{\prime }$ as a linear combination of the old ones,
that is, 
\begin{equation}
\mathbf{a}_{i}^{\prime }=\sum_{j=1}^{n}\beta _{ij}\mathbf{a}_{j}\text{, \ }%
i=1,..,n,  \label{lin}
\end{equation}%
if and only if each 3-block is a \emph{scaling matrix }$B_{ij}=\beta
_{ij}Id, $ where $Id$ denotes the identity matrix\emph{. }Equivalently, $%
\Phi $ commutes with the (diagonal) action of $GL(3)$ on $\mathbb{R}^{3n}$
and hence (by representation theory) belongs to the subgroup $GL(n)$
embedded into $GL(3n)$ by the tensor product action of $GL(3)\times GL(n).$

On the other hand, assuming (\ref{lin}) holds $\Phi$ will be an isometry
(i.e. $I$ is preserved) if and only if the $n\times n$-matrix $(\beta_{ij})$
is \textquotedblright orthogonal\textquotedblright\ in the sense that 
\begin{equation}
(\beta_{ij})\in D^{\prime-1}O(n)D\subset GL(n)  \label{orto}
\end{equation}
where $D=diag(\sqrt{m_{1}},...,\sqrt{m_{n}}),D^{\prime}=diag(\sqrt {%
m_{1}^{\prime}},..,\sqrt{m_{n}^{\prime}}).$ Finally, assuming (\ref{lin})
and (\ref{orto}) it is easy to verify that $\mathbf{\Omega}$ is also
preserved.

\subsection{Equivalent characterizations of Jacobi transformations}

It turns out that the invariance properties (\ref{kinemat2}) are satisfied
solely by demanding invariance of the vector $\mathbf{\Omega.}$ This is
elucidated by the following proposition where the angular momentum
construction is analyzed from a purely algebraic viewpoint.

\begin{proposition}
Consider two n-tuples of vectors in 3-space 
\begin{equation*}
X=(\mathbf{x}_{1},\mathbf{x}_{2},..,\mathbf{x}_{n}),Y=(\mathbf{y}_{1},%
\mathbf{y}_{2},..,\mathbf{y}_{n})
\end{equation*}
and define their angular momentum vector by 
\begin{equation*}
\mathbf{\Omega=\Omega(}X,Y)=\sum_{i=1}^{n}\mathbf{x}_{i}\mathbf{\times y}_{i}
\end{equation*}
Regard $X,Y$ as vectors in $\mathbb{R}^{3n}$ $=\mathbb{R}^{3}\times
....\times\mathbb{R}^{3}$ and consider general linear transformations $\Phi:%
\mathbb{R}^{3n}\rightarrow\mathbb{R}^{3n},$ 
\begin{equation*}
X\rightarrow X^{\prime}=(\mathbf{x}_{1}^{\prime},\mathbf{x}_{2}^{\prime },..,%
\mathbf{x}_{n}^{\prime}),\text{ \ \ }Y\rightarrow Y^{\prime}=(\mathbf{y}%
_{1}^{\prime},\mathbf{y}_{2}^{\prime},..,\mathbf{y}_{n}^{\prime})
\end{equation*}
Then $\Phi$ preserves angular momentum, that is, for all $X,Y$ 
\begin{equation*}
\mathbf{\Omega(}X,Y)=\mathbf{\Omega(}X^{\prime},Y^{\prime})
\end{equation*}
if and only if $\Phi$ is an $O(3)$-equivariant isometry.
\end{proposition}

\begin{proof}
The $3\times3$-matrices 
\begin{equation*}
S_{k}=(s_{ij}^{(k)}),\text{ \ }k=1,2,3,
\end{equation*}
whose only non-zero entries are $s_{ij}^{(k)}=-s_{ji}^{(k)}=1$ whenever $%
(i,j,k)$ is an even permutation of $(1,2,3),$ constitute a basis for the Lie
algebra $so(3)$ of all skew-symmetric matrices $S$. The action of $SO(3)$ on 
$\mathbb{R}^{3n}$ embeds each $S$ as a \textquotedblright diagonal
block\textquotedblright\ matrix 
\begin{equation*}
\Delta(S)\in so(3n)\ 
\end{equation*}
with $n$ copies of $S$ along the diagonal. Observe that the matrices $%
\Delta(S_{k})$ actually represent the three components $\Omega_{k}$ of the
vector $\mathbf{\Omega}$ ,\ as skew-symmetric bilinear forms on $\mathbb{R}%
^{3n}$ expressed as matrix products, namely 
\begin{equation*}
\Omega_{k}(X,Y)=X^{t}\Delta(S_{k})Y
\end{equation*}
where $X,Y$ are regarded as column vectors with 3n components.

Let us identify a transformation $\Phi$ with its matrix in $GL(3n).$ Those
transformations leaving $\Omega_{k}$ invariant constitute a matrix group $%
G_{k}$ defined by the constraint 
\begin{equation*}
\Phi^{t}\Delta(S_{k})\Phi=\Delta(S_{k})
\end{equation*}
and hence the group leaving $\mathbf{\Omega}$ invariant is the intersection $%
G=G_{1}\cap G_{2}\cap G_{3}$ defined by the constraint 
\begin{equation}
\Phi^{t}\Delta(S)\Phi=\Delta(S)\text{, \ for all }S\in so(3)  \label{inf}
\end{equation}
The proof amounts to show $G=O(n)\subset O(3n)$, namely that $G$ is the
group of isometries commuting with $O(3)$ (or equivalently with $SO(3)$).

If we knew $\Phi$ is orthogonal, then we could have applied the exponential
function to (\ref{inf}) to conclude that $\Phi$ commutes with $SO(3)$ and
hence belongs to $O(n).$ We claim, however, the isometry condition is itself
a consequence of (\ref{inf}). To see this, write a typical matrix of the Lie
algebra $L(G)$ as an $n\times n$-matrix $(A_{ij})$ with blocks $A_{ij}$, see
(\ref{blocks}). Then the infinitesimal version of (\ref{inf}) simply reads 
\begin{equation*}
A_{ij}^{t}S+SA_{ij}=0,\text{ \ for all i,j and all }S\in so(3),
\end{equation*}
or equivalently 
\begin{equation*}
A_{ij}=-A_{ji},\text{ \ \ }A_{ij}=\alpha_{ij}Id\text{ \ \ \ \ (scaling
matrix)}
\end{equation*}
However, these are precisely the conditions defining the Lie subalgebra $%
so(n)$ and consequently $G$ and $O(n)$ have the same connected component $%
SO(n).$

Finally, to show $G=O(n)$ we use the fact that $G\ $lies in the normalizer
of $SO(n)$ in $GL(3n)$, namely in the image of $GL(3)\times O(n)$ by the
tensor product representation. Therefore, we may assume $\Phi=\Delta(B)$
where $B\in GL(3)$ and hence by (\ref{inf}) 
\begin{equation*}
B^{t}SB=S\ \text{\ \ for all }S\in so(3)
\end{equation*}
This condition says $B$ leaves invariant all skew-symmetric bilinear forms
in $\mathbb{R}^{3}$, and consequently $B=\pm Id.$ This completes the proof.
\end{proof}

The basic Jacobi transformations (\ref{Jacobi1}), (\ref{Jacobi3}), (\ref%
{permutation}) are maps between n-configuration spaces with possibly
different mass distribution. However, using normalizing transformations $%
\rho_{0}$, as in (\ref{Jacobi3}), any composition may be \textquotedblright
pulled back\textquotedblright\ to a transformation $M(3,n)\rightarrow M(3,n)$
and with all masses $m_{i}=1$. In this interpretation, the (i,j)-basic
Jacobi transformation $\rho_{ij}\ $will be a (2-dimensional) rotation $\hat{%
\rho }_{ij}\in O(n)$ and it is easy to check that $\hat{\rho}_{12},\hat{\rho}%
_{23},\hat{\rho}_{13}$ generate a dense subgroup of $SO(3)\subset
SO(3)\times SO(n-3)\subset O(n).$ Therefore, all rotations $\hat{\rho}_{ij}$
together with the permutations (\ref{permutation}) generate a dense subgroup
of $O(n).$

As a consequence of the above observations we now state the following result
on the characterization of \emph{Jacobi transformations}, a terminology
justified by (iii) below. In view of Section 1.3 we shall not attempt to
define the angular momentum vector $\mathbf{\Omega}$ in the broad category
of n-body spaces, but in the following theorem we refer to n-configuration
spaces as defined in Section 1.1.

\begin{theorem}
The following four classes of linear transformations $\Phi$ between
n-configuration spaces are identical :

(i) $\ \ \Phi$ is an isometry which preserves the angular momentum vector $%
\mathbf{\Omega}$.

(ii) $\ \Phi$ preserves $\mathbf{\Omega}$ .

(iii) $\ \Phi$ is in the closure of the set of transformations generated by
the basic Jacobi transformations.

(iv) $\ \Phi$ is an $O(3)$-equivariant isometry. \ 
\end{theorem}

The Jacobi transformations on a fixed (n-body or n-configuration) space are
also its symmetries (cf. Section 1.3), and they constitute a group
isomorphic to $O(n).$ In particular, the symmetry group for our standard
model, namely the matrix space $M(3,n)$, is the group $O(n)$ acting by
matrix multiplication on the right side.

Transformations which preserve angular momentum and kinetic energy are
certainly useful in quantum mechanics. Here one tries to keep operators
\textquotedblright separable\textquotedblright, that is, with no cross terms
(cf. e.g. \cite{B-J}, \S 10.1). As a consequence of the above results Jacobi
transformations provide all possible linear combinations $\mathbf{a}%
_{j}^{\prime}$ of the vectors $\mathbf{a}_{i}$ which preserve the kinetic
energy operator, in the sense that 
\begin{equation}
T=-\frac{\ \hslash^{2}}{2}\ \left[ \frac{1}{m_{1}}\nabla_{\mathbf{a}%
_{1}}^{2}+...+\frac{1}{m_{n}}\nabla_{\mathbf{a}_{n}}^{2}\right] =-\frac {\
\hslash^{2}}{2}\ \left[ \frac{1}{m_{1}^{\prime}}\nabla_{\mathbf{a}%
_{1}^{\prime}}^{2}+...+\frac{1}{m_{n}^{\prime}}\nabla_{\mathbf{a}%
_{n}^{\prime }}^{2}\right]  \label{T}
\end{equation}
A reduction of variables is achieved by choosing the transformation $\Phi$
so that $\mathbf{a}_{n}^{\prime}$ becomes the center of mass vector and
hence vanishes relative to a center of mass coordinate system. Then, for
example, the (time dependent) Schr\"{o}dinger equation with potential energy 
$V$ reduces to 
\begin{equation*}
i\hslash\frac{\partial}{\partial t}\Psi=\left[ -\frac{\ \hslash^{2}}{2}%
(\sum_{i=1}^{n-1}\frac{1}{m_{i}^{\prime}}\nabla_{\mathbf{a}%
_{i}^{\prime}}^{2})+V(\mathbf{a}_{1}^{\prime},...,\mathbf{a}_{n-1}^{\prime})%
\right] \Psi
\end{equation*}
where $\Psi=\Psi(\mathbf{a}_{1}^{\prime},...,\mathbf{a}_{n-1}^{\prime},t)$.
In the following subsection an explicit construction of such coordinates is
given, and by usage of the symmetry group $O(n)$ they can be modified to
satisfy additional properties for a specific purpose.

On the other hand, in the study of atomic structures one of the masses, say $%
m_{n}$ (the nucleus), may be relatively large and then non-Jacobian
coordinates may also turn out to be useful. However, in that case \emph{mass
polarization terms} cannot be avoided, namely mixed terms $\nabla _{\mathbf{a%
}_{i}^{\prime}}\cdot\nabla_{\mathbf{a}_{j}^{\prime}}$ will appear in the
kinetic energy operator (see e.g. Appendix 8 in \cite{B-J1}).

\subsection{Construction of Jacobi vectors}

Here we shall construct an explicit Jacobi transformation (\ref{coordinate
map}), or rather its inverse 
\begin{equation}
\Phi=\Psi^{-1}:X=(\mathbf{x}_{1},\mathbf{x}_{2},...,\mathbf{x}_{n-1}\mathbf{%
)\rightarrow(a}_{1},\mathbf{a}_{2},..,\mathbf{a}_{n-1},-\frac{1}{m_{n}}%
\sum_{i=1}^{n-1}m_{i}\mathbf{a}_{i})  \label{equivariant1}
\end{equation}
from $M(3,n-1)$ to the centered configurations space $M_{n}$. By definition, 
$\Phi$ must be an $O(3)$-equivariant isometry, and hence in analogy with (%
\ref{isometry1}) 
\begin{align}
i)\text{ \ }\Phi(gX) & =g\Phi(X)\text{, \ }g\in O(3)  \label{equivariant2} \\
ii)\ \text{\ }\left\vert \Phi(X)\right\vert & =\left\vert X\right\vert \text{%
, \ }X\in M(3,n-1)  \notag
\end{align}
and let us also add the following "normalizing" condition\ : 
\begin{equation}
(\mathbf{x}_{1},\mathbf{x}_{2},..,\mathbf{x}_{k},0,...,0\mathbf{)\rightarrow
(a}_{1},\mathbf{a}_{2},..,\mathbf{a}_{k},\mathbf{b}^{(k)},..,\mathbf{b}%
^{(k)})\text{, \ for }k<n-1  \label{naturality1}
\end{equation}
Namely, the Jacobi vectors $\mathbf{x}_{i}$ should vanish for \ $i>k$ if and
only if the particles $P_{i}$ \textquotedblright collide\textquotedblright\
for $i>k$, that is, they occupy the same position $\mathbf{b}^{(k)}$.
Equivalently, the Jacobi vectors define an orthogonal decomposition 
\begin{equation*}
M_{n}=\mathbb{R}_{1}^{3}\oplus\mathbb{R}_{2}^{3}\oplus...\oplus\mathbb{R}%
_{n-1}^{3}
\end{equation*}
where $\mathbb{R}_{k}^{3}$ $\simeq\mathbb{R}^{3}$ is the image of the
imbedding 
\begin{equation}
(0,..,0\mathbf{,x}_{k},0,..,0\mathbf{)\rightarrow}\text{ }(0,..,0\mathbf{,a}%
_{k}-\mathbf{b}^{(k-1)},\mathbf{b}_{(k)}\mathbf{,..,b}_{(k)}\mathbf{)}
\label{3-body1}
\end{equation}
which actually represents a collinear $3$-body system in the sense that $%
P_{1},P_{2},...,P_{k-1}$ are located at the origin and $P_{k+1,}....,P_{n}$
at the common position vector 
\begin{equation*}
\mathbf{b}_{(k)}=\mathbf{b}^{(k)}-\mathbf{b}^{(k-1)}.
\end{equation*}
The condition that $\mathbf{X}=(\mathbf{a}_{1},..,\mathbf{a}_{k},\mathbf{b}%
^{(k)},...,\mathbf{b}^{(k)})$ in (\ref{naturality1}) belongs to $M_{n}$
yields the formula 
\begin{equation}
\mathbf{b}^{(k)}=\frac{-1}{m^{(k)}}\sum_{i=1}^{k}m_{i}\mathbf{a}_{i},\text{
\ \ \ }\mathbf{b}^{(0)}=0\mathbf{,}\text{ \ }\   \label{b/upper/k}
\end{equation}
where by definition 
\begin{equation}
m^{(k)}=m_{k+1}+...+m_{n},\ \text{\ \ \ }\ m^{(0)}=\sum m_{i}=\bar{m}  \notag
\end{equation}
As a consequence of (\ref{naturality1}) we also note that the matrix $\left[
\Phi\right] $ formally representing $\Phi$ by matrix multiplication 
\begin{equation}
\left[ \Phi\right] \left( 
\begin{array}{c}
\mathbf{x}_{1} \\ 
\mathbf{x}_{2} \\ 
: \\ 
\mathbf{x}_{n-1}%
\end{array}
\right) =\left( 
\begin{array}{c}
\mathbf{a}_{1} \\ 
\mathbf{a}_{2} \\ 
: \\ 
\mathbf{a}_{n-1}%
\end{array}
\right)  \label{matrixL}
\end{equation}
is lower triangular.

By combining (\ref{3-body1}) with the $O(3)$-equivariance of $\Phi$ we
obtain the following Jacobi vectors 
\begin{equation}
\mathbf{x}_{k}=\ \zeta_{k}(\mathbf{a}_{k}-\mathbf{b}^{(k-1)})=\zeta _{k}(%
\mathbf{a}_{k}+\frac{1}{m^{(k-1)}}\sum_{i=1}^{k-1}m_{i}\mathbf{a}_{i})
\label{xk}
\end{equation}
for suitable constants $\zeta_{k}\neq0,$ whose square is determined by the
isometry condition ii) of (\ref{equivariant2}), namely 
\begin{equation}
\zeta_{k}^{2}=\frac{m_{k}m^{(k-1)}}{m^{(k)}},\text{ }1\leq k\leq n-1
\end{equation}
For example, for $k=1$ we have $\mathbf{x}_{1}=\zeta_{1}\mathbf{a}_{1}$ with 
\begin{equation*}
\left\vert \mathbf{x}_{1}\right\vert ^{2}=\zeta_{1}^{2}\left\vert \mathbf{a}%
_{1}\right\vert ^{2}=\left\vert (\mathbf{a}_{1},\mathbf{b}^{(1)},...,\mathbf{%
b}^{(1)})\right\vert ^{2}=\left\vert \mathbf{a}_{1}\right\vert ^{2}(m_{1}+%
\frac{m_{1}^{2}}{(\bar{m}-m_{1})})
\end{equation*}
and hence $\zeta_{1}^{2}=m_{1}\bar{m}/(\bar{m}-m_{1}).$

Now it is easy to describe all $\Phi $ satisfying the condition (\ref%
{naturality1}), namely $\Phi $ is uniquely determined by the choice of sign
of the numbers $\zeta _{k}$, $1\leq k\leq n-1.$ There are $2^{n-1}$ such
choices, and a natural choice is to take all $\zeta _{k}>0$, namely our
distinguished $\Phi =\Phi _{0}$ is defined by taking 
\begin{equation}
\zeta _{k}=\sqrt{\frac{m_{k}m^{(k-1)}}{m^{(k)}}},\text{ }1\leq k\leq n-1
\label{alfak}
\end{equation}%
Its inverse $\Psi _{0}=\Phi _{0}^{-1}$ will be regarded as our \emph{standard%
}\textit{\ }Jacobi transformation, and hence the \emph{standard }Jacobi
vectors are the columns of the matrix $X=$ $\Psi _{0}(\mathbf{X).}$ Then the
corresponding lower triangular matrix $\left[ \Psi _{0}\right] =$ $\left[
\Phi _{0}\right] ^{-1}$ (see ($\ref{matrixL}))$, has entries 
\begin{equation}
\left[ \Psi _{0}\right] _{ik}=\zeta _{i}.\left\{ 
\begin{array}{cc}
1 & i=k \\ 
\frac{m_{k}}{m^{(i-1)}}, & 1\leq k<i\leq n-1 \\ 
0 & 1\leq i<k\leq n-1%
\end{array}%
\right.  \label{matrixL0inv}
\end{equation}%
derived from the relation (\ref{xk}). On the other hand, the entries of $%
\left[ \Phi _{0}\right] $ $=(l_{ik})$ are $\ $%
\begin{equation}
\left[ \Phi _{0}\right] _{ik}=l_{ik}=\zeta _{k}^{-1}.\left\{ 
\begin{array}{cc}
1 & i=k \\ 
-\frac{m_{k}}{m^{(k)}}, & 1\leq k<i\leq n-1 \\ 
0 & 1\leq i<k\leq n-1%
\end{array}%
\right.  \label{matrixL0}
\end{equation}%
where in each column the entries below the diagonal are identical. Moreover,
the square sum for each row is given by the simple formula 
\begin{equation*}
l_{k1}^{2}+...+l_{kk}^{2}=\frac{1}{\bar{m}}\frac{\bar{m}-m_{k}}{m_{k}},\text{
\ \ }1\leq k\leq n-1
\end{equation*}%
$\ $

\begin{theorem}
\label{teorem1}The following Jacobi transformation 
\begin{equation*}
\tilde{\Psi}_{0}=\rho _{0}\circ \rho _{1n}\circ \rho _{2n}\circ ...\circ
\rho _{n-1,n}\ :\hat{M}_{n}\rightarrow M(3,n)\ 
\end{equation*}%
maps $(\mathbf{a}_{1},...,\mathbf{a}_{n})$\ to $(\mathbf{x}_{1},.\ ..,%
\mathbf{x}_{n})$, where $\mathbf{x}_{n}$ $=\bar{m}^{-1}\sum m_{i}\mathbf{a}%
_{i}$ is the center of mass and the remaining $\mathbf{x}_{i}$ are given by
the expressions (\ref{xk}), (\ref{alfak}). In particular, the restriction of 
$\tilde{\Psi}_{0}$ to the centered configuration space yields $\mathbf{x}%
_{n}=0$ and it coincides with the standard Jacobi transformation 
\begin{equation*}
\Psi _{0}:M_{n}\rightarrow M(3,n-1):(\mathbf{a}_{1},...,\mathbf{a}%
_{n})\rightarrow (\mathbf{x}_{1},.\ ..,\mathbf{x}_{n-1})
\end{equation*}
\end{theorem}

The proof follows by an explicit calculation of $\mathbf{x}_{k}$ as a linear
combination of $\mathbf{a}_{1},...,\mathbf{a}_{k},$ which shows that $%
\mathbf{x}_{k}$ coincides with the vector in (\ref{xk}) for $k<n$, and
furthermore, $\mathbf{x}_{n}$ is the center of mass of the $n$-configuration 
$\mathbf{X}$ $=(\mathbf{a}_{1},...,\mathbf{a}_{n}).$\ \ \ 

The above transformation $\Psi _{0}:M_{n}\rightarrow M(3,n-1)$ turns out, in
fact, to be identical with the coordinate transformation described in
Theorem 1 of Hsiang \cite{Hsiang1}, derived by a similar but different
"naturality" principle. For convenience, explicit formulae (\ref{xk}) for
the \emph{standard} Jacobi vectors when $n=3,4$ are listed below :$\mathbf{%
\qquad }$%
\begin{align}
n=3:\text{ \ }& \mathbf{x}_{1}=\sqrt{\frac{m_{1}\bar{m}\ }{m_{2}+m_{3}}}%
\mathbf{a}_{1},\text{ \ \ }\mathbf{x}_{2}=\sqrt{\frac{m_{2}(m_{2}+m_{3})}{\
m_{3}}}(\mathbf{a}_{2}+\frac{m_{1}}{m_{2}+m_{3}}\mathbf{a}_{1})\text{ \ \ }\ 
\notag \\
n=4:\text{\ \ }& \mathbf{x}_{1}=\sqrt{\frac{m_{1}\bar{m}}{m_{2}+m_{3}+m_{4}}}%
\mathbf{a}_{1}\text{\ \ }  \label{vectors} \\
& \mathbf{x}_{2}=\sqrt{\frac{m_{2}(m_{2}+m_{3}+m_{4})}{\ m_{3}+m_{4}}}(%
\mathbf{a}_{2}+\frac{m_{1}}{m_{2}+m_{3}+m_{4}}\mathbf{a}_{1})\text{\ \ } 
\notag \\
& \mathbf{x}_{3}=\sqrt{\frac{m_{3}(m_{3}+m_{4})}{m_{4}}}\left[ \mathbf{a}%
_{3}+\ \frac{1}{m_{3}+m_{4}}(m_{1}\mathbf{a}_{1}+m_{2}\mathbf{a}_{2})\right]
\ \ \ \ \ \ \ \ \ \ \ \ \   \notag
\end{align}

\section{ Orthogonal transformation groups on matrix spaces}

By applying a fixed Jacobi transformation $\Psi :M_{n}\rightarrow M(3,n-1)$,
quantities and constructions in the centered configuration space $M_{n}$ are
transported to its matrix model $M(3,m)\ $with $m=n-1$, as explained in
Section 1.4. In this setup we may and shall, however, discuss more generally
the matrix space 
\begin{equation*}
M=M(d,m)\simeq \mathbb{R}^{d}\otimes \mathbb{R}^{m}
\end{equation*}%
with its Euclidean norm (\ref{norm1}) and orthogonal transformation group 
\begin{equation*}
G=G_{1}\times G_{2}=O(d)\times O(m)
\end{equation*}%
acting by the tensor product $\rho _{d}\otimes \rho _{m}$, namely $(\psi
,\varphi )\in G$ acts on matrices $X$ by 
\begin{equation}
((\psi ,\varphi ),X)\rightarrow \psi X\varphi ^{-1}\text{ (matrix
muliplication) \ \ \ }\   \label{tensorproduct}
\end{equation}%
Physically, we have in mind $M$ as the configuration space for $m$ free
bodies (or $m+1$ bodies with fixed center of mass) in $d$-dimensional
Euclidean space, and the orbit space $\bar{M}=M/G_{1}$ is the internal
configuration space with the induced action of $G_{2}$ as the internal
symmetries. Therefore, for further investigation of the topology of $\bar{M}$
it is natural to analyze its induced $G_{2}$-orbital structure. Thus we
shall investigate two successive equivariant systems $(G_{1},M)$ and $(G_{2},%
\bar{M})$, where by an \emph{equivariant system }$(K,Y)$ we mean in general
a (compact) Lie group $K$ acting on a space $Y$. We also say $(K,Y)$ is a 
\emph{transformation group }or we simply say $Y$ is a $K$-space, and it is
convenient to refer to the orbit space construction and the orbit map 
\begin{equation}
\pi :Y\rightarrow Y/K  \label{fibration}
\end{equation}%
as a \emph{K-orbital fibration}. In (\ref{fibration}) the "fibers" are the
orbits, and therefore $\pi $ may not be the projection of a fiber bundle (or
fibration) in the usual sense since in general the orbit type is not unique
(see below).

\subsection{ Compact transformation groups and orbital decomposition}

First we shall recall some basic facts concerning equivariant systems $(K,%
\mathcal{M})$ where $K$ is a compact Lie group acting smoothly on a manifold 
$\mathcal{M}$, that is, the action map $K\times \mathcal{M\rightarrow M}$ is 
$C^{\infty }$-smooth. As a good reference on this topic we propose, for
example, the book Bredon$\cite{Bredon}$. Then we specialize to the compact
linear groups, and of particular interest are the natural transformation
groups on the matrix spaces.

\subsubsection{Basic definitions and constructions}

The space $\mathcal{M}$ splits into a disjoint union of $K$-orbits; these
are minimal (or homogeneous) $K$-spaces in the obvious sense. The isotropy
(or stability) groups $K_{p}$ at points $p$ along the same orbit constitute
a single conjugacy class $(K_{\alpha })$ of subgroups representing the \emph{%
type }of the orbit. For simplicity we say the orbit is of type $K_{\alpha }$%
, and as a $K$-space such an orbit is naturally equivalent to the coset
space $K/K_{\alpha }.$ By grouping together orbits of the same type, $%
\mathcal{M}$ decomposes as a union of orbit strata which are smooth
submanifolds $\mathcal{M}_{\alpha }$, all of which have dimension less than $%
\mathcal{M}$ except the \emph{principal} stratum $\mathcal{M}_{\omega }$,
which is an open and dense submanifold. The corresponding principal type $%
K_{\omega }$ is characterized as the unique smallest in the sense that (up
to conjugation) $K_{\omega }\subset K_{\alpha }$ for each $\alpha .$

The orbit map 
\begin{equation}
\pi :\mathcal{M}\rightarrow \mathcal{\bar{M}=M}/K  \label{orbitproj}
\end{equation}%
induces a smooth\ functional structure on the orbit space, in the sense that
a function $f$ on $\mathcal{\bar{M}}$ is called \emph{smooth}\ if the
composition $f\circ \pi $ is a smooth function on $\mathcal{M}$. Then $%
\mathcal{\bar{M}}$ is a \emph{differential} \emph{space}, but it may not be
even locally Euclidean and hence it is not a manifold in general. However,
still it has the nice and rich structure of a \emph{stratified smooth
manifold} with the image sets $\mathcal{\bar{M}}_{\alpha }=\pi (\mathcal{M}%
_{\alpha })$ as strata, and they are actually smooth manifolds. In this way
we may regard $\mathcal{M}$ as a finite union of smooth \emph{orbit bundles }%
denoted by\emph{\ } 
\begin{equation}
K/K_{\alpha }\hookrightarrow \mathcal{M}_{\alpha }\rightarrow ^{\pi _{\alpha
}}\mathcal{\bar{M}}_{\alpha }\text{ }  \label{bundle}
\end{equation}%
whose fibers are those orbits of a fixed type $K_{\alpha }$ and $\pi
_{\alpha }$ is, in fact, the projection of an actual fiber bundle.

For convenience, let us recall the general notion of a (locally trivial)
fiber bundle, typically illustrated by a sequence like (\ref{bundle}) 
\begin{equation}
F\hookrightarrow E\rightarrow^{\pi}B\   \label{bundle2}
\end{equation}
where $\pi$ is the projection map, $E$ (resp. $B)$ is the total (resp. base)
space, and $F$ denotes the typical fiber, that is, the fibers $E_{b}=\pi
^{-1}(b)$ are homeomorphic to $F$ for each $b\in B$. The simplest example is
the product bundle, where $E=B\times F$ and $\pi$ is the obvious projection.
An isomorphism between two bundles over $B$ is given by a fiber preserving
homeomeorphism $\varphi:E_{1}\rightarrow E_{2}$ between their total spaces,
and a bundle is \emph{trivial} if it is isomorphic to a product bundle. The
local triviality property (which is part of the definition) means that $B$
is covered by open sets $U$ such that the portion $\pi^{-1}(U)$ over $U$ is
a trivial bundle, that is, $\pi^{-1}(U)\simeq U\times F$ . Finally, the
bundle is smooth if all the above spaces are smooth manifolds and the maps
involved are smooth.

In particular, a $K$-orbit bundle like (\ref{bundle}) is defined by an
equivariant system $(K,E)$ with a single orbit type, and moreover, $B$ is
the orbit space $E/K$ and hence the fibers are the $K$-orbits. A \emph{%
principal} \emph{bundle} (or principal fibration) is the special case 
\begin{equation}
K\hookrightarrow E\rightarrow E/K=B  \label{princip}
\end{equation}
of (\ref{bundle}) where $K$ acts freely (i.e. all isotropy groups are
trivial).\ 

When $K$ acts by isometries on a Riemannian manifold $\mathcal{M}$, there is
an induced Riemannian metric on $\mathcal{\bar{M}}_{\alpha }$ so that the
projection map $\pi _{\alpha }$ in (\ref{bundle}) is a so-called Riemannian
submersion. Namely, the tangent map $d\pi _{\alpha }$ at any point $p\in 
\mathcal{M}_{\alpha }$ restricts to an isometry between the \emph{horizontal}
space $\mathcal{H}_{p}$ (i.e. vectors perpendicular to the orbit $\bar{p}%
=K(p))$ and the tangent space of $\mathcal{\bar{M}}_{\alpha }$ at the point $%
\bar{p}$. In this way the total orbit space $\mathcal{\bar{M}}$ in (\ref%
{orbitproj}) becomes a stratified Riemannian space; in particular, the
strata are Riemannian manifolds. Moreover, $\mathcal{\bar{M}}$ has also a
global orbital distance metric which measures the distance between orbits in 
$\mathcal{M}$ (c.f. e.g. \cite{Hsiang1}, \cite{Straume1}). This metric is
certainly determined by its restriction to the principal stratum $\mathcal{%
\bar{M}}_{\omega }$, which in turn is derived from the Riemannian metric on
this stratum. Briefly, the Riemannian manifold $\mathcal{\bar{M}}_{\omega }$
determines the Riemannian structure on each $\mathcal{\bar{M}}_{\alpha }$.

Next, we turn to the so-called \emph{slice theorem} which reduces the local
study of $(K,\mathcal{M})$ to linear representation theory, as follows. The 
\emph{local representation} of the isotropy group $H=K_{p}$ at a given point 
$p$, denoted by $(local)_{p}$ in (\ref{split}), is the induced linear action
(via differentiation) on the tangent space $T_{p}\mathcal{M}$ of $\mathcal{M}
$ at $p$. This splits into the \emph{isotropy} and \emph{slice }%
representation and we write 
\begin{equation}
T_{p}\mathcal{M=}T_{p}K(p)\oplus \mathcal{H}_{p}\text{, \ \ }%
(local)_{p}=(Iso)_{p}+(slice)_{p},  \label{split}
\end{equation}%
to indicate that the isotropy representation is the $H$-action on the
tangent space $T_{p}K(p)$ of the orbit $K(p)$, and the slice representation
is the induced action on some $H$-invariant complementary subspace $\mathcal{%
H}_{p}$ (which exists since $H$ is compact). For example, when $\mathcal{M}$
is Riemannian we take $\mathcal{H}_{p}$ to be the horizontal space, that is,
the orthogonal complement of $T_{p}K(p)$.

Now, the slice theorem says the orbit $K(p)$ has a $K$-invariant \emph{%
tubular neighborhood} which has the structure of a \emph{twisted\ product} $%
K\times _{H}\mathcal{H}_{p}$, consisting of equivalence classes $[k,v]$ in $%
K\times \mathcal{H}_{p}$ modulo the relation $[k,v]=[kh^{-1},hv]$, for $k\in
K,h\in H,v\in $ $\mathcal{H}_{p}$. The "tube" $K\times _{H}\mathcal{H}_{p}$
has the structure of a vector bundle over $K(p)\simeq K/H$ with fiber $%
\mathcal{H}_{p}$, and moreover, as a $K$-space it has the following left
action 
\begin{equation*}
k^{\prime }[k,v]=[k^{\prime }k,v]\text{, \ }k^{\prime }\in K
\end{equation*}%
and consequently the isotropy groups at points in the tubular neighborhood
of\ $K(p)$ looks like 
\begin{equation*}
K_{[k,v]}=kH_{v}k^{-1}
\end{equation*}

In particular, by calculating the slice representation $(K_{p},\mathcal{H}%
_{p})$ as the difference 
\begin{equation}
(slice)_{p}=(local)_{p}-(Iso)_{p}  \label{slicerep}
\end{equation}
one can determine the orbit types that occur in the neighborhood. Moreover,
there is the following isomorphism of orbit spaces 
\begin{equation}
(K\times_{K_{p}}\mathcal{H}_{p})/K\simeq\mathcal{H}_{p}/K_{p}  \label{local}
\end{equation}
which enables one to analyse the local smooth structure of orbit spaces by
successive application of the slice theorem.

Finally, we recall the notion of a \emph{fundamental domain} for $(K,%
\mathcal{M})$, or equivalently a \emph{cross-section} for the orbit map (\ref%
{orbitproj}). This is a closed subset $\Sigma $ $\subset $ $\mathcal{M}$%
\emph{\ }which intersects each orbit in a unique point, and consequently
there is an identification 
\begin{equation}
\Sigma \simeq \mathcal{\bar{M}}  \label{orbitspace4}
\end{equation}%
In the special case that $\mathcal{M}$ is Riemannian and $\Sigma $ is an
orthogonal cross-section, and hence the submanifold $\Sigma $ is also
perpendicular to each orbit, the identification (\ref{orbitspace4}) is in
fact an isometry. We remark, however, that fundamental domains in the above
strict sense exist only in special cases -- but fortunately including those
orthogonal transformation groups that we shall investigate below.

\subsubsection{Compact linear groups on Euclidean spaces\ }

Now, let us consider pairs $(K,V)$ and the corresponding orbit map 
\begin{equation}
\pi :V\rightarrow \bar{V}=V/K  \label{projmap}
\end{equation}%
where $K$ is a compact Lie group acting orthogonally on a Euclidean space $V$%
. Since the orbits are compact it is a well known fact that they can be
separated by $K$-invariant polynomial functions $\mathfrak{p}_{i}.$ Namely,
we can choose a Hilbert basis or \textquotedblright sufficiently
many\textquotedblright\ separating and invariant polynomial functions $%
\mathfrak{p}_{i}$ as the components of a map 
\begin{equation*}
\mathfrak{p}=(\mathfrak{p}_{1},\mathfrak{p}_{2},...,\mathfrak{p}%
_{N}):V\rightarrow \mathbb{R}^{N}
\end{equation*}%
which induces an embedding 
\begin{equation}
\mathfrak{\bar{p}}:\bar{V}\ \rightarrow \mathfrak{p}(V)\subset \mathbb{R}^{N}
\label{orbitmap}
\end{equation}%
and hence identifies the orbit space with its image $\mathfrak{p}(V)$. The
latter is a semi-algebraic subset, that is, it is defined by polynomial
identities and inequalities in $\mathbb{R}^{N}$.

Concerning smoothness, we recall that $\bar{V}$ has, on the one hand, the
structure of a \emph{differential space }such that a function $f$ on $\bar{V}
$ is \textquotedblright smooth\textquotedblright\ if and only if the
composition $f\circ\pi$ is a smooth function on $V$. On the other hand,
there is another approach defining a function on $\mathfrak{p}(V)$ to be
\textquotedblright smooth\textquotedblright\ if it is the restriction of a
smooth function on $\mathbb{R}^{N}.$ Fortunately, the two notions of
smoothness are identical, and therefore we may regard the bijective
correspondence $\mathfrak{\bar{p}}$ in (\ref{orbitmap}) to be a
diffeomorphism, cf. Schwarz \cite{Schwarz}. Consequently, $f$ is
\textquotedblright smooth\textquotedblright\ if and only if $f\circ\pi=\bar{f%
}\circ\mathfrak{p}$ \ for some smooth function $\bar{f}$ on $\mathbb{R}^{N}.$
Note, however, the orbital distance metric on $\bar{V}$ $\simeq\mathfrak{p}%
(V)$ is certainly not induced from the Euclidean space in (\ref{orbitmap}).

In the Euclidean space $V$ \emph{size}\ and \emph{shape}\ are quantities
preserved by orthogonal transformations. The norm $\rho=\left\vert
X\right\vert $ is a size\ function which measures the distance from the
origin $O$, namely the arc-length along rays emanating from $O$. A nonzero
vector can be uniquely scaled to unit size\ by a homothety transformation $%
X\rightarrow cX$ $(c>0)$, and by expressing the Euclidean metric $ds^{2}$ of 
$V$ in \emph{polar coordinates}\ we actually describe $V$ as a Riemannian
cone over its unit sphere $V^{1}:$ $\ $ 
\begin{equation}
(V^{1},d\theta^{2})\subset(V,ds^{2})\text{, \ \ }ds^{2}=d\rho^{2}+\rho
^{2}d\theta^{2}  \label{cone1}
\end{equation}
where $d\theta^{2}$ denotes the induced spherical metric on $V^{1}$.

On the other hand, the $K$-orbit space $\bar{V}$ and its subspace $V^{\ast
}=V^{1}/K$ are stratified Riemannian spaces as explained above. Moreover,
since the norm function $\rho$ is $K$-invariant it is also a size\ function
on $\bar{V}$ and then $V^{\ast}=(\rho=1)$ becomes the unit distance
\textquotedblright sphere\textquotedblright. Furtermore, in analogy with (%
\ref{cone1}) $\bar{V}$ inherits the structure of a stratified Riemannian
cone over its \textquotedblright sphere\textquotedblright\ $V^{\ast}$ :\ $\
\ $ 
\begin{equation}
(V^{\ast},d\sigma^{2})\subset(\bar{V},d\bar{s}^{2})\text{, \ }d\bar{s}%
^{2}=d\rho^{2}+\rho^{2}d\sigma^{2}  \label{cone2}
\end{equation}
where $d\sigma^{2}$ denotes the orbital distance metric induced from $%
d\theta^{2}$. In this cone the rays emanating from the cone vertex (or base
point), still denoted by $O$, are also geodesics for the metric $d\bar{s}%
^{2} $, and homothety transformations move points $\neq O$ along these rays.

\subsubsection{The configuration space for m-body systems in d-space}

Henceforth, we take $V$ to be the matrix space\ $M=M(d,m)$ with the
orthogonal action of $G=O(d)\times O(m)$ as in (\ref{tensorproduct}). By
regarding $M$ as the configuration space for an m-body system in $\mathbb{R}%
^{d}$, in analogy with the \textquotedblright physical
world\textquotedblright\ case $d=3$ described in Section 1, $G_{1}=O(d)$ and 
$G_{2}=O(m)$ play the roles of congruence and internal symmetry groups
respectively. Then there is the following two-step orbital decomposition of
stratified Riemannian spaces 
\begin{equation}
M\rightarrow \bar{M}=\frac{M}{G_{1}}\rightarrow \frac{\bar{M}}{G_{2}}=\frac{M%
}{G}  \label{twostep1}
\end{equation}%
where the \emph{internal} \emph{configuration space }$\bar{M}=\bar{M}(d,m)$
and its symmetry group constitute a natural equivariant system $(G_{2},\bar{M%
})$. Again, the metric (\ref{cone2}) on $\bar{M}$ will be referred to as the 
\emph{kinematic metric}, since in the physical case $d\leq 3$ it is, in
fact, representing the internal kinetic energy 
\begin{equation*}
\bar{T}=\frac{1}{2}d\bar{s}^{2}=T-T^{\omega }=\frac{1}{2}ds^{2}-T^{\omega },
\end{equation*}%
namely the total kinetic energy $T$ minus the purely rotational kinetic
energy $T^{\omega }$ of the n-body system, see e.g. \cite{Straume1}.

The internal symmetry orbit space $\bar{M}/G_{2}$, which is a cone in $%
\mathbb{R}^{\mu }$, $\mu =\min \left\{ d,m\right\} $, is actually a Weyl
chamber of Cartan type $B_{\mu }$. Indeed, the $G$-representation on $M$ is
the isotropy representation of the symmetric space $O(d+m)/O(d)\times O(m)$,
which is of type $B_{\mu }$, and as such it is well understood in terms of
the classical invariant theory of matrices and canonical forms modulo
orthogonal transformation groups.

Let $M^{1}$ be the unit sphere of $M.$ In view of the cone structure (\ref%
{cone1}), (\ref{cone2}), we may faithfully replace (\ref{twostep1}) by its
unit distance "sphere\textquotedblright\ version 
\begin{equation}
M^{1}\rightarrow M^{\ast}=\frac{M^{1}}{G_{1}}\rightarrow\frac{M^{\ast}}{G_{2}%
}=\frac{M^{1}}{G}=M^{\ast\ast}  \label{twostep2}
\end{equation}
where $M^{\ast}=M^{\ast}(d,m)$ will be referred to as the \emph{shape space. 
}Its $G_{2}$-orbit space $M^{\ast\ast}$ is a spherical simplex and it is
explicitly described in (\ref{simplex}) below.

The orbit types of the above transformation groups will be described in
detail. In principle, the topological structure of the shape space $M^{\ast
} $ can be reconstructed and analyzed as the union of well understood orbit
bundles, but in this paper we shall only describe the global topology of $%
M^{\ast }$ in those few cases where it is a msnifold. In particular, the
cases $d=m-1$ or $d=m$ are rather simple since $(G_{2},M^{\ast })$ is
essentially isomorphic to a \emph{linear model}, that is, a sphere $S^{N}$
or disk $D^{N+1}$ with an orthogonal action of $G_{2}$, see Section \ 3.4.3.
Before turning to these topics, let us have a closer look at some important
classical matrix spaces and their natural transformation groups.

\subsection{Transformations of the space of symmetric matrices}

For $\ d\leq m$, consider the following sets of $m\times m$-matrices 
\begin{equation}
\begin{array}{cccccc}
Sym^{+}(m)_{\leq d} & \subset & Sym^{+}(m) & \subset \  & \ \ Sym(m)\subset
M(m,m) &  \\ 
\cup &  & \cup &  & \ \cup &  \\ 
B^{+}(d)\simeq B^{+}(m)_{\leq d} & \subset & B^{+}(m) & \subset & 
B(m)\subset \ Diag(m)\ \simeq \mathbb{R}^{m} & \ 
\end{array}
\label{diagram}
\end{equation}%
where in the first row $Sym(m)$ is the space of symmetric matrices, $%
Sym^{+}(m)$ is the convex cone of positive semidefinite matrices, and $%
Sym^{+}(m)_{\leq d}$ is the subcone of matrices of rank at most $d.$ In the
second row$\ Diag(m)$ is the space of diagonal matrices $diag(\lambda
_{1},...,\lambda _{m})$, containing the cone 
\begin{equation}
B(m)=\left\{ Y=diag(\lambda _{1},\lambda _{2},...,\lambda _{m});\text{ \ }%
\lambda _{1}\geq \lambda _{2}\geq \text{. .. }\geq \lambda _{m}\right\} ,
\label{B(m)}
\end{equation}%
and the subcones $B^{+}(m)$ and $B^{+}(m)_{\leq d}$ are given by the
additional constraint $\lambda _{m}$ $\geq 0$, and $\lambda _{i}=0$ for all $%
i>d$, respectively.

With the usual inner product on $M(m,m)$ 
\begin{equation*}
Y_{1}\cdot Y_{2}=trace(Y_{1}^{t}Y_{2})\text{, \ \ \ \ cf. (\ref{norm1}) \ \ }
\end{equation*}
where $Y^{t}$ means the transpose of $Y$, $O(m)$ acts orthogonally by
conjugation

\begin{equation}
(\varphi,Y)\rightarrow\varphi Y\varphi^{-1},\text{ \ \ \ }\varphi\in
O(m),Y\in M(m,m)  \label{conjugation}
\end{equation}
and the subsets in the first row of (\ref{diagram}) are clearly invariant.
Thus the transformation group $(O(m),Sym(m))$ is the second symmetric tensor
product $S^{2}\rho_{m}$ of the standard representation $\rho_{m}=$ $(O(m),%
\mathbb{R}^{m})$, and let us recall the classical result about
diagonalization of symmetric matrices. Namely, every symmetric matrix is
conjugate to a unique matrix\ - its \emph{canonical form} - in the cone $%
B(m) $, which is therefore a fundamental domain (or cross section) of the
orbit map 
\begin{equation}
Sym(m)\rightarrow\frac{Sym(m)}{O(m)}\simeq\frac{\ \mathbb{R}^{m}}{S_{m}}%
\simeq B(m)\subset Sym(m)  \label{domain}
\end{equation}
As indicated, $B(m)$ is also a fundamental domain (or cross section) for $%
(S_{m},\mathbb{R}^{m})$, where $S_{m}$ is the symmetric group acting on $%
\mathbb{R}^{m}$ by permuting the coordinates $\lambda_{i}.$

Fundamental domains for the $O(m)$-action restricted to invariant subspaces
of $Sym(m)$ are the corresponding subsets of $B(m)$ in the second row of (%
\ref{diagram}), consequently 
\begin{equation}
B^{+}(m)\simeq\frac{\ \mathbb{R}^{m}}{B_{m}}\simeq\frac{Sym^{+}(m)}{O(m)}%
\supset\frac{Sym^{+}(m)_{\leq d}}{O(m)}\simeq\frac{\ \mathbb{R}^{d}}{B_{d}}%
\simeq B^{+}(d)  \label{fundamental}
\end{equation}
As indicated, we recognize $B^{+}(k)$ as a Weyl chamber for the compact
connected Lie group $SO(2k+1)$, namely a fundamental domain of the canonical
representation of the Weyl group $B_{k}\supset S_{k}$ acting on $\mathbb{R}%
^{k}$ as a group generated by reflections.

Note that $Sym(m)$ is not an irreducible $O(m)$-space. In fact, there is the
orthogonal and invariant decomposition 
\begin{equation}
Sym(m)=Sym^{0}(m)\oplus\mathbb{R}^{1}  \label{Symnull}
\end{equation}
where the trivial summand $\mathbb{R}^{1}$ is spanned by the identity
matrix, and the first summand, which consists of matrices of trace zero,
gives the irreducible representation $S^{2}\rho_{m}-1$ with fundamental
domain $B^{0}(m)$ $\subset B(m)$ defined by $\sum\lambda_{i}=0:$ 
\begin{equation}
B^{0}(m)\simeq\frac{\ \mathbb{R}^{m-1}}{S_{m}}\simeq\frac{Sym^{0}(m)}{O(m)}\ 
\label{B(m)0}
\end{equation}
Here we have also recognized $B^{0}(m)$ as a fundamental domain of the
canonical representation of the Weyl group $A_{m-1}=S_{m}$ of $\ SU(m).$

In addition to its Euclidean norm where $\left\vert Y\right\vert =\sqrt{%
trace(Y^{2})}$, the positive cone $Sym^{+}(m)$ has another $O(m)$-invariant
size function $\rho^{\ast}(Y)=trace(Y)$, with corresponding
\textquotedblright unit sphere\textquotedblright\ 
\begin{equation}
Sym^{\ast}(m)=\left\{ Y\in Sym^{+}(m)\text{ ; }trace(Y)=1\right\}
\label{shape}
\end{equation}
For later reference, we note that its intersection with the Euclidean unit
sphere in $Sym(m)$ is the subset 
\begin{equation}
Sym^{\ast}(m)\cap Sym(m)^{1}=Sym^{\ast}(m)_{1}=\left\{ Y\in Sym^{\ast }(m)%
\text{ ; }rk(Y)=1\right\}  \label{Sym1}
\end{equation}

Let $Y\in B(m).$ The tangent space at $Y$ of the $O(m)$-orbit through $Y$ in 
$Sym(m)$ is the subspace 
\begin{equation*}
\left[ so(m),Y\right] =\left\{ SY-YS\text{ ; }S\text{ is skew-symmetric}%
\right\} \subset Sym(m),
\end{equation*}
and it is perpendicular to $B(m)$ since the inner product with any $%
Y^{^{\prime}}\in B(m)$ is 
\begin{equation*}
(SY-YS)\cdot Y^{\prime}=trace((SY-YS)Y^{\prime})=0
\end{equation*}
A direct consequence of this observation is as follows :

\begin{remark}
\label{ortosec}The fundamental domain $B(m)$ of $(O(m),Sym(m))$ is an \emph{%
orthogonal} cross section, in the sense that it is perpendicular to the $%
O(m) $-orbits. Therefore, the orbit space of $Sym(m)$ (with the orbital
distance metric) is isometric to $B(m).$ Consequently, orbit spaces of
various $O(m)$- invariant subspaces of $Sym(m)$ in (\ref{diagram}) are
isometric to the corresponding subsets of $B(m).$
\end{remark}

Finally, let us determine the orbit types of the $O(m)$-action on symmetric
matrices. To this end, consider a matrix $Y\in B(m)$ and let the "strings"
of equal entries $\lambda_{i}$ in (\ref{B(m)}) have length $m_{1},m_{2},...,$
so that 
\begin{equation*}
m=m_{1}+m_{2}+...+m_{p}\text{ \ \ }(m_{i}>0,\text{ }p>0)
\end{equation*}
is a partition of $m$ and $p$ is the number of different $%
\lambda_{i}^{\prime }s$. It is easy to see that the isotropy group at $Y$ is
the \textquotedblright block\textquotedblright\ orthogonal matrix group 
\begin{equation}
O(m_{1},m_{2},...,m_{p})=O(m_{1})\times O(m_{2})\times...\times
O(m_{p})\subset O(m)  \label{isotropy1}
\end{equation}
and therefore the partition function $\pi(m)$ of $m$ enumerates the
different orbit types. The orbits are\ in fact connected and therefore they
are also $SO(m)$-orbits, that is, 
\begin{equation*}
\frac{Sym(m)}{SO(m)}=\frac{Sym(m)}{O(m)}\simeq B(m)
\end{equation*}

\subsection{Algebraic realization of orbit spaces and orbital stratification}

With the results from the previous subsection as a basis, the construction
of the orbit spaces in (\ref{twostep1}) or (\ref{twostep2}) as semialgebraic
subsets is based upon the following two simple but fundamental properties of
matrices :

\begin{itemize}
\item The polynomial map 
\begin{equation}
\mathfrak{p}:M(d,m)\ \text{ \ }\rightarrow\ \ Sym(m),\text{\ \ \ \ }%
X\rightarrow Y=X^{t}X  \label{polynomial1}
\end{equation}
preserves the matrix rank, and its image consists of all positive
semidefinite symmetric matrices $Y\ $of rank $\leq\min\left\{ d,m\right\} . $
Moreover, for $\psi\in O(d),\varphi\in O(m),$%
\begin{equation}
\mathfrak{p}(\psi X\varphi^{-1})=\varphi\mathfrak{p}(X)\varphi^{-1}\
=\varphi Y\varphi^{-1}  \label{p1p2}
\end{equation}
and the entries of $Y$ are the inner products $\mathbf{x}_{i}\cdot \mathbf{x}%
_{j}\ $of the column vectors $\mathbf{x}_{i}$ of $X$.

\item Consider a finite collection of vectors $\mathbf{x}_{j}\ $in a
Euclidean space $\mathbb{R}^{d}.$ Then the collection is uniquely
determined, up to $O(d)$-congruence, by the inner products $\mathbf{x}%
_{i}\cdot\mathbf{x}_{j}$. For a proof we refer to Weyl\cite{Weyl}, page 52.
\end{itemize}

The map $\mathfrak{p}$ in (\ref{polynomial1}) is clearly constant on each $%
O(d)$-orbit in $M=M(d,m)$ and induces a surjective map 
\begin{equation}
\mathfrak{\bar{p}}:\bar{M}(d,m)\ \rightarrow Sym^{+}(m)_{\leq d}\text{ }%
\subset Sym(m)\ \   \label{pi-bar}
\end{equation}
which is, moreover, injective by the above mentioned result of Weyl. In
other words, the components of $\mathfrak{p}$, namely the collection of
inner products $\mathbf{x}_{i}\cdot\mathbf{x}_{j}$, constitute a Hilbert
basis in the sense of Section 3.1.1. Consequently, $\mathfrak{\bar{p}}$ is a
diffeomorphism between differential spaces which identifies $\bar{M}=M/O(m)$
with the set of positive semidefinite symmetric $m\times m$-matrices of rank
at most $d.$ In particular, 
\begin{equation}
\bar{M}(d,m)=\bar{M}(m,m)\simeq Sym^{+}(m)\text{ \ if \ }d\geq m,
\label{d-large}
\end{equation}
so we will henceforth assume without loss of generality that $d\leq m.$ Thus
we have explained why all orbit spaces involved are realizable as
appropriate spaces of symmetric matrices, amenable to the setting in Section
3.2.

\subsubsection{Cross sections and canonical foms}

In general, the inclusions $M(k,m)\subset M(k+1,m)$, defined by taking the
last row to be zero, induce the increasing filtration 
\begin{equation}
\bar{M}(1,m)\subset\bar{M}(2,m)\subset...\subset\bar{M}(m,m)\ \simeq
Sym^{+}(m)\text{ }  \label{filtration}
\end{equation}
and this coincides with the matrix rank filtration of the positive cone $%
Sym^{+}(m)$, namely for $k+i\leq m$ 
\begin{equation*}
\bar{M}(k,m)=\bar{M}(k+i,m)_{\leq k}\simeq Sym^{+}(m)_{\leq k}\text{ }
\end{equation*}

On the other hand, $G_{2}=O(m)$ acts on the spaces in (\ref{pi-bar}). The
action on $\bar{M}=\bar{M}(d,m)$ is induced from the action (\ref%
{tensorproduct}), and $G_{2}$ acts on $Sym(m)$ by conjugation (\ref%
{conjugation}). From the equivariance property (\ref{p1p2}) it follows that
these actions commute with the map $\mathfrak{\bar{p}}$, that is, $\mathfrak{%
\bar{p}}$ is an isomorphism of $G_{2}$-spaces and hence induces a
diffeomorphism between $G_{2}$-orbit spaces 
\begin{equation}
\mathfrak{\tilde{p}}\ :\frac{M}{G}=\frac{\bar{M}}{G_{2}}\text{ }\rightarrow 
\frac{Sym^{+}(m)_{\leq d}}{G_{2}}\simeq B^{+}(d)\subset Sym^{+}(m)_{\leq
d}\simeq \bar{M}  \label{orbitspace2}
\end{equation}%
The cross section $B^{+}(d)\subset $ $\bar{M}$ in (\ref{orbitspace2}), being
transversal to the $G_{2}$-orbits in $\bar{M}$, further lifts to a cross
section $M^{+}(d)\ $of the composite orbit map $M\rightarrow M/G,$ as
follows : 
\begin{equation}
B^{+}(d)\simeq M^{+}(d)\subset \mathbf{\ }M:\left\{ 
\begin{array}{c}
\text{ \ }X=diag(r_{1},r_{2},...,r_{d})\text{ } \\ 
\text{ }r_{1}\geq r_{2}\geq ...\geq r_{d}\geq 0\text{ }%
\end{array}%
\right.  \label{cross section}
\end{equation}%
where $X=(x_{ij})$ has $x_{ii}=$ $r_{i}\ $and zero entries otherwise. Thus
we have also established the diagonalization procedure saying that every
matrix in $M=M(d,m)$ can be transformed by the action (\ref{tensorproduct})
to a unique matrix\ - its canonical form - in the subset $M^{+}(d)$. As a
(simplicial) Euclidean cone in the $\left\{ r_{i}\right\} $-coordinate space 
$\mathbb{R}^{d}$, $M^{+}(d)$ is also the Weyl chamber of the Weyl group $%
B_{d}$, cf. (\ref{fundamental}).

\subsubsection{Orthogonality of cross sections}

In view of Remark \ref{ortosec}, the cone $M^{+}(d)$ in (\ref{cross section}%
) is orthogonal to the $G$-orbits and is therefore an orthogonal cross
section for $(G,M)$. Hence, we have an isometry $M/G\simeq M^{+}(d)$ of
Euclidean cones with the metric 
\begin{equation}
\ ds^{2}=\sum_{i=1}^{d}dr_{i}^{2}\ =\frac{1}{4}\sum_{i=1}^{d}\frac{1}{%
\lambda_{i}}d\lambda_{i}^{2}  \label{metric2}
\end{equation}
Here $\left\{ r_{i}\right\} $ from (\ref{cross section}) and $\left\{
\lambda_{i}\right\} $ from (\ref{B(m)}) are coordinate systems for the cone
and are related by the polynomial map\ $\mathfrak{p}$ in (\ref{polynomial1}%
), which restricts to a diffeomorphism 
\begin{equation}
\mathfrak{p}:M^{+}(d)\rightarrow B^{+}(d)\text{ }\subset\ Sym^{+}(m)\text{,
\ \ }r_{i}\rightarrow r_{i}^{2}=\lambda_{i}  \label{diffeo}
\end{equation}
Thus, the $G$-orbit of a given matrix $X\in M=M(d,m)$ has coordinates $%
\lambda_{i}$ interpreted as the eigenvalues of $X^{t}X$. In the special case
that $d=m$ and $X$ is symmetric, the numbers $r_{i}$ are the absolute values
of the eigenvalues of $X.$

\begin{remark}
At this point, observe that the internal configuration space $\bar{M}=\bar
{%
M}(d,m)$ has two metrics, namely the kinematic metric (i.e. the $O(d)$%
-orbital distance metric) and the induced \textquotedblright
fake\textquotedblright\ metric as a subset $Sym^{+}(m)_{\leq d}$ of the
Euclidean space $Sym(m)$, cf. (\ref{pi-bar}). For both metrics the symmetry
group $O(m)$ is actually an isometric transformation group with $B^{+}(d)$
as a fundamental domain (i.e. cross section). In the \textquotedblright
fake\textquotedblright\ metric $B^{+}(d)$ inherits the Euclidean metric\ $%
\sum d\lambda_{i}^{2}$ and is by Remark \ref{ortosec} an orthogonal cross
section, whereas in the kinematic (but still\ Euclidean) metric\ (\ref%
{metric2}) of $\bar{M}$ the orthogonality property of $B^{+}(d)$ fails. See
also Remark \ref{metric5} below.
\end{remark}

\subsubsection{Rank and subrank stratification of matrix spaces}

The \emph{subrank stratification} of the matrix space $M=M(d,m)$ is a
natural refinement of the usual \emph{rank stratification}. The latter
coincides with the orbit type stratification of the action of $G_{1}$(or $%
G_{2})$ on $M$ and the mentioned refinement coincides with the orbit type
stratification of the action of the full group 
\begin{equation*}
G=G_{1}\times G_{2}=O(d)\times O(m)
\end{equation*}
In order to describe the combinatorial structure involved we turn to the
two-step orbital decomposition (\ref{twostep2}), where the first orbit space
is the shape space 
\begin{equation}
M^{\ast}=M^{\ast}(d,m)\simeq Sym^{\ast}(m)_{\leq d}\ ,\text{ cf. (\ref{shape}%
)}\   \label{shape2}
\end{equation}
and the final $G$-orbit space is 
\begin{equation*}
M^{\ast\ast}=M^{\ast\ast}(d,m)=M^{\ast}/G_{2}=M^{1}/G\ 
\end{equation*}
which we may identify with the following fundamental domain in the sphere $%
M^{1}$: 
\begin{equation}
\Delta^{d-1}=M^{+}(d)\cap M^{1}:r_{1}\geq r_{2}\geq...\geq r_{d}\geq0,\sum
r_{i}^{2}=1  \label{simplex}
\end{equation}
This is a spherical simplex whose structure, indeed, identifies it with the
spherical Weyl chamber of type $B_{d}$ (see (\ref{fundamental})); in
particular, it is homeomorphic to the closed disk $D^{d-1}$.

Thus we also conclude that the above simplex (\ref{simplex}) is a cross
section for the transformation group $(G,M^{1})$, and hence any matrix in $%
M^{1}$ is $G$-equivalent to a unique matrix\ - its \emph{canonical form} -
in the simplex. On the other hand, a typical matrix of rank $k$ in (\ref%
{simplex}) may be characterized by the following numerical data

\begin{equation}
X:\left\{ 
\begin{array}{c}
r_{1}=r_{k_{1}}>r_{k_{1}+1}=r_{k_{1}+k_{2}}>r_{k_{1}+k_{2}+1}=....=r_{k}>0=r_{k+1}\ 
\\ 
1\leq \sum_{i=1}^{p}k_{i}=k\leq d\text{ , \ }k_{i}>0,\text{ }p>0,%
\end{array}%
\right.  \label{subrank3}
\end{equation}%
and then its \emph{subrank }is defined to be the corresponding (unordered)
partition of $k$ into $p$ positive integers 
\begin{equation}
\kappa =\left( k_{1},k_{2},...,k_{p}\right) ,\text{ }\left\vert \kappa
\right\vert =\ k_{1}+\text{\ }...+k_{p}=k  \label{multirank}
\end{equation}%
which records the length of the strings of equal\ numbers in (\ref{subrank3}%
).

We label the rank (resp. subrank) stratum by the subscript $k$ (resp. $%
\kappa $) of the corresponding sets. Thus $X$ belongs to $M_{k}^{1}$ if and
only if its $G_{1}$-orbit has the type $O(d-k)$ (cf. also (\ref{regular})
below). Moreover, by (\ref{simplex}) the image of $M_{k}^{1}$ in $M^{\ast
\ast }$ is the "semi-open" simplicial disk 
\begin{equation*}
M_{k}^{\ast \ast }\simeq \Delta _{k}^{k-1}\subset \Delta
_{k}^{d-1}:r_{1}\geq r_{2}\geq ...\geq r_{k}>r_{k+1}=0,\sum r_{i}^{2}=1,
\end{equation*}%
which is subdivided into its various subrank strata 
\begin{equation}
M_{\kappa }^{\ast \ast }\simeq \Delta _{\kappa }^{k-1}\subset \Delta
_{k}^{k-1}\subset \Delta ^{d-1}\simeq M^{\ast \ast }\ \   \label{strata}
\end{equation}%
In particular, the principal stratum has subrank $\kappa _{0}=\left(
1,1,...,1\right) $ with $\left\vert \kappa _{0}\right\vert =d$, and $%
M_{\kappa _{0}}^{\ast \ast }\simeq \Delta _{\kappa _{0}}^{d-1}$ is the\
interior of the spherical simplex $\Delta ^{d-1}$ and is therefore
homeomorphic to $\mathbb{R}^{d-1}$.

To see why the subrank stratification actually coincides with the $G$-orbit
type stratification, consider the isotropy group $G_{X}$ of the matrix in (%
\ref{subrank3}), using the following notation for groups : 
\begin{align}
\Delta H & =diag(H\times H)\simeq H\text{ \ (diagonal embedding)}  \notag \\
O(k_{1},..,k_{p}) & =O(k_{1})\times...\times O(k_{p})\subset O(k)
\label{isotropy} \\
\ G(k_{1},..,k_{p}) & =O(d-k)\times\Delta O(k_{1},..,k_{p})\times
O(m-k)\subset G  \notag
\end{align}
Then it is not difficult to show by direct calculations with matrices that 
\begin{equation}
G_{X}=\left\{ (g_{1},g_{2})\in O(d)\times O(m)\text{ };g_{1}Xg_{2}^{-1}=X%
\right\} =G(k_{1},..,k_{p})  \label{isotropy2}
\end{equation}
(cf. Table 1, \#10, in \cite{Straume2}), and hence the isotropy types
uniquely characterize the subrank strata, as claimed.

We are particularly interested in the stratification of $M^{\ast}$, but now
there are two natural options, namely the induced subrank stratification and
the $G_{2}$-orbit type stratification. Our next claim, however, is that they
are identical, and hence one can describe $M^{\ast}$ is a union of $G_{2}$%
-orbit bundles lying over the various strata (\ref{strata}).

To calculate the $G_{2}$-orbit types and how they correspond to the subrank
strata we proceed as follows. Let $X^{\ast}\in M^{\ast}$ be the image of a
matrix $X$ of rank $k$ and consider the "large" orbit $G(X)\subset M$. As a $%
G_{1}$-space all the orbits of $G(X)$ have the same type as the "small"
orbit $G_{1}(X)$. Moreover, the orbit space $G(X)/G_{1}\subset M^{\ast}$
coincides with the $G_{2}$-orbit through $X^{\ast}$, namely the set $%
G_{2}(X^{\ast})$. Thus, the \textquotedblright large\textquotedblright\ $G$%
-orbit through $X$ has the structure of a $G_{1}$-orbit bundle, and in terms
of homogeneous spaces (i.e. coset spaces of groups) this fiber bundle can be
described as follows : 
\begin{equation}
\begin{array}{cccccc}
G_{1}(X) & \rightarrow & G(X)\  & \rightarrow & G_{2}(X^{\ast}) &  \\ 
\shortparallel &  & \shortparallel &  & \shortparallel &  \\ 
\frac{O(d)}{O(d-k)} & \rightarrow & \frac{G}{G(k_{1},k_{2},\ ..,k_{p})} & 
\rightarrow & \frac{O(m)}{O(k_{1},k_{2},\ ..,k_{p})\times O(m-k)} & 
\end{array}
\label{orbits}
\end{equation}
The base space $G_{2}(X^{\ast})$ of the fiber bundle is in turn a fiber of
another bundle, namely the multistratum $M_{\kappa}^{\ast}$ which as a $%
G_{2} $-orbit bundle fibers as follows 
\begin{equation}
\frac{O(m)}{O(k_{1},k_{2},..,k_{p})\times O(m-k)}\rightarrow
M_{\kappa}^{\ast }\rightarrow
M_{\kappa}^{\ast\ast}\simeq\Delta_{\kappa}^{k-1}  \label{fiber}
\end{equation}
In this way the subrank stratification of the shape space $\ $ 
\begin{equation*}
M^{\ast}(d,m)=\bigcup_{k=1}^{d}M_{k}^{\ast}=\bigcup_{k=1}^{d}\bigcup
_{\left\vert \kappa\right\vert =k}M_{\kappa}^{\ast}
\end{equation*}
has strata which are fiber bundles of type (\ref{fiber}), and there are
althogether 
\begin{equation}
\ \pi(1)+\pi(2)+...+\pi(d)  \label{partition}
\end{equation}
such bundles (or strata), where $\pi(k)\emph{\ }$is the partition function
of $k.$ The principal stratum $M_{\kappa_{0}}^{\ast}$ constitutes an open
and dense subset of $M^{\ast}$, and the principal orbit bundle 
\begin{equation}
\frac{O(m)}{O(1)^{d}\ \times O(m-d)}\rightarrow M_{\kappa_{0}}^{\ast
}\rightarrow M_{\kappa_{0}}^{\ast\ast}\approx\mathbb{R}^{d-1}
\label{principal}
\end{equation}
is actually trivial (e.g. since the base space is contractible). The orbit
type, either for the $G_{2}$-action on $M^{\ast}$ or the $G$-action on $%
M^{1} $, for a stratum of subrank\ $\kappa=\left(
k_{1},k_{2},...,k_{p}\right) $ can be read off from (\ref{orbits}).

\begin{example}
\label{triang} $d=3:$ $M(3,n-1)\simeq M_{n}$ is the centered configuration
space for n-body systems in 3-space. See Section 3.4.4 for a brief
description of the shape spaces $M_{n}^{\ast}$ when $n$ is small. In
general, for $n>3$ their $G_{2}$-orbit space 
\begin{equation*}
\frac{M_{n}^{\ast}}{G_{2}}=M^{\ast\ast}(3,n-1)\simeq\Delta^{2}
\end{equation*}
is the spherical triangle on the sphere $r_{1}^{2}+r_{2}^{2}+r_{3}^{2}=1$
with vertices 
\begin{equation*}
A=(1,0,0),\ \text{\ }B=(1/\sqrt{2},1/\sqrt{2},0),\text{ \ }C=(1/\sqrt {3},1/%
\sqrt{3},1/\sqrt{3})
\end{equation*}
On the circle $r_{3}=0$ lies the closed circular arc of length $\pi/4$%
\begin{equation*}
\left[ A,B\right] :r_{1}\geq r_{2}\geq0
\end{equation*}
which consists of those points of rank $\leq2$. It has three strata, namely
the vertex $A$ (of rank 1), the vertex $B$ and the open arc $(A,B)$ (both of
rank 2). However, the open arcs 
\begin{equation*}
(A,C):r_{1}>r_{2}=r_{3}>0,\text{ \ \ \ }(B,C):r_{1}=r_{2}>r_{3}>0
\end{equation*}
constitute the same stratum of subrank $\kappa=(2,1)$. Thus, although the
triangle decomposes into $7=3+3+1$ strata components, namely vertices, open
edges and the interior of $\Delta^{2}$, there are only 6 multistrata (since
one of them has $2$ components), in accordance with the enumeration formula (%
\ref{partition}).
\end{example}

\begin{remark}
\label{metric5}In general, the kinematic metric on the shape space $M^{\ast}$
is uniquely determined by its restriction to the principal stratum, which is
a Riemannian manifold ($M_{\kappa_{0}}^{\ast},d\sigma^{2})$. There is a
general procedure, in the setting of equivariant differential geometry, for
the calculation (or description) of the Riemannian connection on such a
principal orbit bundle, see for example \cite{Back}. The simplest case $d=3$%
, as in the above example, is analyzed in Hsiang\cite{Hsiang1}, but similar
calculations can also be done for $d>3$. We shall, however, leave the
geometric issues here and concentrate on the topological structures in the
sequel.
\end{remark}

\subsection{Topology of the shape space and related spaces}

For fixed $m$ the largest shape space (with $d=m)$ is 
\begin{equation}
M^{\ast}(m,m)\simeq Sym^{\ast}(m)  \label{universal}
\end{equation}
and by (\ref{shape2}) it contains any other shape space 
\begin{equation*}
M^{\ast}(d,m)=M^{\ast}(m,m)_{\leq d}\simeq Sym^{\ast}(m)_{\leq d}\text{ }\ 
\end{equation*}
as the union of those strata of matrix rank at most $d$. Therefore, we shall
refer to the space (\ref{universal}) as the \emph{m-universal shape space}.
Thus there is the increasing and $O(m)$-invariant rank filtration 
\begin{equation}
M^{\ast}(1,m)\subset M^{\ast}(2,m)\subset...\subset M^{\ast}(m,m)\ \ 
\label{filtration1}
\end{equation}
and corresponding $O(m)$-orbit spaces $\approx\Delta^{d-1},$ $1\leq d\leq m,$
$\ $

\begin{equation}
M^{\ast\ast}(1,m)\subset M^{\ast\ast}(2,m)\subset...\subset M^{\ast\ast
}(m,m)=\Delta^{m-1}  \label{filtration2}
\end{equation}
where the first three spaces in the chain (\ref{filtration2}) are, indeed,
the inclusions of simplices 
\begin{equation*}
\left\{ A\right\} =\Delta^{0}\subset\lbrack A,B]=\Delta^{1}\subset\lbrack
A,B,C]=\Delta^{2}
\end{equation*}
from Example \ref{triang}.

By inspection the chain in (\ref{filtration1}) starts with 
\begin{align}
M^{\ast}(1,m) & =S^{m-1}/O(1)=\mathbb{R}P^{m-1}\   \notag \\
M^{\ast}(2,m) & =S^{2m-1}/O(2)=\mathbb{C}P^{m-1}/\mathbb{Z}_{2}  \label{CP2}
\end{align}
where 
\begin{equation*}
\mathbb{R}P^{m-1}\subset\mathbb{C}P^{m-1}=S^{2m-1}/SO(2)
\end{equation*}
are the real and complex the projective (m-1)-space, and $\mathbb{Z}%
_{2}=O(2)/SO(2)$ acts on $\mathbb{C}P^{m-1}$ by complex conjugation with $%
\mathbb{R}P^{m-1}$ as fixed point set. In particular, the vertex $\Delta
^{0}=M^{\ast\ast}(1,m)$ of the simplex $\Delta^{m-1}$ in (\ref{filtration2})
is the single $O(m)$-orbit $M^{\ast}(1,m)$, with the topology of a real
projective space as noted above.

The topology of the next shape space $M^{\ast}(2,m)$ is more difficult to
describe. One approach is to utilize the fact that it is a $O(m)$-space of 
\emph{cohomogeneity one}, that is, with one-dimensional orbit space. This
holds since 
\begin{equation*}
\frac{M^{\ast}(2,m)}{O(m)}=M^{\ast\ast}(2,m)\simeq\Delta^{1}=[A,B]\text{ }
\end{equation*}
is an interval. We also know the orbit types $(K_{1}),(K_{2}),(H)$
corresponding to the three strata $A,B$ and $(A,B)$ respectively. Therefore,
the space can be described topologically in terms of its orbit types by the
construction (cf. e.g. \cite{Straume3}, Vol.1, Chap. IV) known as the \emph{%
equivariant union} 
\begin{equation*}
M^{\ast}(2,m)=M(\pi_{1})\cup M(\pi_{2})
\end{equation*}
of the mapping sylinders $M(\pi_{i})$ of the canonical projections $\pi
_{i}:O(m)/H\rightarrow O(m)/K_{i}$, where in our case 
\begin{align}
K_{1} & =O(1)\times O(m-1)\text{, \ }K_{2}=O(2)\times O(m-2)\text{, \ }
\label{iso} \\
H & =K_{1}\cap K_{2}=O(1)^{2}\times O(m-2)\text{ }  \notag
\end{align}

The space $M^{\ast }(2,2)$ is a 2-disk and $M^{\ast }(2,3)\ $is actually
homeomorphic to $S^{4}$, see (\ref{homeo}) and Section 3.4.4). For $m>3$ $%
M^{\ast }(2,m)$ fails to be a manifold in a neighborhood of the singular $%
O(m)$-orbit $\ $ 
\begin{equation*}
A=O(m)/K_{1}=M^{\ast }(1,m)\simeq \mathbb{R}P^{m-1},
\end{equation*}%
but two copies of $M^{\ast }(2,m)$ glued together along $A$ yields, indeed,
the differentiable manifold $S^{2m-1}/SO(2)=\mathbb{C}P^{m-1}$. Below we
will return to this construction and further investigate the topology of the
spaces $M^{\ast }(d,m)$ for $2<d\leq m.$

\subsubsection{Local and global topology}

By referring to Section 3.1 concerning compact transformation groups and the
slice theorem, let us first recall\ a nice property of \emph{regular}
representations of classical groups such as $O(d)$. Namely,\ we consider the
linear action of $O(d)$ on $\mathbb{R}^{N}$ by some representation of type $%
\Phi =m\rho _{d}+\tau _{q}$, $N=md+q$, which is $m$ copies of the standard
representation $\rho _{d}$ plus a $q$-dimensional trivial summand $\tau _{q}$%
. Then the orbit types constitute the following "connected" string of \emph{%
regular} subgroups $\ $ 
\begin{equation}
O(k)\text{ ; \ }\ \text{ }d-m\ \leq k\leq d  \label{regular}
\end{equation}%
where $O(k)=1$ if $k\leq 0$. To calculate the slice representation $\Phi
_{k} $ at a point with isotropy group $O(k)$ we calculate the difference as
in (\ref{slicerep}), with the local representation equal to $\Phi $
restricted to $O(k)$, that is, $m\rho _{k}+(trivial)$. Moreover, the
isotropy representation $Iso(d,k)$ of $O(k)$ is the "linearized" action on
the Stiefel manifold $O(d)/O(k)=SO(d)/SO(k)$ at the base point, and a simple
calculation (involving the adjoint representation of $O(d)$) yields the
representation $(d-k)\rho _{k}+(trivial)$, consequently \ 
\begin{equation}
\Phi _{k}=\Phi |_{O(k)}-Iso(d,k)\ \equiv (m-d+k)\rho _{k}\text{ (mod trivial
)}\   \label{slice}
\end{equation}%
From this we conclude that each slice representation inherits the \emph{%
regularity} property of the original action $\Phi $.

Now we turn to the \textquotedblright left side\textquotedblright\ action
of\ $O(d)$ on the matrix space $M=M(d,m)\simeq \mathbb{R}^{md}$ , which is
just the regular representation $\Phi =m\rho _{d}$. However, we shall rather
consider the restricted action on the unit sphere $M^{1}$, noting that the
only change in the above calculations is that the local representations lose
a trivial summand $\tau _{1}$. The orbit types still constitute the string (%
\ref{regular}) and also (\ref{slice})\ holds, except that $k<d$ in (\ref%
{regular}) since $O(d)$ has no fixed point on the sphere $M^{1}$.

\begin{lemma}
(i) The m-universal shape space $M^{\ast}(m,m)$ $\simeq Sym^{\ast}(m)$ is a
manifold with boundary $\ $ 
\begin{equation*}
\partial M^{\ast}(m,m)\ =M^{\ast}(m-1,m)\simeq Sym^{\ast}(m)_{\leq m-1}\ 
\end{equation*}

(ii) For $p>0,$ $M^{\ast}(d,d+p)$ is a manifold (and with no boundary) if
and only if $p=1$ or $d=1$.
\end{lemma}

\begin{proof}
By (\ref{local}), the local topology of $M^{\ast }(m,m)$ around an $O(m)$%
-orbit of type $O(k)$ is the topology of the orbit space of the slice
representation $(O(k),\Phi _{k})$ in $M^{1}$, with $m=d$ in (\ref{slice}).
Hence, part (i) of the lemma follows by induction on $m\geq 1$, since at the
initial step $m=1$ the orbit space of $(O(1),\rho _{1})$ is the half-line $%
\mathbb{R}_{+}^{1}=\left[ 0,\infty \right) .$

For the proof of (ii) we may assume $d>1$, and the slice representation (\ref%
{slice}) of $O(k)$ is $\Phi_{k}=(k+p)\rho_{k}$ (mod trivial). By induction
the proof reduces to the crucial singular case $(O(1),(1+p)\rho _{1})$,
where $O(1)$ acts by inversion $v\rightarrow-v,$ and here the orbit space is
the cone $\frac{\mathbb{R}^{p+1}}{O(1)}=C(\mathbb{R}P^{p})\ $over the real
projective p-dimensional space.$\ $This is a manifold if and only if $p=1$,
in which case the orbit space is homeomorphic to the Euclidean plane 
\begin{equation}
\frac{\mathbb{R}^{2}}{O(1)}=C(\mathbb{R}P^{1})\approx C(S^{1})=\mathbb{R}^{2}
\label{cone3}
\end{equation}
\ \ \ \ \ \ 
\end{proof}

Next, let us actually determine the topological type of the compact manifold 
$M^{\ast}(m,m)\simeq Sym^{\ast}(m)$ and its boundary. By the $O(m)$-action
on $M(m,m)$ every $X=(x_{ij})$ can be mapped to the subset 
\begin{equation*}
T^{+}(m)\subset M(m,m)
\end{equation*}
of upper triangular matrices with diagonal entries $x_{ii}\geq0,$ and
therefore the restriction of the polynomial map in (\ref{polynomial1}) 
\begin{equation}
\mathfrak{p:}\ T^{+}(m)\rightarrow Sym^{+}(m)\   \label{triangular}
\end{equation}
is still surjective. The subset $T^{+}(m)_{m}$ of matrices $X$ with all $%
x_{ii}>0$, i.e. of maximal rank $m$, is clearly diffeomorphic to the
Euclidean space $\mathbb{R}^{m(m+1)/2}$, and it is easy to verify that
different matrices $X$ lie on different $O(m)$-orbits. Therefore the
following lemma must hold. $\ $

\begin{lemma}
The polynomial map in (\ref{triangular}) further restricts to a
diffeomorphism $\ $ 
\begin{equation*}
\mathfrak{p}:\text{ }T^{+}(m)_{m\ }\simeq Sym^{+}(m)_{m}\ \simeq \mathbb{R}%
^{m(m+1)/2}
\end{equation*}
with the set of positive definite symmetric matrices. Hence, the interior of 
$Sym^{\ast}(m)$ is the open \textquotedblright
sphere-octant\textquotedblright\ 
\begin{equation*}
Sym^{\ast}(m)_{m}\simeq T^{+}(m)_{m\ }\cap M^{1}(m,m)\ :\sum
x_{ij}^{2}=1,x_{ii}>0
\end{equation*}
which is diffeomorphic to $\mathbb{R}^{m(m+1)/2-1}$.
\end{lemma}

We know from the previous two lemmas that $Sym^{\ast}(m)$ is a compact
manifold with boundary and its interior is an open disk. Using some manifold
theory this information actually suffices to conclude that $Sym^{\ast}(m)$
is a closed disk. For example, one may apply the so-called collar
neighborhood theorem and the h-cobordism theorem, cf. Milnor\cite{Milnor}).
Anyhow, we have the topological types 
\begin{align}
M^{\ast}(m,m) & \simeq Sym^{\ast}(m)\approx D^{m(m+1)/2-1}  \label{homeo} \\
M^{\ast}(m-1,m) & \simeq Sym^{\ast}(m)_{\leq m-1}\approx S^{m(m+1)/2-2} 
\notag
\end{align}
and taking the cone over these spaces yields the topological types 
\begin{align*}
\ \bar{M}(m,m) & \approx\mathbb{R}^{m(m+1)/2-1}\times\left[ 0,\infty \right) 
\text{ \ \ (half-space)} \\
\bar{M}(m-1,m) & \simeq\partial\bar{M}(m,m)\ \approx\mathbb{R}^{m(m+1)/2-1}%
\text{ \ \ (Euclidean space)}
\end{align*}

\begin{remark}
The linear model construction in Section 3.4.3 provides another proof of the
homeomorphisms (\ref{homeo}).
\end{remark}

\subsubsection{Branched coverings and generalized Hopf fibrations}

The usage of $SO(d)$ rather than $O(d)$ as the congruence group acting on $%
M(d,m)$ may lead to a different "shape space", and we shall explain this
distinction below. First, observe that for $k\geq1$ we have connected and
hence equal orbits 
\begin{equation*}
O(d)/O(k)\simeq SO(d)/SO(k)\text{ }
\end{equation*}
and consequently the orbit spaces of $SO(d)$ and $O(d)$ will be identical as
long as $d>m,$ see (\ref{regular}). However, due to (\ref{d-large}) we have
been assuming $d\leq m$ and therefore $k=0$ also occurs in the string (\ref%
{regular}), so the two orbit spaces cannot be identical. In order to explain
the difference the following discussion will be helpful.

Consider a space $\tilde{Q}$ with a given involution $\sigma$ (i.e. a
transformation of order two) and hence we shall regard $\mathbb{Z}%
_{2}=\left\{ Id,\sigma\right\} $ as a transformation group on $\tilde{Q}$.
We denote the orbit space of $\mathbb{Z}_{2}$ by $Q$ and the fixed point set
by $\Sigma$. We also assume $\Sigma\neq\varnothing$ and regard it as a
subset of both $\tilde{Q}$ and $Q$. Then the orbit map 
\begin{equation}
\pi:\tilde{Q}\rightarrow\frac{\tilde{Q}}{\mathbb{Z}_{2}}=Q  \label{ramif}
\end{equation}
is an example of a \emph{ramified double covering} which is ramified along $%
\Sigma$. Conversely, starting from $Q$ and $\Sigma$ we may reconstruct $%
\tilde{Q}$ as the \emph{double }of $Q$ 
\begin{equation*}
2Q=Q\cup_{\Sigma}Q\simeq\tilde{Q}
\end{equation*}
by taking two copies of $Q$ and identify (or glue together) their "singular"
set $\Sigma$. However, the pair $(Q,\Sigma)$ does not always lead to a
unique double space, so we shall rather have in mind a specified ramified
covering as in (\ref{ramif}). Thus, we define for $d\leq m$ 
\begin{equation}
2M^{\ast}(d,m)=M^{\ast}(d,m)\cup_{M^{\ast}(d-1,m)}M^{\ast}(d,m)
\label{double}
\end{equation}
and refer to the orbit map 
\begin{equation}
2M^{\ast}(d,m)\simeq\frac{M^{1}(d,m)_{\ }}{SO(d)}\ \ \rightarrow\frac {%
M^{1}(d,m)_{\ }}{O(d)}=M^{\ast}(d,m)  \label{2-covering}
\end{equation}
of the action of $O(d)/SO(d)=$ $\mathbb{Z}_{2}$, which is, indeed, a double
covering ramified along the fixed point set $M^{\ast}(d-1,m)$ of $\mathbb{Z}%
_{2}$.

\begin{lemma}
For $p\geq0$ and $d>1$, the space $2M^{\ast}(d,d+p)$ is a manifold (and with
no boundary) if and only if $p=0$ or $d=2$.
\end{lemma}

\begin{proof}
For $d=2$ the $SO(2)$-orbit space is the complex projective space $\mathbb{C}%
P^{p+1}$, so let us assume $d>2$. Using again the slice theorem and an
inductive argument, the crucial case will be the topology of the orbit space
of $(SO(2),(p+2)\rho_{2})$, namely the cone 
\begin{equation*}
\frac{\mathbb{R}^{2(2+p)}}{SO(2)}=\ C(\frac{S^{2(2+p)-1}}{SO(2)})=C(\mathbb{C%
}P^{p+1})
\end{equation*}
over the complex projective space $\mathbb{C}P^{p+1}$. It is well known that
this is a manifold (i.e. locally Euclidean) if and only if $p=0$, in which
case it is a cone homeomorphic to Euclidean 3-space 
\begin{equation}
C(\mathbb{C}P^{1})\approx C(S^{2})=\mathbb{R}^{3}  \label{cone4}
\end{equation}
\end{proof}

\begin{lemma}
\label{cone}The induced smooth functional structure on the cone (\ref{cone3}%
) (resp. (\ref{cone4})) is a refinement of the standard differentiable
structure of $\mathbb{R}^{2}$ (resp. $\mathbb{R}^{3}$). The structures are
identical away from the origin (cone vertex), where the cone fails to be a
smooth manifold.
\end{lemma}

\begin{proof}
Consider the case (\ref{cone4}), namely the orbit space of $%
(SO(2),2\rho_{2}) $ acting on $\mathbb{R}^{2}\times\mathbb{R}^{2}$. To
describe it algebraically, let $\mathbf{u,v}$ denote vectors in $\mathbb{R}%
^{2}$ and define the polynomial map 
\begin{equation*}
\mathfrak{p}=(X,Y,Z,W):\mathbb{R}^{2}\times\mathbb{R}^{2}\rightarrow \mathbb{%
R}^{4}\text{ \ , see (\ref{orbitmap})}
\end{equation*}
where $X=$ $\left\vert \mathbf{u}\right\vert ^{2}-\left\vert \mathbf{v}%
\right\vert ^{2},Y=2\mathbf{u\cdot v},Z=2(\mathbf{u\times v)\cdot k,}\
W=\left\vert \mathbf{u}\right\vert ^{2}+\left\vert \mathbf{v}\right\vert
^{2} $, which identifies the orbit space with the cone in 4-space 
\begin{equation*}
X^{2}+Y^{2}+Z^{2}=W^{2},\text{ \ }W\geq0
\end{equation*}
By projecting the cone onto the coordinate $(X,Y,Z)$-space we identify it
with $\mathbb{R}^{3}$ as in (\ref{cone4}), with (standard) coordinate
functions $X,Y,Z$ which are certainly smooth in the induced orbital
functional structure. However, at the origin the function $f=\sqrt{%
X^{2}+Y^{2}+Z^{2}}$ is not smooth with respect to the standard structure,
although its composition with the orbit map is by our construction the
smooth function $W.$ The case (\ref{cone3}) is similar but simpler, using $%
\mathfrak{p}=\ (X,Y,W)$ and $(u,v)\in\mathbb{R}^{1}\times\mathfrak{\ }%
\mathbb{R}^{1}.$
\end{proof}

According to (\ref{homeo}) and (\ref{2-covering}) we have constructions of
orbital fibrations of spheres over spheres involving two types of orthogonal
transformation groups, as follows :

\begin{description}
\item \textbf{Case 1 : }Consider the regular representation $(O(m-1),m\rho
_{m-1})$, acting on the unit sphere $M^{1}(m-1,m)=S^{(m-1)m-1}$ with orbit
space as in the second line of (\ref{homeo}), 
\begin{equation}
S^{(m-1)m-1}\rightarrow\frac{S^{(m-1)m-1}}{O(m-1)}\approx S^{m(m+1)/2-2}%
\text{, \ \ }m=2,3,4,...  \label{Hopf2}
\end{equation}
Here the initial case $m=2$ is the trivial Hopf fibration or double covering 
\begin{equation}
O(1)\hookrightarrow S^{1}\rightarrow^{\pi}\ \mathbb{R}P^{1}=\frac{S^{1}}{O(1)%
}\simeq S^{1}  \label{Hopf1}
\end{equation}

\item \textbf{Case 2 : }Consider the\ regular representation $(SO(m),m\rho
_{m})$, acting on the unit sphere $M^{1}(m,m)=S^{m^{2}-1}$. The orbit space
coincides with the \textquotedblright double\textquotedblright\ construction
(\ref{double}) applied to the first line of (\ref{homeo})$\ $ \ 
\begin{equation*}
\ \frac{M^{1}(m,m)}{SO(m)}=2M^{\ast}(m,m)%
\approx2D^{m(m+1)/2-1}=S^{m(m+1)/2-1}\ \ 
\end{equation*}
and there is the orbit map 
\begin{equation}
S^{m^{2}-1}\rightarrow\frac{S^{m^{2}-1}}{SO(m)}\approx S^{m(m+1)/2-1}\text{,
\ }m=2,3,4,....  \label{Hopf3}
\end{equation}
which for $m=2$ is the well known Hopf fibration 
\begin{equation}
SO(2)\hookrightarrow S^{3}\rightarrow^{\pi}\ \mathbb{C}P^{1}=\frac{S^{3}}{%
SO(2)}\simeq S^{2}  \label{Hopf}
\end{equation}
\end{description}

\begin{remark}
\label{artif}With the induced differential structure the above \emph{%
quotient spheres} are, of course, stratified differentiable manifolds (cf.
Section 3.1) which are also locally Euclidean spaces since they are spheres
in the topological sense. In fact, for $m=2$ they are the standard sphere
(i.e. $S^{1}$ or $S^{2}$) in the differentiable sense as well. However, for $%
m>2$ there are more than one orbit type, and with the induced differential
structure they are actually not differentiable manifolds. However, this
artificial situation can be remedied by "relaxing" the differential
structure so that the quotient sphere becomes the standard sphere while the
map $\pi$ is still differentiable, see Section 3.4.3.
\end{remark}

\begin{remark}
The above orbital fibrations are natural maps between spheres with specific
properties, with possible future applications in physics. Indeed, the Hopf
fibration (\ref{Hopf}) is a simple special case with several well known
applications. For example, it describes the geometry of a magnetic monopole,
and Dirac made the major discovery that the fibration could explain (in a
modern language) the quantization of electric charge. Here $S^{3}$ is the
unit sphere of $\mathbb{R}^{4}=\mathbb{C}^{2}$, the action of $SO(2)\simeq
U(1)$ extends to $\mathbb{C}^{2}$ by complex scalar multiplication, and the
point magnetic source is at the origin of $\mathbb{C}^{2}/U(1)\approx\mathbb{%
R}^{3}$. The 4-space $\mathbb{R}^{4}$ in this setting is also referred to as
the Kaluza-Klein model of the Dirac monopole.
\end{remark}

Arnold\cite{Arnold2} has considered the three special cases $m=2,3,5$ of (%
\ref{Hopf2}) from a different viewpoint. Namely, they fit into the following
unifying pattern with projective spaces over $\mathbb{R},\mathbb{C}$ or $%
\mathbb{H}$ as intermediate quotient spaces : 
\begin{align}
i)\text{ } & \text{\ }S^{1}\rightarrow\mathbb{R}P^{1}\simeq S^{1}  \notag \\
ii)\text{ \ } & S^{5}\rightarrow\mathbb{C}P^{2}\rightarrow\mathbb{C}%
P^{2}/O(1)\approx S^{4}\text{\ }  \label{success1} \\
iii)\text{ } & \text{\ }S^{19}\rightarrow\mathbb{H}\text{\ }P^{4}\rightarrow%
\mathbb{H}\text{\ }P^{4}/\left\langle Aut\mathbb{H},\mathbb{Z}%
_{2}\right\rangle \text{\ }\approx S^{13}  \notag
\end{align}
where $O(1)\simeq\mathbb{Z}_{2}$ acts by (complex) conjugation and $Aut%
\mathbb{H}\simeq SO(3)$ is the automorphism group of $\mathbb{H}$
(quaternions).The last two cases are two-step orbital fibrations defined by
groups $H\rightarrow K\rightarrow K/H$, that is, $K$ acts on the sphere on
the left side in (\ref{success1}), and the intermediate (projective) space
is the orbit space of the normal subgroup $H$, with the induced action of $%
K/H$. The groups corresponding to the last two cases of (\ref{success1}) are
as follows 
\begin{equation*}
ii)\text{ }SO(2)\rightarrow O(2)\rightarrow O(1)\text{, \ \ }iii)\text{ }%
Sp(1)\rightarrow O(4)\rightarrow O(3)
\end{equation*}
where $O(3)=SO(3)\times\mathbb{Z}_{2}$ $\simeq\left\langle Aut\mathbb{H},%
\mathbb{Z}_{2}\right\rangle $. However, from this viewpoint it is also
natural to include the cases $m=2,4$ of (\ref{Hopf3}), where there is no
group $\mathbb{Z}_{2}$. Case $m=2$ is the Hopf fibration (\ref{Hopf}) and
the new case $m=4$ reads 
\begin{equation}
S^{15}\rightarrow\mathbb{H}\text{\ }P^{3}\rightarrow\mathbb{H}\text{\ }%
P^{3}/Aut\mathbb{H}\text{ }\approx S^{9}  \label{success2}
\end{equation}
with the corresponding groups $Sp(1)\rightarrow SO(4)\rightarrow SO(3)$.

\subsubsection{ The m-universal linear model}

We shall describe another prperty of the m-universal shape space $M^{\ast
}(m,m)\simeq Sym^{\ast }(m),$ which by (\ref{homeo}) is known to be a disk
whose boundary sphere is $M^{\ast }(m-1,m)$. In fact, what is remarkable is
that the equivariant system $(O(m),Sym^{\ast }(m))$ has a \emph{linear model}%
, namely it "resembles" closely a Euclidean disk with $O(m)$ acting as an
orthogonal transformation group. To make this correspondence precise, let us
first inquire what is the appropriate linear model.

Recall from Section 3.2 that $O(m)$ acts orthogonally on the space $Sym(m)$
of symmetric matrices of dimension $m$, namely by the symmetric tensor
product representation $S^{2}\rho _{m}$. Let $Sym^{0}(m)$ be the subspace of
matrices of trace zero, where $O(m)$ acts by the irreducible representation $%
S^{2}\rho _{m}-1$, and let 
\begin{equation*}
D^{0}(m)\subset Sym^{0}(m)\text{ \ \ \ }\ \text{cf. (\ref{Symnull})}
\end{equation*}%
be the unit disk centered at the origin. Then we make the following
definition which will be justified below.

\begin{definition}
The above equivariant system $(O(m),D^{0}(m))$ is the \emph{linear model} of 
$(O(m),M^{\ast}(m,m))$, and the restriction to the boundary sphere $\partial
D^{0}(m)$ is the\emph{\ linear} \emph{model} of $\partial
M^{\ast}(m,m)=M^{\ast}(m-1,m)$.
\end{definition}

Now, we shall construct a 1-1 correspondence between $Sym^{\ast }(m)$ and
the disk $D^{0}(m)$ which is smooth and $O(m)$\emph{-equivariant}, that is,
the map commutes with the action of $O(m)$. To each matrix $Y\in Sym^{\ast
}(m)$, namely a positive semidefinite matrix with trace 1, we associate the
matrix 
\begin{equation*}
Y_{0}=\frac{1}{r_{m}}(Y-\frac{1}{m}Id)\in Sym^{0}(m)\text{, \ \ where \ }%
r_{m}=\sqrt{1-\frac{1}{m}}\text{ ,}
\end{equation*}%
and note that $Y_{0}$ is perpendicular to the identity $Id$ and 
\begin{equation*}
\ \left\vert Y_{0}\right\vert ^{2}=\frac{1}{r_{m}^{2}}(\left\vert
Y\right\vert ^{2}-\frac{1}{m})=\frac{1}{r_{m}^{2}}(trace(Y^{2})-\frac{1}{m}%
)\leq \frac{1}{r_{m}^{2}}(trace(Y)^{2}-\frac{1}{m})\ =1\ 
\end{equation*}%
Therefore, the $O(m)$-equivariant affine transformation 
\begin{equation*}
Sym(m)\rightarrow Sym(m):Y\rightarrow \frac{1}{r_{m}}(Y-\frac{1}{m}Id)\
=Y_{0}
\end{equation*}%
restricts to an embedding 
\begin{equation}
Sym^{\ast }(m)\rightarrow D^{0}(m)\text{ ,\ \ }Y\rightarrow Y_{0}\ 
\label{linear}
\end{equation}%
between disks of the same dimension, which maps the\emph{\ geometric center} 
$\frac{1}{m}Id$ to the origin, i.e., the center of $D^{0}(m)$.

Thus, by a translation\ and homothety inside the Euclidean space $Sym(m)$
the convex subset $Sym^{\ast }(m)$ becomes, somehow, an \textquotedblright
inward\textquotedblright\ equivariant deformation of $D^{0}(m)$, which by (%
\ref{Sym1}) fixes the subset $Sym^{\ast }(m)_{1}\simeq \mathbb{R}P^{m-1}$
lying on the boundary sphere $\partial D^{0}(m)$. In the simplest case $m=2$
we have more explicitly 
\begin{equation}
Sym^{\ast }(2)=\left\{ \left( 
\begin{array}{cc}
u & v \\ 
v & 1-u%
\end{array}%
\right) \text{ ; \ }(u-\frac{1}{2})^{2}+v^{2}\leq \frac{1}{4}\right\} \simeq
D^{2}  \label{sym2}
\end{equation}%
and the embedding (\ref{linear}) is actually a diffeomorphism.

However, the embedding (\ref{linear}) is not surjective for $m>2$, so let us
explain how to further deform equivariantly to make the embedding fill the
whole unit disk. First of all, by the convexity of $Sym^{\ast }(m)$ it
follows that each ray in $D^{0}(m)$ from the origin passes through a unique
point of the embedded sphere $\partial Sym^{\ast }(m)$. In particular, by an
additional scaling we obtaing a 1-1 correspondence between the boundary
spheres of the two disks, and the final composition 
\begin{equation}
\psi _{m}:\partial Sym^{\ast }(m)\rightarrow \partial D^{0}(m)=S^{m(m+1)/2-2}%
\text{ , \ \ }Y\rightarrow Y_{0}\rightarrow \frac{1}{\left\vert
Y_{0}\right\vert }Y_{0}  \label{model2}
\end{equation}%
is certainly an $O(m)$-equivariant and smooth homeomorphism.

On the other hand, the above map may be extended to the whole disk, as
follows. First of all, each ray from the origin intersects the embedded disk 
$Sym^{\ast }(m)$ in a segment. So, let us stretch the segment outward along
the ray until it has unit length, but with no stretching in a neigborhood of
the origin. Moreover, the stretching must be specified by a function on the
orbit space since $O(m)$-equivalent segments must be stretched in the same
way to make $\psi _{m}$ equivariant. \ Following this procedure we certainly
obtain an $O(m)$-equivariant and smooth homeomorphism 
\begin{equation}
\psi _{m}:Sym^{\ast }(m)\rightarrow ^{\approx }D^{0}(m)  \label{model1}
\end{equation}%
which extends the map in (\ref{model2}). Such an equivariant and smooth
homeomorhism is not unique, of course. The ambiguity lies in the group $%
Diff_{O(m)}(D^{0}(m))$ consisting of all equivariant diffeomorphisms of the
linear model. Yet, another construction of equivariant homeomorphisms like $%
\psi _{m}$ is decribed briefly at the end of this subsection.

In summary, we arrive at the following result :

\begin{theorem}
\label{model3}There is an $O(m)$-equivariant and differentiable
homeomorphism $\psi_{m}$ from the m-universal shape space $M^{\ast}(m,m)\
\simeq Sym^{\ast }(m)$ to its linear model $D^{0}(m)\simeq D^{m(m+1)/2-1}$.
For $m=2$ this map is a diffeomorphism.
\end{theorem}

By referring to the following diagram%
\begin{equation}
\begin{array}{ccccc}
&  & M^{1}(m,m) &  &  \\ 
&  & \downarrow \pi &  &  \\ 
\mathbb{R} & \longleftarrow ^{g} & Sym^{\ast }(m) & \longrightarrow ^{\psi
_{m}} & D^{0}(m)%
\end{array}
\label{diagram2}
\end{equation}%
let us explain the differential structure of $Sym^{\ast }(m)$ induced via
the orbit map $\pi $. Namely, the function $g$ is said to be smooth if the
composed map $g\circ \pi $ is smooth. Moreover, smoothness of $\psi _{m}$
means the composed map $\psi _{m}\circ \pi $ is smooth, and therefore we
actually know (by our construction) that $\psi _{m}$ is smooth.

On the other hand, for $m>2$ there also exists a smooth map $g$ such that $%
g\circ\psi_{m}^{-1}$ is not smooth, as a function on the Euclidean disk $%
D^{0}(m)$, and consequently $\psi_{m}^{-1}$ cannot be smooth. Here we also
refer to the discussion at the end of Section 3.4.4, together with Lemma \ref%
{cone} and Remark \ref{artif}. Briefly, the differential structure on $%
Sym^{\ast}(m)$ induced via $\pi$ is a strict "refinement" of the \emph{%
standard }structure which $Sym^{\ast}(m)$ inherits from the Euclidean disk
(via\ $\psi_{m}$).

\begin{remark}
\label{induced}We have seen that the m-universal shape space $Sym^{\ast }(m)$
has two naturally induced smooth structures, namely induced via $\pi $ and $%
\psi _{m}^{-1}$ respectively, and they are different when $m>2$. In fact, $%
Sym^{\ast }(m)$ is not a smooth manifold in the first case. But in both
cases $\pi $ and $\psi _{m}$ are smooth maps, but the standard structure
(induced via $\psi _{m}^{-1}$) is the only one that makes $\psi _{m}$ an $%
O(m)$-equivariant diffeomorphism.
\end{remark}

Finally, let us turn to the natural simplicial structures of the orbit
spaces of $Sym^{\ast }(m)$ and the disk $D^{0}(m)$, namely the following two
spherical simplices 
\begin{align}
\ \ \text{ \ \ }\frac{Sym^{\ast }(m)}{O(m)}& \simeq \Delta ^{m-1}=\frac{%
S^{m-1}}{B_{m}}:r_{1}\geq r_{2}\geq ...\geq r_{m}\geq 0,\text{ }\sum
r_{i}^{2}=1  \label{orbitspace3} \\
\ \ \frac{D^{0}(m)}{O(m)}& \simeq \bar{\Delta}^{m-1}=\frac{D^{m-1}}{S_{m}}%
:\lambda _{1}\geq \lambda _{2}\geq ...\geq \lambda _{m},\text{ }\sum \lambda
_{i}=0,\text{ }\sum \lambda _{i}^{2}\leq 1  \notag
\end{align}%
Despite the above theorem, which implies that\ $\psi _{m}$ induces an $O(m)$%
-orbit type strata preserving homeomorphism $\bar{\psi}_{m}:\Delta
^{m-1}\rightarrow \bar{\Delta}^{m-1}$, the two structures in (\ref%
{orbitspace3}) are conspicuously different. The reason is that the
simplicial structure in the first case of reflects the induced subrank
stratification of $Sym^{\ast }(m)$ (cf. Section 3.3.3), namely the common
refinement of the rank and $O(m)$-orbit type stratification, whereas the
simplicial structure in the second case is merely reflecting the pure $O(m)$%
-orbital stratification of the orthogonal transformation group $O(m)$ on $%
D^{0}(m)$. Thus, by passing from $Sym^{\ast }(m)$ to its linear model $%
D^{0}(m)$ the notion of rank is seemingly lost. Therefore, let us also
investigate how the rank strata can be recognized in the linear model itself.

Recall that $M^{\ast }(k,m)\subset Sym^{\ast }(m)$ is the subspace lying
above $\Delta ^{k-1}:r_{k+1}=...=r_{m}=0$, and the rank $k$ stratum is
defined by the subset $\Delta _{k}^{k-1}\subset \Delta ^{k-1}:r_{k}>0$. This
stratum has all the $O(m)$-orbit types 
\begin{equation}
O(k_{1},k_{2},..,k_{p})\times O(m-k),\text{ \ }\sum k_{i}=k\text{ , cf. (\ref%
{isotropy})}  \label{iso1}
\end{equation}%
labelled by the various subranks $\kappa =(k_{1},..,k_{p})$ which record the
strings of equalities among the numbers $r_{i}>0$, see (\ref{subrank3}), (%
\ref{multirank}).

On the other hand, in the linear disk model any pattern of equality strings
among the $\lambda _{i}$'s corresponds in the same way to a tuple $\kappa
^{\prime }=(k_{1},..,k_{p},k_{p+1})$ of positive integers, where $k_{1}$ is
the number of $\lambda _{i}$'s in (\ref{orbitspace3}) equal to $\lambda _{1}$%
, $k_{2}$ is the number of $\lambda _{i}$'s equal to $\lambda _{k_{1}+1}$
etc., and the factor $O(m-k)$ in (\ref{iso1}) is replaced by $O(k_{p+1})$.
In any case, the last factor $O(m-k)$ (resp. $O(k_{p+1})$) of the isotropy
group (\ref{iso1}) is no more special than the other factors $O(k_{i})$, and
this clearly explains why the rank is not determined by the $O(m)$-orbit
type.

We claim, however, the rank $k$ is determined in the linear model by the
identity $k_{p+1}=m-k$ provided the obvious condition $\sum \lambda
_{i}^{2}=1$ holds. The latter condition merely says $M^{\ast }(k,m)$ embeds
as a subset of the sphere $\partial D^{0}(m)$ and hence away from the
interior $D^{0}(m)$ of the disk. The interior is, of course, given by $\sum
\lambda _{i}^{2}<1$ and here the rank is $k=m$. Moreover, by removing $%
k_{p+1}$ from the tuple $\kappa ^{\prime }$ we are left with the correct
subrank tuple $\kappa $. The above claim about $k$ will be settled below.

Any $O(m)$-equivariant homeomorphism 
\begin{equation*}
\psi :Sym^{\ast }(m)\rightarrow ^{\approx }D^{0}(m)
\end{equation*}%
such as $\psi _{m},$ for example, induces an $O(m)$-orbital strata
preserving homeomorphism $\bar{\psi}:\Delta ^{m-1}\rightarrow \bar{\Delta}%
^{m-1}$. Conversely, let us see how to start from $\bar{\psi}$ and construct
an appropriate lifting $\psi $ as above. The idea is to construct $\bar{\psi}
$ as a map between fundamental domains with the $O(m)$-\emph{isovariant}
property, that is, a point and its image point have the same isotropy group.
Then $\psi $ will be the unique equivariant extension to all of $Sym^{\ast
}(m)$. To choose appropriate fundamental domains, first observe that $\bar{%
\Delta}^{m-1}\subset D^{0}(m)$ is a fundamental domain if the numbers $%
\lambda _{i}$ in (\ref{orbitspace3}) are regarded as the entries of a
diagonal matrix. Similarly, the following subset of diagonal matrices 
\begin{equation*}
\tilde{\Delta}^{m-1}\subset Sym^{\ast }(m):\lambda _{1}\geq \lambda _{2}\geq
...\geq \lambda _{m}\geq 0,\text{ }\sum \lambda _{i}=1
\end{equation*}%
is a fundamental domain in $Sym^{\ast }(m)$. Then the $O(m)$-orbit map
projection $\tilde{\Delta}^{m-1}\rightarrow $ $\Delta ^{m-1}$ is just $%
\lambda _{i}\rightarrow \sqrt{\lambda _{i}}=r_{i}$, see (\ref{diffeo}), and
we note that $\partial \tilde{\Delta}^{m-1}$ $\simeq \tilde{\Delta}^{m-2}$
is the subset with $\lambda _{m}=0$.

For example, the above equivariant homeomorphism $(\ref{model1})$ between
the boundary spheres corresponds to the $O(m)$-isovariant\emph{\ }map

\begin{equation}
\frac{S^{m-2}}{B_{m-1}}\simeq\tilde{\Delta}^{m-2}=\partial\tilde{\Delta}%
^{m-1}\rightarrow^{\bar{\psi}_{m}}\partial\bar{\Delta}^{m-1}\simeq \frac{%
S^{m-2}}{S_{m}}  \label{isov}
\end{equation}
which sends $Y\ =diag(\lambda_{1},..,\lambda_{m-1},0)$ to the matrix

\begin{equation}
\ \frac{1}{\left\vert Y_{0}\right\vert }Y_{0}=\frac{1}{\lambda }diag(\lambda
_{1}-1/m,..,\lambda _{m-1}-1/m,-1/m)\text{ \ }\   \label{map}
\end{equation}%
where $\lambda =\sqrt{\lambda _{1}^{2}+...+\lambda _{m-1}^{2}-1/m}$.

The map (\ref{isov}) is isovariant since the inequality pattern among the
entries of the matrix $Y$ and its image is preserved. From this it is also
clear that $\lambda _{k+1}=0$ if and only if the $m-k$ last entries in (\ref%
{map}) are identical. This also settles the above claim concerning the rank $%
k$ recognition in the linear model.

\subsubsection{$\ $The simplest shape spaces}

We return to the shape spaces $M^{\ast }(3,m)$ together with their $SO(3)$%
-version for $m=2,3,4$. Recall from Section 1 that $M^{\ast }(3,n-1)=$ $%
M_{n}^{\ast }$ is the shape space for the n-body problem, and for $n=3,4,5$
the topological classification of these spaces are as follows : 
\begin{align}
n& =3:M_{3}^{\ast }\simeq \frac{M^{1}(3,2)}{O(3)}=\frac{M^{1}(3,2)}{SO(3)}=%
\frac{M^{1}(2,2)}{O(2)}=D^{2}  \notag \\
n& =4:M_{4}^{\ast }\simeq \frac{M^{1}(3,3)}{O(3)}\approx D^{5},\text{ }\frac{%
M^{1}(3,3)}{SO(3)}\approx 2D^{5}=S^{5}  \label{shape3} \\
n& =5:M_{5}^{\ast }\simeq \frac{M^{1}(3,4)}{O(3)}\approx S^{8},\text{ }\frac{%
M^{1}(3,4)}{SO(3)}\approx 2S^{8}=S^{8}\cup _{P}S^{8}\ \text{\ }(P=\mathbb{C}%
P^{3}/\mathbb{Z}_{2})  \notag
\end{align}

We shall make some further comments on the cases $n=3,4.$

\begin{description}
\item $n=3:$ With the induced metric the disk $D^{2}$ is a hemisphere of the
base space 
\begin{equation}
2D^{2}=\frac{M^{1}(2,2)}{SO(2)}=\mathbb{C}P^{1}=S^{2}(1/2)  \label{shape4}
\end{equation}%
of the Hopf fibration (\ref{Hopf}), which is a round sphere of radius $1/2$.
Its equator circle $M^{\ast }(1,2)=\mathbb{R}P^{1}=S^{1}(1/2)$ represents
the shapes of degenerate (i.e. collinear) 3-configurations. But in the study
of the 3-body problem it is, in fact, natural to use the whole sphere as the
shape space. The reason is that a non-degenerate 3-configuration in 3-space
is geometrically a triangle which can be oriented in two different ways, by
the ordering of the vertices. Then the two hemispheres in (\ref{shape4})
represent triangles with opposite orientation, cf. \cite{Hsiang2}, \cite%
{Hsiang4}. In this way a 3-body motion corresponds to a continuous motion of
a mass triangle whose orientation changes when the motion passes through an
eclipse, that is, when the shape curve crosses the equator circle.

\item $n=4:$ The boundary of the disk $D^{5}$ in (\ref{shape3}) is the shape
space of coplanar 4-configurations, that is, the sphere 
\begin{equation}
M^{\ast }(2,3)=\mathbb{C}P^{2}/\mathbb{Z}_{2}\approx S^{4}\text{ \ (cf. (\ref%
{Hopf2}) with }m=3)  \label{distort}
\end{equation}%
and the action of its isometry group $O(3)$ is equivalent (by Theorem\ \ref%
{model3}) to that of its linear model $(O(3),S^{4},S^{2}\rho _{3}-1)$.
Namely, the linear model is the space of symmetric 3$\times $3-matrices with
zero trace and unit norm, with the natural action of $O(3)$ by conjugation.
\end{description}

The fact that the quotient space $\mathbb{C}P^{2}/\mathbb{Z}_{2}$ is
homeomorphic to $S^{4}$ was already known to L.S. Pontryagin in the 1930's,
according to Arnold\cite{Arnold2}, and we refer to \cite{Arnold1}, \cite%
{Kuiper}, \cite{Massey} for different proofs of this specific result. Massey
also observed that the\ induced $SO(3)$-action on $\mathbb{C}P^{2}/\mathbb{Z}%
_{2}$ has the same orbit structure as that of the above linear model $%
(SO(3),S^{4})$, and the existence and construction of an equivariant
homeomorphism (such as (\ref{model2})) was, in fact, a problem stated by
Massey. Moreover, Arnold\cite{Arnold1} has constructions which are very
close to our linear model construction.

Arnold also discusses the differentiable structure of $S^{4}$, as a quotient
space $\mathbb{C}P^{2}/\mathbb{Z}_{2}$, and it is stated that $\mathbb{C}%
P^{2}/\mathbb{Z}_{2}$ is diffeomorphic to $S^{4}$ (cf. \S 1 in \cite{Arnold1}%
). However, what is shown is that the composite map%
\begin{equation}
M^{1}(2,3)=S^{5}\rightarrow \mathbb{C}P^{2}\rightarrow \mathbb{C}P^{2}/%
\mathbb{Z}_{2}\rightarrow ^{\approx }S^{4}  \label{comp}
\end{equation}%
from $S^{5}$ to $S^{4}$ is differentiable (in the usual sense). Both $%
\mathbb{C}P^{2}$ and $\mathbb{C}P^{2}/\mathbb{Z}_{2}$ have the induced
smooth functional structure as quotient spaces of $S^{5}$, but only $\mathbb{%
C}P^{2}$ becomes a differentiable manifold in this way. So, Arnold "relaxes"
the differential structure on $\mathbb{C}P^{2}/\mathbb{Z}_{2}$ so that the
last map in (\ref{comp}) becomes a diffeomorphism, that is, he defines the
differential structure to be the \emph{standard} structure mentioned in
Remark \ref{induced}. This makes the last map in (\ref{comp}) a
diffeomorphism while the composed map $S^{5}\rightarrow S^{4}$ in (\ref{comp}%
) is still a smooth map. The same applies, of course, to all the
constructions $S^{p}\rightarrow S^{p}/K\rightarrow ^{\approx }S^{q}$ in
Section 3.4.2, where none of the \textquotedblright quotient
spheres\textquotedblright\ $S^{p}/K$ is really a smooth manifold when $q\geq
4$.

Finally, let us explain why the "quotient spheres" are not smooth manifolds.
Since all cases are analogous we consider again the simplest case 
\begin{equation}
M^{\ast }(2,3)=S^{5}/O(2)=\mathbb{C}P^{2}/\mathbb{Z}_{2}\approx S^{4}
\label{S4}
\end{equation}%
viewed as an $O(2)$-orbit space with the induced smooth structure. Then our
claim is that the orbit space in (\ref{S4}) is not a differentiable manifold
in a neighborhood of the subset $A=M^{\ast }(1,3)\simeq \mathbb{R}P^{2}$. To
see this we shall apply the slice theorem (cf. Section 3.1.1), according to
which each $O(2)$-orbit in $S^{5}$ belongs to a tubular neighborhood $U$
whose image $\bar{U}=U/O(2)$ in the orbit space is an open set of type (\ref%
{local}), namely diffeomorphic to the orbit space of the slice
representation. The set $A$ represents those orbits of type $O(1)$, and the
slice representation of $O(1)$ acts on $\mathbb{R}^{4}\ $with the
eigenvalues $(-1,-1,1,1)$, consequently any point on $A$ has an open
neighborhood diffeomorphic to 
\begin{equation}
\bar{U}=\frac{O(2)\times _{O(1)}(\mathbb{R}^{2}\oplus \mathbb{R}^{2})}{O(2)}%
\simeq \frac{\mathbb{R}^{2}\oplus \mathbb{R}^{2}}{O(1)}\simeq \frac{\mathbb{R%
}^{2}}{O(1)}\times \mathbb{R}^{2}  \label{U1}
\end{equation}%
where the first factor is of type (\ref{cone3})\ and is transversal to the
set $A$ and the second is $\simeq \bar{U}\cap A$. However, the product space
in (\ref{U1}) is homeomorphic to $\mathbb{R}^{4}$, but in view of Lemma \ref%
{cone} it is not a differentiable manifold.

\section{Geometric invariants of n-body systems}

By a \emph{geometric invariant }on the centered configuration space 
\begin{equation*}
M_{n}:\sum m_{i}\mathbf{a}_{i}=\mathbf{0}
\end{equation*}%
we mean a polynomial function $F(\mathbf{a}_{1},...,\mathbf{a}_{n})$ which
is invariant under congruence and internal symmetries. We will describe the
ring of all these invariants by calculating the ring of invariants for
matrix spaces $M(d,m)$ in general. Denote a typical matrix by 
\begin{equation*}
X=(\mathbf{x}_{1},...,\mathbf{x}_{m})=\ \left[ \mathbf{x}_{1}^{\ast },...,%
\mathbf{x}_{d}^{\ast }\right]
\end{equation*}%
where the $\mathbf{x}_{i}$ and $\mathbf{x}_{j}^{\ast }$ are the column and
row vectors respectively. The product group $GL(d)\times GL(m)$ acts on $%
M(d,m)$ by matrix multiplication 
\begin{equation*}
\ X\rightarrow \psi X\varphi ^{t}\text{, \ \ \ }(\psi ,\varphi )\in
GL(d)\times GL(m)\ ,
\end{equation*}%
and $O(d)\times O(m)\ $is the subgroup leaving invariant the standard metric
form \ \ \ \ \ \ \ 
\begin{equation}
I=\left\vert X\right\vert ^{2}=trace(X^{t}X)=\sum_{i=1}^{m}\left\vert 
\mathbf{x}_{i}\right\vert ^{2}=\sum_{j=1}^{d}\left\vert \mathbf{x}_{j}^{\ast
}\right\vert ^{2},\   \label{metric6}
\end{equation}%
On the other hand, for fixed mass distribution $\mu =(m_{1},..,m_{m}),$ $%
m_{i}>0,$ let us also consider the mass dependent metric defined by 
\begin{equation}
I^{(\mu )}=trace(DX^{t}XD)=\sum_{i=1}^{m}m_{i}\left\vert \mathbf{x}%
_{i}\right\vert ^{2},\text{ \ \ \ }\   \label{Imass}
\end{equation}%
where $D=diag(\sqrt{m_{1}},\sqrt{m_{2}},..,\sqrt{m_{m}})$ and the
corresponding isometry subgroup is $O(d)\times O^{\ast }(m)\subset $\ $%
GL(d)\times GL(m)$, where$\ $ 
\begin{equation}
O^{\ast }(m)=DO(m)D^{-1}\ \subset GL(m)  \label{Ostar}
\end{equation}%
is the subgroup leaving invariant the metric form $\tsum m_{i}x_{i}^{2}$\ on 
$\mathbb{R}^{m}$ (space of row vectors with right side action of $GL(m)$).

\begin{lemma}
\bigskip The invariant ring on $M=M(d,m)$, $d\leq m,$ under the action of $%
O(d)\times O^{\ast}(m)$, is the polynomial ring with generators 
\begin{equation}
I_{k}=\sum_{i_{1}<i_{2}....<i_{k}}m_{i_{1}}m_{i_{2}}...m_{i_{k}}.\left| 
\mathbf{x}_{i_{1}}\wedge\mathbf{x}_{i_{2}}\wedge...\wedge\mathbf{x}%
_{i_{k}}\right| ^{2}\text{\ , \ \ }1\leq k\leq d,  \label{invariants5}
\end{equation}
where\ $I_{1}=I^{(\mu)}$ and the exterior product space $\wedge^{k}\mathbb{R}%
^{m}$ has the standard norm.
\end{lemma}

\begin{proof}
First assume $m_{i}=1$ for all i. It is easy to verify the identity 
\begin{equation}
I_{k}=\sum_{i_{1}<i_{2}....<i_{k}}\left\vert \mathbf{x}_{i_{1}}\wedge 
\mathbf{x}_{i_{2}}\wedge ...\wedge \mathbf{x}_{i_{k}}\right\vert
^{2}=\sum_{i_{1}<i_{2}....<i_{k}}\left\vert \mathbf{x}_{i_{1}}^{\ast }\wedge 
\mathbf{x}_{i_{2}}^{\ast }\wedge ...\wedge \mathbf{x}_{i_{k}}^{\ast
}\right\vert ^{2}  \label{Invariants3}
\end{equation}%
for each $k=1,..,d.$ Then, from the column vector version it is clear that $%
I_{k}$ is invariant under the action of $O(d),$ and similarly the row vector
expression is invariant under the action of $O(m)$. Hence, $I_{k}$ is an
invariant of $G.$

Let us apply the \emph{reduction principle} for orthogonal transformation
groups (cf. e.g. \cite{Straume4}). Namely, we first calculate the \emph{%
reduced group} of the action of $G=O(d)\times O(m)$ on $M$, which is the
quotient group 
\begin{equation*}
\bar{G}=N_{G}(H)/H\simeq B_{d}
\end{equation*}
and is finite in our case, where 
\begin{equation*}
H=\Delta O(1)^{d}\times O(m-d)\subset G,\text{ \ cf. (\ref{isotropy})\ }
\end{equation*}%
is the principal isotropy group and $N_{G}(H)$ is its normalizer in $G$.
Next we determine the fixed point set of $H$ 
\begin{equation*}
\mathbb{R}^{d}=F(H,M)\ :\text{ }X=diag(r_{1},r_{2},..,r_{d})\text{, \ \ cf. (%
\ref{cross section})\ }
\end{equation*}%
which contains the fundamental domain 
\begin{equation*}
M^{+}(d)\subset \mathbb{R}^{d}\subset M:r_{1}\geq r_{2}\geq ....\geq
r_{d}\geq 0\text{ \ \ }
\end{equation*}%
of the $G$-action, identified with the orbit space of $G$ in Section 3.3.1.
Moreover, for the $G$-orbit through $X$, $r_{i}^{2}=$ $\lambda _{i}$ are the
eigenvalues of $X^{t}X$.

Finally, we consider the induced action of\ $\bar{G}$ on $\mathbb{R}^{d}$,
which is a group generated by reflections and as such it is the Weyl group $%
B_{d}$, having the above cone $M^{+}(d)$ as a fundamental domain. Now, the
reduction principle says that the restriction of polynomials induces an
isomorphism between the invariant rings of $(G,M)$ and $(\bar{G},\mathbb{R}%
^{d})$. But the latter ring is certainly generated by the elementary
symmetric functions 
\begin{equation*}
I_{1}=\tsum \lambda _{i}\ ,\text{ }I_{2}=\tsum_{i<j}\lambda _{i}\lambda _{j},%
\text{ }I_{3}=\tsum_{i<j<k}\lambda _{i}\lambda _{j}\lambda _{k},\text{ . . .
.}
\end{equation*}%
which are also the restrictions of the functions in (\ref{Invariants3}), and
this proves the lemma for masses $m_{i}=1.$

Next, let $M(d,m)^{(\mu)}$ be the matrix space with the mass dependent
metric (\ref{Imass}) and consider the transformation 
\begin{equation*}
M(d,m)^{(\mu)}\rightarrow M(d,m),\text{ \ \ }X\rightarrow XD=(\sqrt{m_{1}}%
\mathbf{x}_{1},\sqrt{m_{2}}\mathbf{x}_{2},...,\sqrt{m_{m}}\mathbf{x}_{m})
\end{equation*}
where the left and right hand space have metrics (\ref{Imass}) and (\ref%
{metric6}) respectively. The transformation is a $G$-equivariant isometry,
where $O(m)$ acts on $M(d,m)^{(\mu)}$ by first applying the isomorphism 
\begin{equation*}
O(m)\rightarrow DO(m)D^{-1}=O^{\ast}(m),\text{ \ cf. (\ref{Ostar})}
\end{equation*}
Hence, the transformation also induces an isomorphism of the corresponding
invariant rings. In effect, the invariant ring of $M(d,m)^{(\mu)}$ is
obtained by replacing each vector $\mathbf{x}_{i}$ in (\ref{Invariants3}) by 
$\sqrt{m_{i}}\mathbf{x}_{i}.$
\end{proof}

\qquad

In particular, the centered configuration space $M_{n}\simeq M(3,n-1)$ has
the following basic geometric invariants, in terms of the Jacobi vectors and
classical vector operations : 
\begin{equation}
I_{1}=\tsum_{\ }\left\vert \mathbf{x}_{i}\right\vert ^{2}\text{, }%
I_{2}=\tsum_{i<j}\left\vert \mathbf{x}_{i}\times \mathbf{x}_{j}\right\vert
^{2}\text{, }I_{3}=\tsum_{i<j<k}\left\vert (\mathbf{x}_{i}\times \mathbf{x}%
_{j})\cdot \mathbf{x}_{k}\right\vert ^{2}\text{\ }  \label{Invariants2}
\end{equation}%
for all $n\geq 4.$ (For $n=3$ the generators are $I_{1}$ and $I_{2}$\ ). The
expressions for $I_{k}$ hold for any choice of Jacobi vectors, but still it
remains to express them as functions of $\mathbf{a}_{1},...,\mathbf{a}_{n}.$
To this end, consider the free n-configuration space or matrix space 
\begin{equation*}
\hat{M}_{n}=M(3,n)^{(\mu )}=\left\{ X=(\mathbf{a}_{1},\mathbf{a}_{2},...,%
\mathbf{a}_{n})\right\} \text{ }
\end{equation*}%
with the metric form $I^{(\mu )}=\tsum m_{i}\left\vert \mathbf{a}%
_{i}\right\vert ^{2}$ as in (\ref{Imass}. By the above lemma its invariant
ring with respect to the group $O(3)\times O^{\ast }(n)$ has the basic
generators 
\begin{align}
I_{1}& =\tsum_{\ }m_{i}\left\vert \mathbf{a}_{i}\right\vert ^{2}\text{ , \ \ 
}I_{2}=\tsum_{i<j}m_{i}m_{j}\left\vert \mathbf{a}_{i}\times \mathbf{a}%
_{j}\right\vert ^{2}\text{ ,\ }  \label{Invariants4} \\
\text{\ }I_{3}& =\tsum_{i<j<k}m_{i}m_{j}m_{k}\left\vert (\mathbf{a}%
_{i}\times \mathbf{a}_{j})\cdot \mathbf{a}_{k}\right\vert ^{2}  \notag
\end{align}%
and these are still complete and independent as invariants of the subgroup $%
O(3)\times O^{\ast }(n-1)$ acting on the subspace $M_{n}\subset \hat{M}_{n}$%
, as long as $3\leq n-1.$ Here $O^{\ast }(n-1)$ $\subset O^{\ast }(n)$ \ is
the internal symmetry group of $M_{n}$ and is the subgroup leaving $M_{n}$
invariant, or equivalently, the subgroup of $O^{\ast }(n)$ which under the
action on $\mathbb{R}^{n}$ (row vectors) fixes the vector $%
(m_{1},m_{2},...,m_{n}).$ This proves the following result.

\begin{theorem}
The ring of geometric invariants of the centered n-configuration space $%
M_{n} $ is the polynomial ring generated by $I_{1},..,I_{q}$ in (\ref%
{Invariants4}), $q=\min \{3,n-1\}$.
\end{theorem}

\begin{remark}
Starting from the Jacobi vector expressions (\ref{Invariants2}) of the
invariants, a Jacobi transformation may transform them to expressions
involving the vectors $\mathbf{a}_{i}$. In this way, however, we may arrive
at a non-symmetric expression since $\tsum m_{i}\mathbf{a}_{i}=0$. Below we
shall give examples to illustrate the non-uniqueness of the symmetrization
procedure and also give a geometric interpretation of $I_{n-1}$ for $n=3$ or 
$4.$
\end{remark}

\begin{itemize}
\item $n=3:$ Let $\mathcal{A}$ be the area of the triangle spanned by the
vector triple $(\mathbf{a}_{1},\mathbf{a}_{2},\mathbf{a}_{3})$ with $\tsum
m_{i}\mathbf{a}_{i}=\mathbf{0.}$ By simple trigonometry 
\begin{equation*}
2\mathcal{A}=\frac{\bar{m}}{m_{3}}\left\vert \mathbf{a}_{1}\times \mathbf{a}%
_{2}\right\vert =\frac{\bar{m}}{m_{1}}\left\vert \mathbf{a}_{2}\times\mathbf{%
a}_{3}\right\vert =\frac{\bar{m}}{m_{2}}\left\vert \mathbf{a}_{3}\times%
\mathbf{a}_{1}\right\vert \ \ 
\end{equation*}
On the other hand, from the Jacobi vector formula (\ref{xk}), 
\begin{equation*}
\mathbf{x}_{1}\times\mathbf{x}_{2}=\zeta_{1}\zeta_{2}\mathbf{a}_{1}\times%
\mathbf{a}_{2}
\end{equation*}
and consequently 
\begin{equation*}
\ \left\vert \mathbf{x}_{1}\times\mathbf{x}_{2}\right\vert =\zeta_{1}\zeta
_{2}\left\vert \mathbf{a}_{1}\times\mathbf{a}_{2}\right\vert =2\sqrt {\frac{%
m_{1}m_{2}m_{3}}{\bar{m}}}\text{ }\mathcal{A}
\end{equation*}
and 
\begin{align*}
\ I_{2} & =\left\vert \mathbf{x}_{1}\times\mathbf{x}_{2}\right\vert ^{2}=%
\bar{m}\frac{m_{i}m_{j}}{m_{k}}\left\vert \mathbf{a}_{i}\times \mathbf{a}%
_{j}\right\vert ^{2} \\
& =\ \bar{m}\frac{m_{1}m_{2}}{3m_{3}}\left\vert \mathbf{a}_{1}\times \mathbf{%
a}_{2}\right\vert ^{2}+\ \bar{m}\frac{m_{2}m_{3}}{3m_{1}}\left\vert \mathbf{a%
}_{2}\times\mathbf{a}_{3}\right\vert ^{2}\ +\bar{m}\frac{m_{3}m_{1}}{3m_{2}}%
\left\vert \mathbf{a}_{3}\times\mathbf{a}_{1}\right\vert ^{2}
\end{align*}
where $\left\{ i,j,k\right\} =\left\{ 1,2,3\right\} $ and the last sum is a
symmetrization with all three terms equal, in fact. However, the above
expression for $I_{2}$ and the symmetric expression in (\ref{Invariants4})
are identical as a function on $M_{3}.$

\item $n=4:$ Let $\mathcal{V}$ \ be the volume of the tetrahedron spanned by
the vector quadruple $(\mathbf{a}_{1},\mathbf{a}_{2},\mathbf{a}_{3},\mathbf{a%
}_{4})$ satisfying $\tsum m_{i}\mathbf{a}_{i}=0\mathbf{.}$ Then $\pm6%
\mathcal{V}$ \ equals the triple product of $\mathbf{a}_{i}-\mathbf{a}_{1}$, 
$i=2,3,4,$ and 
\begin{equation*}
6\mathcal{V}=\frac{\bar{m}}{m_{4}}\left\vert (\mathbf{a}_{1}\times \mathbf{a}%
_{2})\cdot\mathbf{a}_{3}\right\vert =\frac{\bar{m}}{m_{1}}\left\vert (%
\mathbf{a}_{2}\times\mathbf{a}_{3})\cdot\mathbf{a}_{4}\right\vert \text{ etc.%
}
\end{equation*}
Again, by expressing the Jacobi vectors $\mathbf{x}_{i}$ in terms of the $%
\mathbf{a}_{i}$ it follows 
\begin{equation*}
\ \left\vert \left( \mathbf{x}_{1}\times\mathbf{x}_{2}\right) \cdot \mathbf{x%
}_{3}\right\vert =\zeta_{1}\zeta_{2}\zeta_{3}\left\vert \left( \mathbf{a}%
_{1}\times\mathbf{a}_{2}\right) \cdot\mathbf{a}_{3}\right\vert =3!\sqrt{%
\frac{m_{1}m_{2}m_{3}m_{4}}{\bar{m}}}\text{ }\mathcal{V}
\end{equation*}
and consequently 
\begin{align*}
I_{3} & =\ \left\vert \left( \mathbf{x}_{1}\times\mathbf{x}_{2}\right) \cdot%
\mathbf{x}_{3}\right\vert ^{2}=\bar{m}\frac{m_{i}m_{j}m_{k}}{m_{l}}%
\left\vert (\mathbf{a}_{i}\times\mathbf{a}_{j})\cdot\mathbf{a}%
_{k}\right\vert ^{2} \\
& =\bar{m}\frac{m_{1}m_{2}m_{3}}{4m_{4}}\left\vert (\mathbf{a}_{1}\times%
\mathbf{a}_{2})\cdot\mathbf{a}_{3}\right\vert ^{2}+\bar{m}\frac {%
m_{2}m_{3}m_{4}}{4m_{1}}\left\vert (\mathbf{a}_{2}\times\mathbf{a}_{3})\cdot%
\mathbf{a}_{4}\right\vert ^{2}\text{ }+\text{etc.}
\end{align*}
where $\left\{ i,j,k,l\right\} =\left\{ 1,2,3,4\right\} $ and the four terms
in the last sum are equal. Again, $I_{3}$ in (\ref{Invariants4}) is an
another symmetric expression for the same function on $M_{4}.$
\end{itemize}

\begin{remark}
The cases $n=3$ or $4$ are special since, for example, their shape space $%
M_{n}^{\ast}$ is topologically a disk $D^{2}$ or $D^{5}$ respectively.
Moreover, in these cases $M_{n}^{\ast}$ has a unique \emph{geometric center,}
and this may be characterized in several ways. It is the unique fixed point
of the symmetry group $O(n-1)$, and the point has the same distance to each
of the binary collision varieties. Furthermore, it is also the maximum point
of $I_{2}$ or $I_{3}$ respectively. That is, for a fixed size $I_{1}=1$ the
triangle (resp. tetrahedron) with the largest area (resp. volume) is at the
geometric center. In terms of Jacobi vectors the condition is that $\mathbf{x%
}_{i}\cdot\mathbf{x}_{j}=0$ and $\left\vert \mathbf{x}_{i}\right\vert
=\left\vert \mathbf{x}_{j}\right\vert $ \ for $i\neq j.$
\end{remark}

\section{The weighted root system of an n-body system}

A typical potential function $U(\mathbf{a}_{1},\mathbf{a}_{2},...,\mathbf{a}%
_{n})$ of an n-body problem depends only on the pairwise distances 
\begin{equation*}
r_{ij}=\left\vert \mathbf{a}_{i}-\mathbf{a}_{j}\right\vert ,\text{ \ \ }%
1\leq i<j\leq n,
\end{equation*}
and for certain applications one would like to express them in terms of
Jacobi vectors $\mathbf{x}_{i}$, say 
\begin{equation*}
X=(\mathbf{x}_{1},..,\mathbf{x}_{n-1})\in M(3,n-1)\ 
\end{equation*}
is the matrix associated with a given Jacobi transformation $%
\Psi:M_{n}\simeq M(3,n-1)$. For this purpose we shall define and investigate
the \emph{weighted root system}, depending on the mass distribution, which
is primarily a book-keeping for expanding the vectors $\mathbf{a}_{i}-%
\mathbf{a}_{j}$ as a linear combination of Jacobi vectors. It is naturally a
\textquotedblright weighted\textquotedblright\ version of the root system of
type $A_{n-1}$ in classical Lie theory, and they are identical when the mass
distribution is uniform.

\subsection{Distance functions and the $\Psi$-root system}

The mutual distances $r_{ij}$ have a nice geometric interpretation as
functions on $M_{n}$ which measure the distance from specific subvarieties.
To explain this, consider the linear transformation 
\begin{equation*}
\ M=M(3,n-1)\rightarrow \mathbb{R}^{3}
\end{equation*}%
defined by 
\begin{equation}
X\rightarrow \left[ \mathbf{u,}X\right] =\sum_{i=1}^{n-1}c_{i}\mathbf{x}%
_{i}\ =X\mathbf{u}^{t}  \label{uX}
\end{equation}%
where $\mathbf{u=(}c_{1},c_{2},...,c_{n-1})\in \mathbb{R}^{n-1}$ is a given
row vector of unit length and $\mathbf{u}^{t}$ is the column vctor. Define
the codimension $3$ subspace $B=B_{\mathbf{u}}$ $\subset $ $M$ to be the
kernel 
\begin{equation*}
B:\left[ \mathbf{u,}X\right] =0\mathbf{.}
\end{equation*}%
It is easy to verify that the distance in $M$ from a point $X$ to the
subspace $B$ is given by the function 
\begin{equation}
\delta _{B}(X)=dist(X,B)=\left\vert \left[ \mathbf{u,}X\right] \right\vert
=\left\vert \sum_{k=1}^{3}(\mathbf{u\cdot x}_{k}^{\ast })^{2}\right\vert ^{%
\frac{1}{2}}  \notag
\end{equation}%
where the $\mathbf{x}_{k}^{\ast }$ are the row vectors of $X$ and the inner
product on $\mathbb{R}^{n-1}$ is the standard one.

In particular, the binary collision varieties 
\begin{equation}
B_{ij}\subset M_{n}:\mathbf{a}_{i}=\mathbf{a}_{j},\text{ }i\neq j
\label{Bij}
\end{equation}
have associated distance functions $\delta_{ij}$ of this kind, defined on $M$
via a chosen Jacobi transformation $\Psi:M_{n}\rightarrow M$. Consequently
there is a unit vector (unique up to sign) 
\begin{equation}
\mathbf{u}_{ij}=(c_{ij}^{1},c_{ij}^{2},...,c_{ij}^{n-1})\in\mathbb{R}^{n-1}
\label{u/ij/2}
\end{equation}
depending on $\Psi$ and the mass distribution, so that 
\begin{equation}
\delta_{ij}(X)=\left\vert \sum_{k=1}^{3}(\mathbf{u}_{ij}\mathbf{\cdot x}%
_{k}^{\ast})^{2}\right\vert ^{\frac{1}{2}}\   \label{distance}
\end{equation}

Let us express $\delta _{ij}$ in terms of the n-configuration $\mathbf{X}=(%
\mathbf{a}_{1},\mathbf{a}_{2},...,\mathbf{a}_{n})$. For $\mathbf{X}$ fixed,
let $\mathbf{Y}_{0}=(\mathbf{b}_{1},\mathbf{b}_{2},....\mathbf{b}_{n})$ be
the critical point of the distance function $\delta (\mathbf{Y})=\left\vert 
\mathbf{Y}-\mathbf{X}\right\vert $ defined on $B_{ij}$, where by definition $%
\mathbf{b}_{i}=\mathbf{b}_{j}$. The condition $\nabla \delta =0$ implies 
\begin{equation*}
\mathbf{b}_{i}=\mathbf{b}_{j}=\frac{m_{i}\mathbf{a}_{i}+m_{j}\mathbf{a}_{j}}{%
m_{i}+m_{j}},\text{ \ \ }\mathbf{b}_{k}=\mathbf{a}_{k}\text{ for }k\neq i,j
\end{equation*}%
and consequently in terms of reduced masses (\ref{redmass}) 
\begin{align*}
\delta _{ij}^{2}(\mathbf{X})& =\ \left\vert \mathbf{Y}_{0}-\mathbf{X}%
\right\vert ^{2}=m_{i}\left\vert \mathbf{b}_{i}-\mathbf{a}_{i}\right\vert
^{2}+m_{j}\left\vert \mathbf{b}_{j}-\mathbf{a}_{j}\right\vert ^{2} \\
& =\frac{m_{i}m_{j}}{m_{i}+m_{j}}\left\vert \mathbf{a}_{i}-\mathbf{a}%
_{j}\right\vert ^{2}=\mu _{ij}r_{ij}^{2}\ 
\end{align*}%
that is, 
\begin{equation}
r_{ij}=\frac{1}{\sqrt{\mu _{ij}}}\delta _{ij}\ =\frac{1}{\sqrt{\mu _{ij}}}%
\left\vert \left[ \mathbf{u}_{ij}\mathbf{,}X\right] \right\vert
\label{delta/ij}
\end{equation}%
Therefore, we may define vectors $\mathbf{u}_{ij}=-\mathbf{u}_{ji}$\textbf{\ 
}uniquely by the constraint 
\begin{equation}
\mathbf{a}_{i}-\mathbf{a}_{j}=\frac{1}{\sqrt{\mu _{ij}}}\left[ \mathbf{u}%
_{ij}\mathbf{,}X\right] ,\text{ \ \ }i\neq j,  \label{displace}
\end{equation}%
and clearly they satisfy all identities of type 
\begin{equation}
\frac{1}{\sqrt{\mu _{ij}}}\mathbf{u}_{ij}+\frac{1}{\sqrt{\mu _{jk}}}\mathbf{u%
}_{jk}=\frac{1}{\sqrt{\mu _{ik}}}\mathbf{u}_{ik}\text{ .}  \label{relation1}
\end{equation}

\begin{definition}
\label{def1}The $\Psi$-root system with mass distribution $(m_{1,}...,m_{n})$
is the above collection of $\binom{n}{2}$ vector pairs $\pm\frac{1}{\sqrt {%
\mu_{ij}}}\mathbf{u}_{ij}\in\mathbb{R}^{n-1},1\leq i<j\leq n.$ The
collection $\left\{ \pm\mathbf{u}_{ij}\right\} $ of unit vectors is the
normalized $\Psi$-root system.
\end{definition}

\subsection{The standard weighted root system}

\bigskip It will be useful to have explicit formulas for the $\Psi_{0}$%
-roots, where

\begin{equation*}
\Psi _{0}:M_{n}\rightarrow M(3,n-1)
\end{equation*}%
is the standard Jacobi transformation constructed in Section 2.3. The
notation used for various constants in Section 2.3 is also used below. We
shall also use the notation foBy definition of $\Psi _{0}$ (cf. (\ref{xk}), (%
\ref{matrixL0})) 
\begin{equation*}
\mathbf{a}_{i}-\mathbf{a}_{j}=\frac{\zeta _{i}}{m_{i}}\mathbf{x}_{i}-\frac{1%
}{\zeta _{j}}\mathbf{x}_{j}+\sum_{k=i+1}^{j-1}\frac{\zeta _{k}}{m^{(k-1)}}%
\mathbf{x}_{k}\text{ }=\frac{1}{\sqrt{\mu _{ij}}}\sum_{k=1}^{n-1}c_{ij}^{k}%
\mathbf{x}_{k}\text{ , }i<j,
\end{equation*}%
(where the term $\frac{1}{\zeta _{j}}\mathbf{x}_{j}$ is undefined when $\
j=n $) and the vector $\mathbf{u}_{ij}=(c_{ij}^{1},...,c_{ij}^{n-1})$ have
its nonzero components $c_{ij}^{k}$ for $k$ in the range $i\leq k\leq \min
\left\{ j,n-1\right\} .$ In particular, 
\begin{equation}
\mathbf{u}_{n-1,n}=(0,0,...,1)  \label{u/n-1,n}
\end{equation}%
and for $i<n-1$%
\begin{equation}
\ \frac{1}{\sqrt{\mu _{i,i+1}}}\mathbf{u}_{i,i+1}=\ (0,...,0,\sqrt{\frac{%
m^{(i-1)}}{m_{i}m^{(i)}}},-\sqrt{\frac{m^{(i+1)}}{m_{i+1}m^{(i)}\ }},0,...,0)%
\text{ \ }  \label{u/simple}
\end{equation}%
\begin{equation}
\frac{1}{\sqrt{\mu _{in}}}\mathbf{u}_{in}=(0,..,0,\sqrt{\frac{m^{(i-1)}}{%
m_{i}m^{(i)}}},\sqrt{\frac{m_{i+1}\ }{m^{(i)}m^{(i+1)}}},...,\sqrt{\frac{%
m_{n-1}\ }{m^{(i)}m^{(n-1)}}})  \label{u/i,n}
\end{equation}%
whereas for $2\leq i+1<j<n$ 
\begin{align}
\frac{1}{\sqrt{\mu _{ij}}}\mathbf{u}_{ij}& =(0,..,0,\sqrt{\frac{m^{(i-1)}}{%
m_{i}m^{(i)}}},\sqrt{\frac{m_{i+1}\ }{m^{(i)}m^{(i+1)}}},..,\sqrt{\frac{%
m_{k}\ }{m^{(k-1)}m^{(k)}}},..  \label{u/ij/4} \\
& ,\sqrt{\frac{m_{j-1}\ }{m^{(j-2)}m^{(j-1)}}},-\sqrt{\frac{m^{(j)}\ }{%
m_{j}m^{(j-1)}}},0,..,0)  \notag
\end{align}%
For a fixed mass distribution, all $\Psi $-root systems are, in fact,
orthogonally equivalent to the above $\Psi _{0}$-root system $\left\{ \pm 
\frac{1}{\sqrt{\mu _{ij}}}\mathbf{u}_{ij}\text{, }i<j\right\} $, see the
following subsection.

\begin{example}
$n=3$ : 
\begin{align*}
\mathbf{u}_{12} & =(\sqrt{\frac{m_{2}\bar{m}}{(m_{2}+m_{3})(m_{1}+m_{2})}},-%
\sqrt{\frac{m_{1}m_{3}}{(m_{2}+m_{3})(m_{1}+m_{2})}}) \\
\mathbf{u}_{13} & =(\sqrt{\frac{m_{3}\bar{m}}{(m_{2}+m_{3})(m_{1}+m_{2})}},%
\sqrt{\frac{m_{1}m_{2}}{(m_{2}+m_{3})(m_{1}+m_{2})}})
\end{align*}
\qquad\qquad
\end{example}

\begin{example}
$n\geq3:$%
\begin{align*}
\ n & \geq3:\frac{1}{\sqrt{\mu_{12}}}\mathbf{u}_{12}=(\sqrt{\frac{\bar{m}}{%
m_{1}(m_{2}+..+m_{n})}},-\ \sqrt{\frac{m_{3}+..+m_{n}}{m_{2}(m_{2}+..+m_{n})}%
},0,...) \\
n & \geq4:\frac{1}{\sqrt{\mu_{23}}}\mathbf{u}_{23}=(0,\sqrt{\frac {%
m_{2}+..+m_{n}}{m_{2}(m_{3}+..+m_{n})}},-\sqrt{\frac{m_{4}+..+m_{n}}{%
m_{3}(m_{3}+..+m_{n})}},0,...) \\
n & \geq5:\frac{1}{\sqrt{\mu_{34}}}\mathbf{u}_{34}=(0,0,\sqrt{\frac {%
m_{3}+..+m_{n}}{m_{3}(m_{4}+..+m_{n})}},-\sqrt{\frac{m_{5}+..+m_{n}}{%
m_{4}(m_{4}+..+m_{n})}},0,...)
\end{align*}
$\ \ \ \ $
\end{example}

\subsection{Weighted root systems and their metric invariants}

It follows from (\ref{relation1})-(\ref{u/ij/4}) that the (normalized) $%
\Psi_{0}$-roots $\mathbf{u}_{ik}$ and $\mathbf{u}_{jl}$ are mutually
perpendicular except when $\left\{ i,k\right\} \cap\left\{ j,l\right\} \
\neq\emptyset,$ in which case we define $\alpha_{k}^{i,j}=$ $\alpha _{k}^{j,i%
\text{ }}$ to be the angle between $\mathbf{u}_{ik}$ and $\mathbf{u}_{kj}$,
namely 
\begin{equation}
\cos\alpha_{k}^{i,j}\ =\mathbf{u}_{ik}\cdot\mathbf{u}_{kj}=-\sqrt{\frac {%
m_{i}m_{j}}{(m_{i}+m_{k})(m_{k}+m_{j})}}  \label{alfa/ijk}
\end{equation}
In particular, for equal masses the possible angles between any two
non-collinear vectors are $\alpha=\frac{1}{2}\pi,\frac{1}{3}\pi$ and $\frac {%
2}{3}\pi,$ and the $\Psi_{0}$-roots constitute, indeed, a root system of
type $A_{n-1}$ in the usual sense. For general masses we make the following
definition.

\begin{definition}
\label{root}A weighted root system (of type $A_{n-1})$ with mass
distribution $(m_{1},m_{2},...,m_{n})$ is a collection of nonzero vectors 
\begin{equation*}
\mathbf{w}_{ij}=-\mathbf{w}_{ji},1\leq i<j\leq n,
\end{equation*}
in $\mathbb{R}^{n-1}$ such that i) $\sqrt{\mu_{ij}}\left| \mathbf{w}%
_{ij}\right| =1$ , ii) $\mathbf{w}_{ik}$ and $\mathbf{w}_{jl}$ are
perpendicular iff $\left\{ i,k\right\} \cap\left\{ j,l\right\} \ =\emptyset,$
and iii) the angle between $\mathbf{w}_{ik}$ and $\mathbf{w}_{kj}$ is $%
\alpha_{k}^{i,j}$, as defined by (\ref{alfa/ijk}).
\end{definition}

The $\Psi_{0}$-root system is our prototype of such a root system, and
clearly the subset of n-1 vectors $\left\{ \frac{1}{\sqrt{\mu_{i,i+1}}}%
\mathbf{u}_{i,i+1}\right\} $ constitutes a system of \emph{simple roots} for
obvious reasons, see (\ref{relation1}), (\ref{u/simple}).

\begin{theorem}
For a given mass distribution, each $\Psi$-root system (see Definition \ref%
{def1}) is a weighted root system in the sense of Definition \ref{root}, and
conversely, each weighted root system is the $\Psi$-root system for a unique
Jacobi transformation $\Psi:M_{n}\rightarrow M(3,n-1)$. Moreover, for the
given mass distribution all the weighted root systems are orthogonally
equivalent.
\end{theorem}

This result is a simple consequence of the fact that any $\Psi$ can be
written uniquely as a composition 
\begin{equation*}
\Psi=\varphi\circ\Psi_{0}:M_{n}\rightarrow M(3,n-1)\rightarrow M(3,n-1)
\end{equation*}
where $\varphi\in O(n-1)$ acts on $M(3,n-1)$ by multiplication, that is, $%
\varphi(X)=X\varphi$. Therefore the Jacobi vector matrices $X$ and $%
X^{\prime}$ of $\Psi_{0}$ and $\Psi$, respectively, are related by $%
X^{\prime}=X\varphi$. Let $\left\{ \pm\mathbf{u}_{ij}\right\} $ and $\left\{
\pm\mathbf{u}_{ij}^{\prime}\right\} $ be the normalized $\Psi_{0}$-roots and 
$\Psi$-roots, respectively. The defining relation (\ref{displace}) 
\begin{equation*}
\sqrt{\mu_{ij}}(\mathbf{a}_{i}-\mathbf{a}_{j})=\ \left[ \mathbf{u}_{ij}%
\mathbf{,}X\right] =\left[ \mathbf{u}_{ij}^{\prime}\mathbf{,}X^{\prime }%
\right]
\end{equation*}
implies $\mathbf{u}_{ij}^{\prime}=\mathbf{u}_{ij}\varphi$, and consequently
the $\Psi$-root system is the orthogonally transformed image 
\begin{equation*}
\left\{ \pm\frac{1}{\sqrt{\mu_{ij}}}\mathbf{u}_{ij}^{\prime}\right\}
=\left\{ \pm\frac{1}{\sqrt{\mu_{ij}}}\mathbf{u}_{ij}\right\} \varphi\text{ \
\ (row vectors)}
\end{equation*}
of the $\Psi_{0}$-root system.

On the other hand, a root system is a finite subset of $\mathbb{R}^{n-1}$
with specified inner products of the vectors, and according to a result of
Weyl (see Section 3.3) these numbers determine the subset modulo orthogonal
equivalence. Moreover, since the weighted root system spans all $\mathbb{R}%
^{n-1}$, any two of them are related by a unique $\varphi\in O(n-1)$. In
particular, a weighted root system is the $\Psi$-root system for a unique
Jacobi transformation $\Psi=\varphi\circ\Psi_{0}$.

\subsection{On the role of the mass distribution}

The mass distribution of a n-body system manifests itself through the
various mass dependent quantities constructed above,\ for example :

\begin{itemize}
\item the collection $\left\{ \mu_{ij}\right\} $ of reduced masses (\ref%
{redmass});

\item the weighted and normalized root system $\left\{ \pm \mathbf{u}%
_{ij}\right\} ;$

\item the collection $\left\{ \alpha_{k}^{i,j}\right\} $ of angles between
the vectors of a weighted root system.
\end{itemize}

The mass distribution (modulo scaling )\ may, in fact, be reconstructed from
any of these invariants. The reduced masses, for example, satisfy the
following two conditions \qquad%
\begin{align*}
(i)\text{ \ \ }\mu_{ij}^{-1}+\mu_{jk}^{-1} & >\mu_{ik}^{-1}\text{ \ \ for }%
i,j,k\text{ different} \\
(ii)\text{\ \ \ }\mu_{ij}^{-1}+\text{\ }\mu_{kl}^{-1} & =\text{\ }\mu
_{ik}^{-1}+\text{\ }\mu_{jl}^{-1}=\text{\ }\mu_{il}^{-1}+\text{\ }\mu
_{jk}^{-1}\text{ \ for }i,j,k,l\text{ different}
\end{align*}
and one can define a \textit{reduced mass distribution of order }$n$ to be a
collection of $\binom{n}{2}$ positive numbers $\mu_{ij}=\mu_{ji},$ $1\leq
i<j\leq n$, constrained by the above two conditions. Then there is a 1-1
correspondence between the usual and the reduced mass distributions (modulo
scaling) given by the formulae for $\mu_{ij}$ and their inversion 
\begin{equation*}
\ \text{\ \ }m_{i}=\frac{2}{\mu_{ij}^{-1}+\mu_{ik}^{-1}-\mu_{jk}^{-1}}\text{
\ \ for }i,j,k\text{ different}
\end{equation*}

On the other hand, the congruence space $\bar{M}_{n}$ (and hence the shape
space $M_{n}^{\ast }$), with the kinematic Riemannian structure, is
independent of the mass distribution. The mass distribution determines,
however, the representation of congruence classes as points in $\bar{M}_{n}$%
; in particular, it determines the relative position of the binary collision
varieties. Conversely, we can reconstruct the masses $m_{i}$ (modulo
scaling) from knowledge of the position of these varieties.

To illustrate the last statement, consider the case $n=3,$ where $%
M_{3}^{\ast }$ is a round hemisphere (or sphere, see Section\ 3.4.4) with a
distinguished \emph{equator circle} $S^{1}$ representing the shapes of the
degenerate triangles. On this circle lie the three collision points $%
b_{12},b_{23},b_{13}$, where $b_{ij}$ represents the shape of those
triangles $(\mathbf{a}_{1},\mathbf{a}_{2},\mathbf{a}_{3})$\ with $\mathbf{a}%
_{i}=\mathbf{a}_{j}\neq 0$ and the center of mass at the origin. The mass
distribution determines their relative position, that is, the angles (or
distances) between the points $b_{ij}$, and once their position have been
fixed we know how to determine the position of any shape (cf. \cite{Hsiang4}%
).

To describe the relative positions of the points $b_{ij}$, consider $S^{1}$
as the circumscribed circle of the triangle $\Delta =\Delta
(b_{12},b_{23,}b_{31})$ in a Euclidean plane with origin at the center of $%
S^{1}$. It turns out that $\Delta $ is a \emph{central} triangle, in the
sense that the center of its circumscribed circle lies in its interior.
Moreover, the three central angles 
\begin{equation*}
0<\beta _{k}<\pi ,\text{ \ }\sum \beta _{k}=2\pi
\end{equation*}%
where $\beta _{k}$ is opposite to the vertex $b_{ij}$ ($i,j,k$ different),
are given in terms of the normalized mass distribution by 
\begin{equation*}
\sin \beta _{k}=2\frac{\sqrt{m_{1}m_{2}m_{3}}}{(m_{k}+m_{i})(m_{k}+m_{j})}%
\text{ \ \ }(\sum m_{i}=1)\text{ }
\end{equation*}%
Conversely, any central triangle can be realized in this way and the
inversion formula is 
\begin{equation*}
m_{k}=1-\frac{2\sin \beta _{k}}{\sum_{i=1}^{3}\sin \beta _{i}}\text{ \ \
(cf. \cite{Hsiang2} or \cite{Hsiang4})}
\end{equation*}%
where the positivity of $m_{k}$, indeed, reflects the central property of
the triangle $\Delta $.

The triple of angles $(\beta _{1},\beta _{2},\beta _{3})$ is in 1-1
correpondence with another triple $(\alpha _{1},\alpha _{2},\alpha _{3})$,
where 
\begin{equation}
\alpha _{k}=\pi -\beta _{k}/2,\text{ \ }k=1,2,3  \label{beta-alfa}
\end{equation}%
and these are the central angles of another central triangle, namely the
triangle representing (up to congruence) the weighted root system of the
mass distribution (cf. Section 5.3). More precisely, whereas $\beta
_{k}=\beta _{ij}$ is the angular distance between $b_{ik}$ and $b_{kj}$ on
the circle $S^{1}$, $\alpha _{k}=\alpha _{ij}$ is the angle between the
normalized root vectors $\mathbf{u}_{ik}$ and $\mathbf{u}_{kj}$ since 
\begin{equation*}
\cos \alpha _{k}=\mathbf{u}_{ik}\cdot \mathbf{u}_{kj}=-\sqrt{\frac{m_{i}m_{j}%
}{(m_{i}+m_{k})(m_{k}+m_{j})}},\text{ \ \ \ cf. (\ref{alfa/ijk})}
\end{equation*}

Finally, we recall that the root system of a $k$-body system is a weighted
root system of type $A_{k-1}$. The way different $A_{2}$ systems combine
into higher rank systems $A_{k},$ $k\geq 3,$ is completely parallel to the\
\textquotedblright standard\textquotedblright\ theory of root systems of
type $A_{k}$, and therefore the higher rank case poses no further problem.

\section{Collinear central configurations revisited}

Consider the classical Newtonian potential function 
\begin{equation}
U=\sum_{i<j}\frac{m_{i}m_{j}}{\left\vert \mathbf{a}_{i}-\mathbf{a}%
_{j}\right\vert }\   \label{U}
\end{equation}%
and its gradient field $\nabla U$ with respect to the kinematic metric (\ref%
{metric1}) in $M_{n}$ (or $\hat{M}_{n})$. An n-configuration $\mathbf{X=(a}%
_{1},...,\mathbf{a}_{n})$ is called \emph{central }if $\nabla U(\mathbf{X)=}%
\lambda \mathbf{X}$ for some constant $\lambda ,$ namely 
\begin{equation}
\lambda \mathbf{a}_{i}=\frac{1}{m_{i}}\frac{\partial U}{\partial \mathbf{a}%
_{i}}=\sum_{j\neq i}\frac{m_{j}(\mathbf{a}_{j}-\mathbf{a}_{i})}{\left\vert 
\mathbf{a}_{i}-\mathbf{a}_{j}\right\vert ^{3}},\text{ \ }i=1,...,n
\label{central1}
\end{equation}%
In fact, $\lambda =-I^{-1}U(\mathbf{X)}$ where $I=\left\vert \mathbf{X}%
\right\vert ^{2}$, since by a classical result of Euler the homogeneity of $%
U $ implies $\nabla U(\mathbf{X)\cdot X=-}U(\mathbf{X})$. The collection of
central configurations is clearly invariant under similarity
transformations. Thus we may fix a scaling of the vectors, say $I=1$, and
ask for solutions modulo $O(3)$-congruence. Then the solutions are just the
critical points of $U$ as a function restricted to the shape space $%
M_{n}^{\ast }$. For $n=3$ the only solution which is the shape of a
non-degenerate triangle is the equilateral triangle (by Lagrange, 1772),
whereas there are three degenerate triangle solutions (by Euler, 1767) and
they are represented by the three \emph{Euler points} on the equator circle
of $M_{3}^{\ast }$.

Even the enumeration of all critical points in $M_{n}^{\ast }$ is, in fact,
still an open problem for $n\geq 4$, and now the number also depends on the
mass distribution. It is only known to be finite for $n<5$. However, the
number of collinear solutions is known to be $n!/2$ for all $n\geq 3$, and
the first proof was presented by Moulton \cite{Moulton}. More recently,
Smale \cite{Smale} has given a topological proof using elementary Morse
theory. In this subsection we shall give a similar and quite simple proof
using the weighted root system of an n-body system.

As usual, we assume that $U$ in (\ref{U}) is a function of n vector
variables\ $\mathbf{a}_{i}$ linearly related by the condition that the
center of mass is at the origin. Actually this condition is a consequence of
the identities (\ref{central1}). Anyhow, let us express $U$ as a function of
the n-1 Jacobi vectors 
\begin{equation*}
X=(\mathbf{x}_{1},\mathbf{x}_{2},...,\mathbf{x}_{n-1})=\left[ \mathbf{x}%
_{1}^{\ast},\mathbf{x}_{2}^{\ast},\mathbf{x}_{3}^{\ast}\right]
\end{equation*}
defined by a fixed Jacobi transformation 
\begin{equation*}
\Psi:M_{n}\rightarrow M(3,n-1)
\end{equation*}
with associated root system $\left\{ \pm\mathbf{w}_{ij}\right\} $. Take, for
example, the standard transformation $\Psi_{0}$ and the expressions for the
vectors $\mathbf{x}_{i}$ and $\mathbf{w}_{ij}$ in Section 2.3 and Section
5.2. The (row) vectors $\mathbf{w}_{ij}$ and $\mathbf{x}_{k}^{\ast}$ belong
to the same Euclidean space $\mathbb{R}^{n-1}$ and by definition \ \ 
\begin{equation*}
\left\vert \mathbf{a}_{i}-\mathbf{a}_{j}\right\vert ^{2}=\ \sum_{k=1}^{3}(%
\mathbf{w}_{ij}\cdot\mathbf{x}_{k}^{\ast})^{2}\text{, \ \ \ }\mathbf{w}_{ij}=%
\frac{1}{\sqrt{\mu_{ij}}}\mathbf{u}_{ij}\text{\ \ \ \ (cf. Section 5.1)}
\end{equation*}

The gradient field $\nabla U$ is tangential to the subvariety of collinear
configurations and therefore the critical points we seek are also the
critical points of $U$ restricted to the subvariety of collinear shapes,
namely $M^{\ast }(1,n-1)\simeq RP^{n-2}$. Let us represent all shapes of
collinear type by n-configurations $\mathbf{X=(a}_{1},\mathbf{a}_{2},..,%
\mathbf{a}_{n})\ $in $M_{n}$ with position vectors $\mathbf{a}_{i}$ along
the x-axis, and therefore the Jacobi matrix $X$ has row vectors $\mathbf{x}%
_{2}^{\ast }=\mathbf{x}_{3}^{\ast }=0$. Hence, we shall regard $U$ as a
function on the (n -1)-space 
\begin{equation*}
\ \mathbb{R}^{n-1}:\mathbf{x}=\mathbf{x}_{1}^{\ast }=(x_{1},...,x_{n-1})%
\text{ }\ 
\end{equation*}%
and the condition $I=1$ means restriction to the unit sphere $S^{n-2}:\tsum
x_{i}^{2}=1$. When antipodal points on this sphere are identified, we obtain
the above projective space $RP^{n-2}$, see (\ref{antipod}) below.

For convenience, we shall identify the root vector $\mathbf{w}_{ij}$ with
the linear functional 
\begin{equation*}
\omega _{ij}:\mathbb{R}^{n-1}\rightarrow \mathbb{R}\text{ , \ \ }\omega
_{ij}(\mathbf{x)=w}_{ij}\cdot \mathbf{x}\text{, \ }
\end{equation*}%
and regard the family 
\begin{equation*}
\Delta =\left\{ \pm \omega _{ij}\right\} =\Delta _{+}\cup \Delta _{-}\ 
\end{equation*}%
$\ $as the \emph{root system}.\emph{\ }The \emph{positive }and \emph{simple }%
roots are the subfamilies 
\begin{equation*}
\Delta _{+}=\left\{ \omega _{ij},\text{ }i<j\right\} \text{ }\supset \text{ }%
\Pi =\left\{ \omega _{12},\omega _{23},,,\omega _{n-1,n}\right\} \text{,}
\end{equation*}%
respectively, with $\omega _{ij}=-\omega _{ji}$ and $\omega _{ij}+\omega
_{jk}=\omega _{ik}$, and following the \textquotedblright
usual\textquotedblright\ procedure we define hyperplanes 
\begin{equation*}
H_{ij}=\ker \omega _{ij}\subset \mathbb{R}^{n-1}
\end{equation*}%
which subdivide the space into disjoint, open and connected chambers $C_{i}$
whose union is 
\begin{equation*}
C_{1}\cup C_{2}\cup ....=\mathbb{R}^{n-1}-\cup _{i<j}H_{ij}\text{ .}
\end{equation*}%
Clearly, there are $n!$ chambers, and each chamber $C$ is distinguished by a
specific choice of signs $\varepsilon _{ij}=\pm 1$ such that 
\begin{equation*}
\mathbf{x\in }C\text{ }\Leftrightarrow \varepsilon _{ij}\omega _{ij}(\mathbf{%
x)}>0\text{ }\ \text{for all }i<j
\end{equation*}

By changing all signs one obtains the antipodal chamber. In particular, our 
\emph{fundamental} \emph{chamber} $C_{0}$ is the one with all $\varepsilon
_{ij}$ positive, namely 
\begin{equation*}
C_{0}\ :\omega _{ij}(\mathbf{x}\ )>0\text{ \ for all }i<j,
\end{equation*}%
or equivalently in terms of the positions $\mathbf{a}_{i}=a_{i}$ of the
point masses (on the x-axis), since $\omega _{ij}(\mathbf{x}\ )=a_{i}-a_{j},$
\begin{equation}
a_{1}>a_{2}>....>a_{n}  \label{ordering}
\end{equation}%
The above decomposition and combinatorial structure is, of course, similar
to the Weyl chamber decomposition for the Weyl group $A_{n-1}=S_{n}$ of $\
SU(n).$

By (\ref{distance}) and (\ref{delta/ij}), the expression (\ref{U}) may be
written 
\begin{equation}
U(\mathbf{x)=}\sum_{i<j}\frac{m_{i}m_{j}}{\left\vert \omega_{ij}(\mathbf{x)}%
\right\vert }\text{ \ }  \label{Ured}
\end{equation}
which in the fundamental chamber reads 
\begin{equation*}
U(\mathbf{x)=}\sum_{i<j}\frac{m_{i}m_{j}}{\omega_{ij}(\mathbf{x)}}\text{ \ \
for }\mathbf{x}\in C_{0}
\end{equation*}

\begin{lemma}
$U(\mathbf{x)}$ is a convex function on the chamber $C_{0}$, and it has a
unique critical point $\mathbf{x}_{0\text{ }}$(where $U$ has a minimum) in
the spherical chamber $C_{0}\cap S^{n-2}.$
\end{lemma}

\begin{proof}
By straightforward calculations 
\begin{align*}
\frac{\partial U}{\partial x_{k}} & =-\sum_{i<j}m_{i}m_{j}\frac{w_{ij}^{k}}{%
\omega_{ij}(\mathbf{x})^{2}} \\
\frac{\partial^{2}U}{\partial x_{k}\partial x_{l}} & =2\sum_{i<j}m_{i}m_{j}%
\frac{w_{ij}^{k}w_{ij}^{l}}{\omega_{ij}(\mathbf{x})^{3}}
\end{align*}
where\textbf{\ }$\mathbf{w}_{ij}=(w_{ij}^{1},w_{ij}^{2},...,w_{ij}^{n-1})$
and $\mathbf{x=(}x_{1},x_{2},,,x_{n-1})$. Hence, the Hessian of $U$ at $%
\mathbf{x}\in C_{0}$ is the following positive definite quadratic form in
the variable $\mathbf{t=(}t_{1},t_{2},...,t_{n-1})$%
\begin{equation*}
HU(\mathbf{x})(\mathbf{t)=}\frac{1}{2}\sum_{k,l=1}^{n-1}\frac{\partial^{2}U}{%
\partial x_{k}\partial x_{l}}(\mathbf{x)}t_{k}t_{l}=\sum_{i<j}m_{i}m_{j}%
\frac{\omega_{ij}(\mathbf{t)}^{2}}{\omega_{ij}(\mathbf{x})^{3}}
\end{equation*}

On the other hand, $U(\mathbf{x)}>0$ on $C_{0}$ and $U(\mathbf{x)\rightarrow
\infty }$ \ as $\mathbf{x}$ approaches the walls of the chamber and is
bounded away from the origin. It follows that $U$ has a unique critical
point in the spherical chamber, namely a minimumspoint $\mathbf{x}_{0}$.
\end{proof}

The induced congruence group\ action on $\mathbb{R}^{n-1}$ is simply the
group $O(1)=\left\{ \pm 1\right\} $ acting by inversion $\mathbf{%
x\rightarrow -x}$, and therefore antipodal points on the sphere $S^{n-2}$
represent collinear n-configurations with the same shape : 
\begin{equation}
S^{n-2}\rightarrow S^{n-2}/O(1)=\mathbb{R}P^{n-2}\subset M_{n}^{\ast }
\label{antipod}
\end{equation}%
In particular, a pairs of antipodal chambers is mapped to the same chamber
in $\mathbb{R}P^{n-2}$, and consequently the latter space is divided into $%
n!/2$ chambers, each containing a unique critical point of $U$ as a function
on $\mathbb{R}P^{n-2}$. This completes the proof of the enumeration result
originally due to Moulton.

\begin{remark}
The above central configuration solutions $(a_{1,}a_{2},...,a_{n})$ can be
distinguished by the ordering $a_{i_{1}}>a_{i_{2}}>....>a_{i_{n}}$ (modulo
inversion of order), and the minimumspoint $\mathbf{x}_{0\text{ }}$in the
fundamental chamber $C_{0}$ is the solution with the ordering (\ref{ordering}%
). However, for a given mass distribution $(m_{1},m_{2},...,m_{n})$ the
solution satisfying $a_{i_{1}}>a_{i_{2}}>....>a_{i_{n}}$ can also be found
by the same procedure, namely as the minimumspoint $\mathbf{x}_{0\text{ }}$%
in the fundamental chamber $C_{0}\ $corresponding to the permuted mass
distribution $(m_{i_{1}},m_{i_{2}},...,m_{i_{n}})$. In particular, for each
string of equal masses $m_{i}=m_{j}=..=m_{k}$ the set of solutions is
invariant under permutations of $a_{i},a_{j},..,a_{k}$.
\end{remark}

Let us briefly consider the explicit numerical calculation of the collinear
central configurations by calculating the critical points of $U$ on $S^{n-2}$%
. Observe that the identity (\ref{central1})\ with the additional condition $%
I=1$ may be interpreted as an application of the classical Lagrange
multiplier method with a constraint. Similarly, in the present case where we
restrict to collinear configurations, we seek the $n!$ solutions of the
system 
\begin{equation}
\nabla U(\mathbf{x)=}\text{ }\lambda \mathbf{x}\text{ , \ }\left\vert 
\mathbf{x}\right\vert =1  \label{central3}
\end{equation}%
where $\lambda =-U(\mathbf{x)}$ (by Euler's formula) and 
\begin{equation*}
\nabla U(\mathbf{x)=}-\sum_{i<j}m_{i}m_{j}\frac{\mathbf{w}_{ij}}{\omega
_{ij}(\mathbf{x)}^{2}}
\end{equation*}%
The condition for a solution is the vanishing of the component of the
gradient tangential to the sphere, namely 
\begin{equation}
\nabla U(\mathbf{x)\ +\ }U(\mathbf{x})\mathbf{x}=0  \label{tangcomp}
\end{equation}%
As a simpleminded algorithm for finding a solution we construct a sequence
of points $\mathbf{y}_{k}\in S^{n-2}$ by \textquotedblright
moving\textquotedblright\ on the sphere in the direction opposite to the
vector (\ref{tangcomp}), as follows 
\begin{equation*}
\mathbf{y}_{k+1}=\frac{(1-U(\mathbf{y}_{k}))\mathbf{y}_{k}-\nabla U(\mathbf{y%
}_{k}\mathbf{)}}{\left\vert (1-U(\mathbf{y}_{k}))\mathbf{y}_{k}-\nabla U(%
\mathbf{y}_{k}\mathbf{)}\right\vert },
\end{equation*}%
starting from an initial point $\mathbf{y}_{1}$ in a given chamber. In
general, the sequence will stay in the chamber and converge to the
minimumspoint of $U.$ The algorithm can certainly be made more effictive,
but we shall not discuss these matters.

Another approach is to introduce new variables $t_{i}$ linearly related to $%
x_{1},x_{2},.,x_{n-1}$ via the simple roots. 
\begin{equation*}
t_{i}=\omega _{i,i+1}(\mathbf{x)}=a_{i}-a_{i+1},\text{ \ }i=1,2,...,n-1\ 
\end{equation*}%
For example, in the case $n=3$ the system (\ref{central3}) reads 
\begin{align}
0& =-\frac{m_{1}}{(t_{1})^{2}}+\frac{1-m_{1}}{\ (t_{2})^{2}}+\frac{m_{1}}{%
(t_{1}+t_{2})^{2}}-\lambda t_{2}\   \label{central4} \\
0& =\frac{m_{2}}{(t_{1})^{2}}+\frac{m_{3}}{(t_{1}+t_{2})^{2}}-\lambda
\lbrack (1-m_{1})t_{1}+m_{3}t_{2}]  \notag
\end{align}%
as compared to the initial system (\ref{central1})\ which has three
dependent equations similar to the above ones. In fact, the first equation
of system (\ref{central1}) and (\ref{central4}) are identical. By
elimination of $\lambda $ we obtain a homogeneous equation of degree 5 which
has a unique solution (modulo scaling) with both $t_{i}>0$. Thus it can be
reduced to an inhomogeneous equation of degree 4, whereas Euler's approach
led to the solution of a 5th order inhomogeneous equation $P(\omega )=0$ (cf.%
\cite{Siegel}, \S 14). The latter equation is, in fact, obtained from the
system (\ref{central4}) when we write $(t_{1}+t_{2})=t,$ $t_{1}=\omega
t,t_{2}=(1-\omega )t$ and eliminate the resulting multiplier $\lambda
^{\prime }=$ $\lambda t^{3}$.

\end{document}